\documentclass[12pt, a4paper]{article}
\usepackage{amsfonts, amssymb}
\usepackage{graphicx}
\usepackage{latexsym,amsmath,color,array}

\usepackage{natbib}
\usepackage{enumitem}

\usepackage{bbding}
\usepackage{amssymb}
\usepackage{amsmath}
\usepackage{graphicx}
\usepackage{amsmath}
\usepackage{bbm}
\usepackage{mathtools}
 \usepackage{color}
 \usepackage{array}
 \usepackage{multirow}
 \usepackage{makecell}
 \usepackage{colortbl}
 
 \usepackage{subcaption}
\usepackage{booktabs}
\usepackage{algorithm}
\usepackage{textcomp}
\usepackage{appendix}
\usepackage{xcolor}
\definecolor{mygreen}{rgb}{0.13, 0.55, 0.13} 
\definecolor{mycodebrown}{rgb}{0.64, 0.08, 0.08} 
\definecolor{mycodegreen}{rgb}{0.00, 0.50, 0.00} 
\definecolor{mycodeblue}{rgb}{0.00, 0.00, 1.00} 

\usepackage{relsize} 
\usepackage{placeins}
\usepackage{setspace}
\usepackage{afterpage}

\def\1{\mathbb{I}}

\renewcommand{\thesection}{\arabic{section}}

\newcounter{thm}[section]
\newcounter{appen}[section]
\newcounter{assum}[section]

\begin{document}

\title{Variable Selection Using a Smooth Information Criterion for Distributional Regression Models}
\author{Meadhbh O'Neill\footnote{Department of Mathematics and Statistics, University of Limerick, Limerick, Ireland} \hspace{2cm}
Kevin Burke\footnotemark[1]}
\date{}

\maketitle
\begin{abstract}
Modern variable selection procedures make use of penalization methods to execute simultaneous model selection and estimation. A popular method is the LASSO (least absolute shrinkage and selection operator), the use of which requires selecting the value of a tuning parameter. This parameter is typically tuned by minimizing the cross-validation error or Bayesian information criterion (BIC) but this can be computationally intensive as it involves fitting an array of different models and selecting the best one. In contrast with this standard approach, we have developed a procedure based on the so-called ``smooth IC" (SIC) in which the tuning parameter is automatically selected in one step. We also extend this model selection procedure to the distributional regression framework, which is more flexible than classical regression modelling. Distributional regression, also known as multiparameter regression (MPR), introduces flexibility by taking account of the effect of covariates through multiple distributional parameters simultaneously, e.g., mean and variance. These models are useful in the context of normal linear regression when the process under study exhibits heteroscedastic behaviour. Reformulating the distributional regression estimation problem in terms of penalized likelihood enables us to take advantage of the close relationship between model selection criteria and penalization. Utilizing the SIC is computationally advantageous, as it obviates the issue of having to choose multiple tuning parameters.

\smallskip


\end{abstract}

\qquad

\newpage

\section{Introduction} \label{sec:introduction}
Enhancements in data collection technologies have highlighted the importance of modern variable selection techniques. Traditional methods, such as best subset selection, are suboptimal and are computationally expensive when the number of variables is high \citep{fan10}. Modern approaches make use of penalization methods to execute simultaneous model selection and estimation. A popular method is the LASSO (least absolute shrinkage and selection operator) \citep{tibshirani96}, which comprises of an $L_{1}$ penalty but leads to biased estimates. In contrast, the adaptive LASSO \citep{zou06} has adaptive weights, which reduce the bias present in the LASSO estimates. These methods have been developed primarily in the context of normal linear regression and have been extended to generalized linear models (GLMs) \citep{tibshirani96, friedman10} and Cox's proportional hazard models for survival data \citep{tibshirani97}. In these classical models, covariates enter through the location parameter (or the hazard scale in the Cox model case). A more modern and flexible approach is to include covariates in multiple distributional parameters, such as the location and dispersion, simultaneously; this approach is known as ``distributional regression" \citep{stas18gamlss} and ``multiparameter regression" (MPR) \citep{burke20}. The goal of this paper is to expand penalized regression to the flexible MPR setting using a novel differentiable $L_{0}$ penalty that does not require tuning parameter selection, which is especially appealing in the MPR setting where one would typically require multiple tuning parameters (one for each distributional parameter).

Originally, methods such as quadratic programming were used to solve these non-differentiable LASSO-type problems, but \citet{efron04} and \citet{friedman07}, respectively, proposed the least angle regression (LARS) and co-ordinate descent algorithms --- with the latter proving to be particularly fast for problems of this type. These are somewhat ``non-standard" estimation procedures in the context of classical statistical estimation, where non-differentiable objective functions are relatively less common. As an alternative non-gradient based optimization, perturbing the penalty function slightly to render it differentiable \citep{hunter05, lloyd18} enables standard optimization methods to be used. \citet{oelker17} outline a series of approximations of different penalties, which allows for penalized smooth functions. These differentiable penalties can be easily implemented and solved using standard gradient based optimization procedures, i.e., Newton-Raphson. The tuning parameter that controls the strength of the penalty is typically obtained by minimizing the cross-validation error or an information criterion (IC), such as the Akaike IC (AIC) \citep{akaike74} or Bayesian IC (BIC) \citep{schwarz78}. This is a two-step estimation process, which tends to be computationally intensive as it involves fitting an array of different models and selecting the best one. 

\citet{su15} and \citet{su18} present an estimation procedure that is not based on the $L_{1}$ norm, titled ``MIC" (minimum approximated information criterion). They exploit the close connection between model selection criteria and penalization \citep{fan10} and introduce an approximated information criteria in order to avoid the classic two-step estimation process. At its core, the MIC utilizes an approximation of the ``$L_{0}$ norm" with a continuous unit dent function. The $L_0$ norm is discrete in nature and it is preferable to have a penalty function with a level of smoothness for optimization purposes. \citet{su15} describes a ``subtle uprooting" method for variable selection, which involves using a smooth surrogate function for approximating cardinality. This is followed by a second technical step for enhancing sparsity, where the final problem becomes non-differentiable. This approach is extended to GLMs in \citet{su18}. Fixing the tuning parameter at two for the AIC or $\log(n)$ for the BIC is computationally advantageous, as it avoids the tuning parameter selection problem. It is not required to compute the whole regularization path of solutions, nor is it necessary to choose the best tuning parameter using cross-validation, as is typically done.

We propose a more straightforward method of approximating the IC function using a smooth approximation of the $L_{0}$ norm, which can be optimized directly. Instead of performing the reparameterization step as outlined in \citet{su15}, which renders the problem non-differentiable, we achieve sparsity in a different way. Our approach squeezes the coefficient values to zero by optimizing a sequence of objective functions that get successively closer to the non-differentiable one. Consequently, our proposed ``smooth IC" (SIC) function can be optimized directly using standard gradient based optimization techniques. Additionally, we extend this new SIC variable selection procedure for use in the developing area of distributional regression \citep{stas18gamlss}. Our proposed methods are implemented in our publicly available R package ``\texttt{smoothic}" \citep{smoothic_package}. To date, penalized estimation has been primarily applied in the context of classical regression models, where the covariates are allowed to enter the model through a single parameter (e.g.,~a location parameter). Other distributional parameters, such as a dispersion parameter, are typically constant. This ``single parameter regression" (SPR) does not take into account the possible impact of covariates on the other distributional parameters. Distributional regression, which is also referred to as ``multiparameter regression" (MPR), is a more flexible approach where multiple parameters are regressed simultaneously on covariates. For example, covariates can enter the model through the location and dispersion parameters, or scale and shape parameters of the hazard function in the survival context (see \citet{burke17}, \citet{burke20} and references therein), or indeed in various different distributional parameters as in generalized additive models for location, scale and shape (GAMLSS) \citep{rigby05}. \citet{mayr12} address the problem of variable selection by utilizing classical gradient boosting techniques to fit GAMLSS models. More recently, \citet{groll19} suggest implementing a LASSO-type penalization in the GAMLSS framework. This regularization approach to GAMLSS is highly flexible, but it has the added complexity of separate tuning parameters for each regression component. \citeauthor{groll19} state that carrying out the computationally demanding grid search for the optimal tuning parameters is a drawback of their method. In our proposed multiparameter regression with smooth IC (MPR-SIC) procedure, this issue is circumvented as the values of both tuning parameters are known in advance.

The model formulation, including the introduction of the ``smooth $L_0$ norm", the estimation procedure and the optimization algorithm are outlined in Section~\ref{sec:model_formulation}. In Section~\ref{sec:simulation_studies}, the performance of our proposed methods is evaluated in both variable selection and parameter estimation through extensive simulation studies. We consider three real data analyses to demonstrate our proposed methods in Section~\ref{sec:real_data}. Finally, we close with some concluding remarks in Section~\ref{sec:discussion}.

\section{Model Formulation} \label{sec:model_formulation}
\subsection{Preliminaries} \label{sec:preliminaries}
The classic normal linear regression is a single parameter problem that assumes there is constant variance in the errors. The model is
\begin{equation}
    y_i={x_{i}^{T}\beta}+\varepsilon_i
    \label{eq:ols}
\end{equation}
for $i = 1, \ldots,n$, where $y_i$ is the response value and ${x_i} = (1, x_{1i},\ldots,x_{pi})^T$ is a vector of covariates for the $i$th individual over the predictor variables $j=0,1,\ldots,p$. Here, ${\beta}=(\beta_{0},\beta_{1},\ldots,\beta_{p})^{T}$ is the vector of regression coefficients for the location parameter and $\varepsilon_i \sim \text{N}(0, \sigma^2)$ under the homogeneity assumption. For the multiparameter regression (MPR) approach, where covariates appear in multiple distributional parameters simultaneously, the single parameter model in \eqref{eq:ols} is extended to include heterogeneity of error variance:
\begin{equation}
    \text{Var}(\varepsilon_i)=\sigma_i^2=e^{{x_{i}^T\alpha}},
    \label{eq:mpr_err}
\end{equation}
where the log-linear form ensures that $\sigma^2_i$ remains positive. The vector of regression coefficients for the dispersion parameter is ${\alpha}=(\alpha_{0},\alpha_{1},\ldots,\alpha_{p})^{T}$. There may be different (possibly overlapping) sets of covariates impacting the location and dispersion, and although we use $x_i$ for both, a given $\beta$ or $\alpha$ coefficient may be set to zero, which removes the covariate from that model component. Because we apply penalized variable selection, the regression coefficients need to be on a similar scale, and therefore we assume that the predictors are scaled to have unit variance.

The log-likelihood function for the MPR normal model is
\begin{equation}
    \ell(\theta)=-\frac{n}{2}\log(2\pi)-\frac{1}{2}\sum_{i=1}^n{x_{i}^T\alpha}-\frac{1}{2}\sum_{i=1}^{n}e^{-{x_{i}^T\alpha}}(y_i-{x_{i}}^{T}{\beta})^2,
    \label{eq:mpr_normal}
\end{equation}
where $\theta=(\beta^T,\alpha^T)^T=(\beta_0,\ldots,\beta_{p},\alpha_0,\ldots,\alpha_{p})^T$. Our focus is on variable selection in the location and dispersion components, and we note that model selection criteria, such as the AIC and BIC, have a penalized functional form similar to regularization. In the distributional regression framework, an information criterion (IC) can be formulated as
\begin{equation}
    \mathrm{IC}=-2\ell(\theta)+\lambda\big[{\lvert \lvert\tilde\beta\rvert\rvert}_0 + \lvert\lvert\tilde\alpha\rvert\rvert_0 + 2\big],
    \label{eq:IC}
\end{equation}
where $\lambda$ is fixed at $\lambda=2$ or $\lambda=\log(n)$ for the AIC and BIC respectively, and $\tilde\beta=(\beta_1,\ldots,\beta_{p})^T$ and $\tilde\alpha=(\alpha_1,\ldots,\alpha_{p})^T$, i.e., the coefficient vectors with the intercepts omitted; there is an addition of two in the penalty to take into account the estimation of the intercept terms $\beta_0$ and $\alpha_0$. The $L_{0}$ norm,  $\lvert \lvert \theta \rvert \rvert_0 = \mathrm{card}(\theta) = \sum_{j=1}^p \mathit{I} (\theta_{j} \neq 0)$, indicates the cardinality or the number of non-zero elements in $\theta$. This is not truly a norm since $\lvert \lvert c \theta \rvert \rvert_0 \neq c\lvert \lvert \theta \rvert \rvert_0$ when $c \neq 0,1$. The AIC is reported to be asymptotically ``selection inconsistent" and ``loss-efficient" as a variable selection criterion \citep{shao97, yang05, wang09}. As a result of its consistency property and superior empirical performance in variable selection, we employ a BIC-type criterion \citep{wangleng07} where $\lambda=\log(n)$.

Using the likelihood in \eqref{eq:mpr_normal} and arranging \eqref{eq:IC} as an IC-based penalized likelihood results in
 \begin{equation}
    \ell^\text{IC}(\theta)=\ell(\theta)-\frac{ \log(n)}{2}\big[\lvert\lvert\tilde\beta\rvert\rvert_0+\lvert\lvert\tilde\alpha\rvert\rvert_0+2\big].
    \label{eq:smooth_ic_nondiff}
\end{equation}
To enable gradient-based optimization, we define $\lvert \lvert \theta \rvert \rvert_{0, \epsilon} = \sum_{j=1}^{p}\phi_\epsilon(\theta_j)$ as the ``smooth $L_0$ norm", and substitute the $L_0$ norm in \eqref{eq:smooth_ic_nondiff} with $\lvert \lvert \tilde\beta \rvert \rvert_{0, \epsilon}$ and $\lvert \lvert \tilde\alpha \rvert \rvert_{0, \epsilon}$. This results in our proposed approach of MPR with smooth IC (MPR-SIC), which is the maximization of
 \begin{equation}
    \ell^\text{SIC}(\theta)=\ell(\theta)-\frac{ \log(n)}{2}\big[\lvert \lvert \tilde\beta \rvert \rvert_{0, \epsilon}+ \lvert \lvert \tilde\alpha \rvert \rvert_{0, \epsilon}+2\big].
    \label{eq:smooth_ic}
\end{equation}
Therefore, since BIC minimization is intrinsic to this formulation, it obviates the usual need for estimating the model at a range of tuning parameter grid points and then evaluating each of these using an external BIC in a second step. Avoiding this grid search is especially useful in the context of distributional regression. For the more commonly used $L_1$ norm, there is no direct link to the BIC, in which case one must search for the optimal tuning parameter. Moreover, one would typically have a separate tuning parameter for each distributional parameter to account for differing scales in these parameters, and this multidimensional grid search optimization is quite computationally intensive. In contrast, the BIC penalizes all parameters equally: it is $\log(n)$ for all non-zero parameters, irrespective of their size or distributional type (e.g.,~location or dispersion), and it is zero for zero parameters.

\begin{figure}[b!]
\centering
\makebox{\includegraphics[width = 0.8\textwidth]{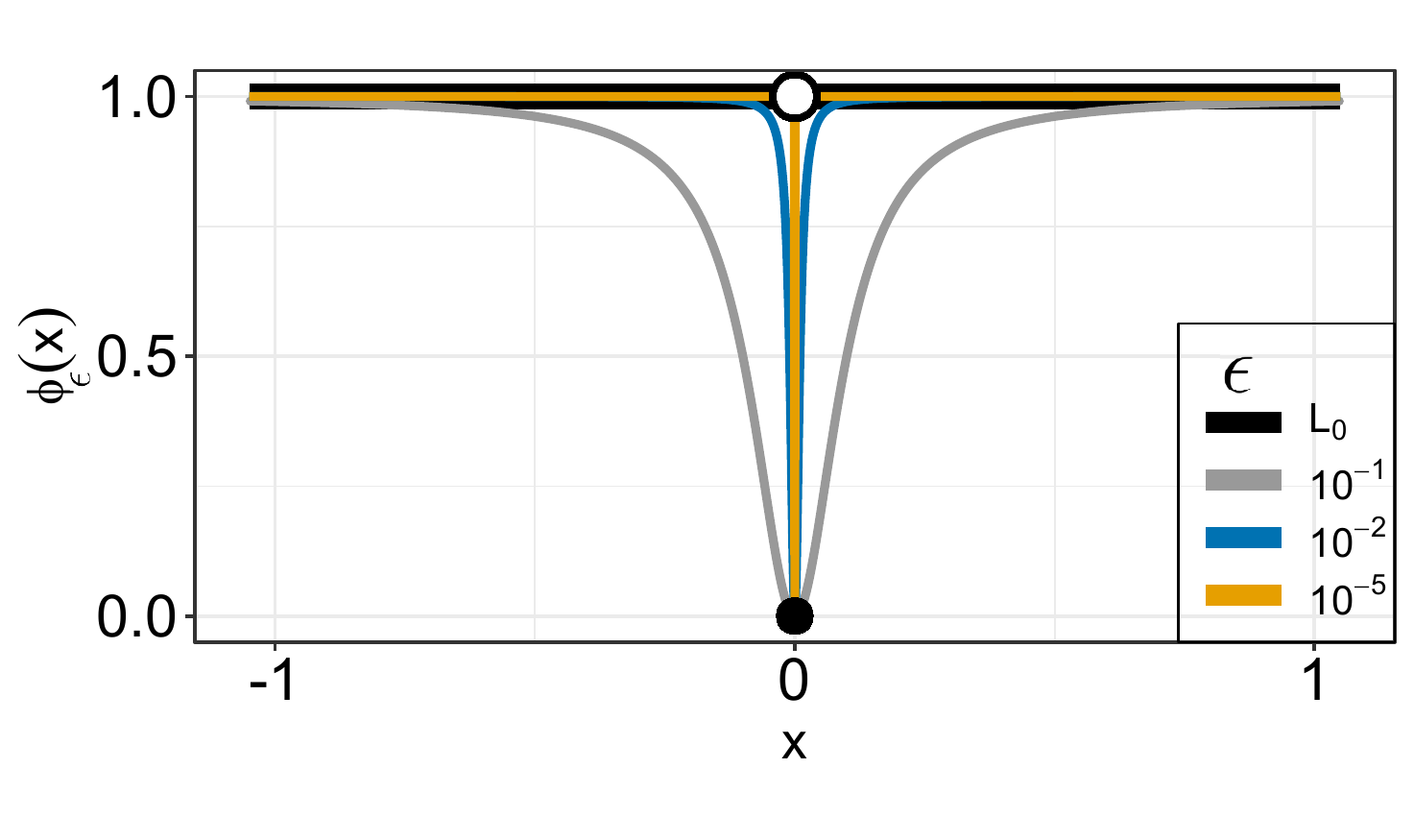}}
\caption{\label{fig:smooth_L0_norm}Smooth $L_0$ norm.}
\end{figure}

\subsection{Smooth $L_0$ Norm} \label{sec:L_0 norm}
Due to the non-differentiability of the $L_0$ norm, we propose a smooth function to approximate it:
\begin{equation}
    \phi_\epsilon(x) = \frac{x^2}{x^2 + \epsilon^2}.
    \label{eq:smooth_L0_norm}
\end{equation}
This is differentiable for $\epsilon>0$ and $\lim_{\epsilon\to 0} \phi_\epsilon(x) = \lvert \lvert x \rvert \rvert_0$. Figure~\ref{fig:smooth_L0_norm} demonstrates how $\phi_\epsilon(x)$ gets closer to $\lvert \lvert x \rvert \rvert_0$ as $\epsilon$ decreases. The smallest value shown ($\epsilon=10^{-5}$) approximates the $L_0$ norm very closely, but it is also near the discontinuity at $x=0$ making it unstable. Ultimately, \eqref{eq:smooth_L0_norm} requires a small $\epsilon$ value to produce shrinkage, but we have found that simply fixing it to a small value from the offset yields poor results due to its closeness to the discontinuity (see the Supplementary Material). Therefore, to create a more stable problem, we recommend the use of a decreasing sequence of $\epsilon$ values (described in Section~\ref{sec:telescoping}). Interestingly, with a fixed ``large'' value of $\epsilon = 1$, this penalty has been referred to as a ``weight elimination penalty'' in the context of neural networks \citep{rumelhart91generalization}. It is noteworthy that \citet{oelker17} develop a general penalized estimation procedure based on smooth approximations to penalties. Within this framework, they consider an $L_0$-norm approximation that is slightly less straightforward than ours, since it is based on a logistic function with two smoothing parameters. \citet{devriendt21} provide an alternative estimation procedure to \citet{oelker17}, which is exact rather than approximate, but which does not include the $L_0$ penalty (albeit they suggest adapting to a stochastic algorithm could potentially handle this). Crucially, however, both of these approaches require a grid search to find the optimal tuning parameter, but this is avoided in our work due to the connection to an information criterion established in Section~\ref{sec:preliminaries}.

Note that the first and second derivatives (for $j>0$) have a simple analytic form and therefore can be used within the gradient based optimization procedure of Section~\ref{sec:estimation_procedure}:
\begin{equation}
    \phi_{\epsilon}^{\prime}(x_{j})= \frac{2x_{j}\epsilon^2}{(x_{j}^2+\epsilon^2)^2}, \qquad \qquad \phi_{\epsilon}^{\prime \prime}(x_{j})=\frac{2\epsilon^2(\epsilon^2-3x_{j}^2)}{(x_{j}^2+\epsilon^2)^3}.
\end{equation}
\FloatBarrier
\subsection{Estimation Procedure} \label{sec:estimation_procedure}
We define the penalized estimates as
\begin{equation*}
    \hat{\theta} = {\arg\max}\big(\ell^{\text{SIC}}(\theta)\big),
\end{equation*}
where $\ell^{\text{SIC}}(\theta)$ is given by \eqref{eq:smooth_ic}. The first and second derivatives with respect to the parameters are
\begin{align} \label{eq:sic_derivs}
            	\begin{split}	
            	\frac{\partial \ell^{\text{SIC}}}{\partial \beta} &= \frac{\partial \ell}{\partial \beta} - \frac{ \log(n)}{2}\nu_{\beta} = X^T z_{\beta} - \frac{ \log(n)}{2}\nu_{\beta},\\
            	\frac{\partial \ell^{\text{SIC}}}{\partial \alpha} &= \frac{\partial \ell}{\partial \alpha} - \frac{ \log(n)}{2}\nu_{\alpha} = X^T z_{\alpha} - \frac{ \log(n)}{2}\nu_{\alpha},
            	\end{split}
         \end{align}
where $X$ is an $n \times (p+1)$ matrix, whose $i$th row is $x_i$, $z_\beta$ and $z_\alpha$ are vectors of length $n$ such that $z_{\beta,i}=e^{-x_{i}^T\alpha}(y_i-x_{i}^T\beta)$ and $z_{\alpha,i}=\big({e^{{-x_{i}^T\alpha}}}{(y_{i}-x_{i}^T\beta)^2}-1\big)/2$, and $\nu_\beta$ and $\nu_\alpha$ are vectors whose $(j+1)$th elements are given by $\phi_\epsilon^{\prime}(\beta_{j})$ and $\phi_\epsilon^{\prime}(\alpha_{j})$ respectively, except whose first elements are zero due to the fact that the intercepts are not penalized.

The matrix of negative second derivatives of $\ell^{\text{SIC}}(\theta)$, i.e., $-\nabla_{\theta}\nabla_{\theta}^T\ell^{\text{SIC}}(\theta)$ is given by
\begin{align*}
I(\theta)
&= I_0(\theta) +
\begin{pmatrix}
             {\log(n)} \Sigma_{\beta}/2
             & 0 \\
             0 & {\log(n)}\Sigma_{\alpha}/2
         \end{pmatrix}\\
&=   \begin{pmatrix}
             X^{T} W_{\beta} X + {\log(n)}\Sigma_{\beta}/2
             & X^{T} W_{\alpha\beta} X \\
             X^{T} W_{\alpha\beta} X & X^{T} W_{\alpha} X + {\log(n)}\Sigma_{\alpha}/2
         \end{pmatrix}
\end{align*}
where $I_0(\theta) = -\nabla_{\theta}\nabla_{\theta}^T\ell(\theta)$ is the observed information matrix of the unpenalized likelihood; $\Sigma_\beta$ and $\Sigma_\alpha$ are diagonal matrices that appear due to the penalties and whose $(j+1)$th diagonal elements are given by $\phi_\epsilon^{\prime\prime}(\beta_{j})$ and $\phi_\epsilon^{\prime\prime}(\alpha_{j})$ respectively, except whose first diagonal elements are zero due to the fact that the intercepts are not penalized; $W_\beta$, $W_\alpha$ and $W_{\beta, \alpha}$ are $n \times n$ diagonal weight matrices whose $i$th diagonal elements are given by $e^{-x_{i}^T\alpha}$, $e^{-x_{i}^T\alpha}(y_i-x_{i}^T\beta)^2/2$ and $e^{-x_{i}^T\alpha}(y_i-x_{i}^T\beta)$ respectively. We employ the ``RS" algorithm \citep{rigby05}, which does not use cross derivatives. This algorithm is motivated by the fact that in many classical models, including location and scale models, the parameters are information orthogonal as discussed in \citet{cox87}. However, \citet{stas07} report that the RS algorithm works well even when the parameters are not information orthogonal.

The resulting system of Newton-Raphson equations can be expressed compactly as
\begin{multline}
\hspace{-0.35cm}
			\begin{pmatrix}
             X^{T} W_{\beta}^{(m)} X + {\log(n)}\Sigma_{\beta}^{(m)}/2
             & 0 \\
             0 & X^{T} W_{\alpha}^{(m)} X + {\log(n)}\Sigma_{\alpha}^{(m)}/2
         \end{pmatrix}
         {
         \begin{pmatrix}
             \beta^{(m+1)} - \beta^{(m)}\\
             \alpha^{(m+1)} - \alpha^{(m)}
         \end{pmatrix}}
         \\=
         \begin{pmatrix}
             X^T z_{\beta}^{(m)} - {\log(n)}\nu_{\beta}^{(m)}/2\\
             X^T z_{\alpha}^{(m)} - {\log(n)}\nu_{\alpha}^{(m)}/2
     \end{pmatrix}.
     \label{eq:nr_eq}
\end{multline}

They are iteratively solved for {$\theta^{(m+1)} = ({\beta^{(m+1)}}^T, {\alpha^{(m+1)}}^T)^T$}, where the elements super-scripted by $(m)$ depend on $\theta^{m}$, but this is excluded for notational convenience. Note that, since the RS algorithm sets the off-diagonal blocks to zero, it is possible to optimize the problem by alternating between the estimation of the mean and variance models; however, this is not considered here. We use the classical ordinary least squares estimates as initial values for the location parameter, i.e., $\beta^{(0)} = (X^T X)^{-1} X^T Y$, where $Y = (y_1,\ldots,y_n)^T$ is the response vector. We fix the starting value for the intercept of the dispersion term at $\log(s^2)$, where the classical residual variance estimator $s^2 = \sum_{i=1}^n (y_i - x_i^T \beta^{(0)})^2/(n - p)$ is used. The remaining elements of the $\alpha^{(0)}$ parameter vector are set to zero \citep{rutemiller68, harvey76}, which gives $\alpha^{(0)}=(\log(s^2), 0, \ldots, 0)^T$. The standard errors of the estimates can be directly acquired by estimating the covariance of the penalized estimates for the true non-zero parameters using the sandwich formula,
\begin{equation}
    \hat{\mathrm{cov}}(\hat{\theta})=\{I(\hat{\theta})\}^{-1} I_0(\hat{\theta})\{I(\hat{\theta})\}^{-1},
    \label{eq:sandwich}
\end{equation}
which has been shown to be accurate for moderate sample sizes \citep{fan01, fan02}.

\subsection{$\epsilon$-telescoping} \label{sec:telescoping}
Although smaller values of $\epsilon$ lead to a better approximation of the $L_0$ norm (see Figure~\ref{fig:smooth_L0_norm}), and hence IC optimization, we have found that the procedure becomes less numerically stable (see Supplementary Material). On the other hand, larger values of $\epsilon$ lead to a more stable optimization procedure, but one that does not yield coefficients close to zero. Therefore, we propose a method that ``telescopes" through a decreasing sequence of $\epsilon$ values and makes use of ``warm starts", whereby the solution to the previous optimization problem is used as the initial point for the current nearby problem. The method can produce final estimates of the true zero coefficients that are extremely close to zero, and, so, can be treated as being equal to zero for practical purposes.

In this paper, we treat values below $10^{-8}$ as being zero. We have found that using a sequence of $T=100$ steps from $\epsilon_1=10$ to $\epsilon_{T}=10^{-5}$ performs well. Of course, applying fewer steps in the sequence from $\epsilon_1$ to $\epsilon_T$ is an option in practice. However, larger values of $T$ help to ensure that the repeated fitting procedure brings the parameters close to zero while avoiding estimation instability. If $T$ is too small (e.g.,~$T=10$), then the variable selection performance declines; simulation results for $T=50$ and $T=10$ are provided in the Supplementary Material. Once an adequate number of steps are used, the performance of the method is not highly influenced by the choice of the sequence. A large enough $\epsilon_1$ must be chosen in order to introduce smoothness and give stable estimates, while the smaller $\epsilon_{T}$ more closely approximates the $L_0$ norm to induce pseudo-sparsity (where we say ``pseudo" since the algorithm produces coefficients that can be made arbitrarily close to zero while not being exactly zero). In addition to this, we propose implementing an exponentially decaying sequence in the form of $\epsilon_1 r^{t-1}$, where $\epsilon_1$ is the starting value, $r \in (0,1)$ is the rate of decay and $t$ is the step number. For our suggested sequence with $T=100$ steps from $\epsilon_1=10$ to $\epsilon_{T}=10^{-5}$, the decay parameter is $r = 0.87$. This is advantageous as the optimization begins with large increments from $\epsilon_1 = 10$, which provides rapid improvements and estimates that are initially close to the unpenalized values. The smaller increments leading to $\epsilon_{T}=10^{-5}$ allow for smaller refinements, especially with regard to squeezing some coefficients to be close to zero.

Although we avoid a grid search over penalty tuning parameters (typically denoted by $\lambda$ in penalized estimation), we instead have a sequence of $\epsilon$ values. However, there is a key distinction between the objectives of these two approaches. In tuning parameter selection, the grid search over $\lambda$ is an optimization procedure, which, as previously discussed, is computationally demanding in the context of distributional regression due to it being a multidimensional grid. Moreover, the position of the optimal solution is unknown and could potentially be missed --- especially if one reduces the number of grid points to combat the aforementioned computational expense. In contrast, our $\epsilon$-telescoping approach is unidimensional and is not itself an optimization procedure since we know in advance, from a mathematical perspective, that $\epsilon$ should effectively be zero. Thus, the role of the $\epsilon$-telescope is to move the problem to an arbitrarily small value of $\epsilon_{T}$ in a stable way. It should be noted that, although we use $\epsilon_{T}=10^{-5}$, it may be that a relationship between $\epsilon_{T}$ and the sample size could be established using asymptotic analysis, e.g., a larger $\epsilon_{T}$ value might be acceptable at smaller sample sizes; however, this is beyond the scope of the current article.

Table~\ref{table:tele_zoom} presents an example of the coefficient estimates for a true zero and true non-zero coefficient for some simulated data. The shrinkage effect due to decreasing $\epsilon$ values is apparent. The value of the true zero coefficient $\beta_1$ drastically reduces in magnitude through each step. It is obvious that the final estimate at $\epsilon_{T} = 10^{-5}$ is extremely close to zero. As a result, it can be treated as a zero coefficient without any issues --- and, indeed, could be shrunk further if desired by reducing $\epsilon_{T}$. The estimate for the true non-zero coefficient $\beta_2$ does not vary greatly over the telescoping steps.

\begin{table}[b!]
\begin{center}
\caption{\label{table:tele_zoom}Coefficient values of $\beta_1$ and $\beta_2$ as the method telescopes through $\epsilon$}%
\begin{tabular}{@{}ccc@{}}
  \toprule
    \multicolumn{1}{l}{$\epsilon$} & $\beta_1 = 0$ & $\beta_2=1$ \\
    \midrule
    $10^{-2}$ & 0.0007631062 & 0.9998 \\[0.05cm] 
    $10^{-3}$ & 0.0000078397 & 0.9998 \\[0.05cm] 
    $10^{-4}$ & 0.0000000815 & 0.9998 \\[0.05cm] 
   $10^{-5}$ & 0.0000000008 & 0.9998 \\
   \bottomrule
\end{tabular}
\end{center}
\end{table}

\begin{figure}[h!]
\centering
\makebox{\includegraphics[width = 0.9\textwidth]{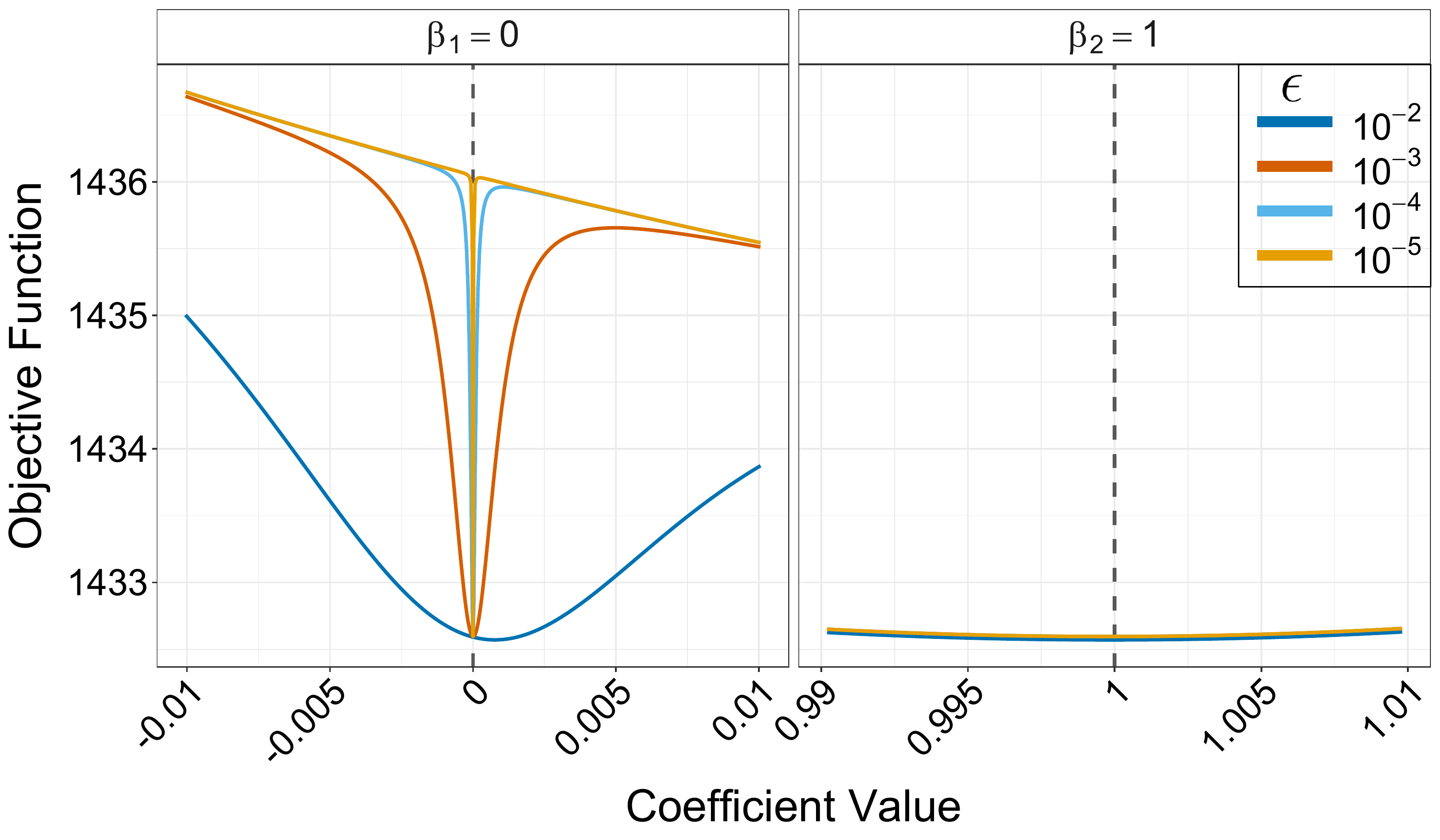}}
\caption{\label{figs:tele}Slice through objective function for different values of $\epsilon$. Dashed vertical lines mark true value.}
\end{figure}

Figure~\ref{figs:tele} provides a visualization of the telescoping effect in terms of the objective function. This is a slice through the objective function, which is plotted as a function of the coefficient value. Different curves are plotted corresponding to the $\epsilon$ values in the telescope sequence. In the case of the true zero coefficient $\beta_1$, it is clear that as $\epsilon$ decreases, the width of the curves become narrower and therefore there is less uncertainty around the estimate. Additionally, it is evident that the minimum of the curve shifts towards zero as $\epsilon$ decreases. For the true non-zero coefficient $\beta_2$, the curves for the different $\epsilon$ values are almost identical, i.e., the telescoping has little impact on the shape of the objective function in this case.

\subsection{Algorithm} \label{sec:algorithm}
The proposed MPR-SIC variable selection method is summarized in Algorithm~\ref{algo:tele}.
\begin{algorithm}[t!]
\caption{Implementation of the MPR-SIC $\epsilon$-telescope Method}
\label{algo:tele}
    \begin{enumerate}[label=\textbf{\arabic*}.]
        \item \textbf{Initialization:} Set $\theta^{(0)}=(\beta^{(0)T},\alpha^{(0)T})^T$, where $\beta^{(0)}$ and $\alpha^{(0)}$ are the initial values for the location and dispersion parameters respectively (see Section~\ref{sec:estimation_procedure}).
        \item \textbf{Telescoping:} Go through the exponentially decaying sequence of telescope values of length $T$ from $\epsilon_1$ to $\epsilon_T$, where $\epsilon_t=\epsilon_1 r^{t-1}$ for step $t=1,\ldots,T$ and $r \in (0,1)$ is the rate of decay (see Section~\ref{sec:telescoping}). 
        \begin{itemize}[leftmargin=*]
        \item\textbf{For $\boldsymbol{t=1,\ldots,T}$:}\\
        \textbf{Optimization:} Maximize $\ell^{\text{SIC}}(\theta)$ in \eqref{eq:smooth_ic} by iteratively re-solving the system of equations in \eqref{eq:nr_eq} with initial values $\theta^{(0)}_{\epsilon_t}$, to obtain $\hat{\theta}_{\epsilon_t}$. Convergence is achieved when $\lvert \theta^{(m+1)}_{\epsilon_t} - \theta^{(m)}_{\epsilon_t}\rvert\leq\omega$ for some small tolerance, e.g., $\omega=10^{-8}$. For warm starts, set $\theta^{(0)}_{\epsilon_{t+1}}$ = $\hat{\theta}_{\epsilon_t}$ so that the obtained estimates are used as initial values for the next step in the telescope. Note that we set $\hat\theta_{\epsilon_0}=\theta^{(0)}$.
        \end{itemize}
        \item \textbf{Output:} At $t = T$, the final estimates $\hat{\theta}_{\epsilon_T}$ are obtained and any estimates that are very close to zero (below $10^{-8}$ for example) can be treated as being zero. The corresponding standard errors are computed by evaluating \eqref{eq:sandwich} at $\hat{\theta}_{\epsilon_T}$. Note that because we are applying penalized variable selection, the predictors are scaled to have unit variance. However, the final estimates are converted back to their original scale.
    \end{enumerate}
\end{algorithm}

\FloatBarrier
\section{Simulation Studies} \label{sec:simulation_studies}
\subsection{Setup}\label{sec:sim_setup}
We have undertaken a simulation study to investigate the performance of our proposed MPR-SIC method. We simulate data from the normal MPR model from Section~\ref{sec:model_formulation}, where $X$ is the matrix of 12 covariates. To achieve a realistic setup, we make use of a range of different regression parameter values and covariate distributions. The regression coefficients, provided in Table~\ref{tab:true_values}, take values of 0, 0.5 or 1, respectively, corresponding to covariates having no effect, a weak effect or a strong effect. Moreover, there are several combinations where covariates enter through: both distributional parameters ($X_1$ to $X_4$); the location only ($X_5$ and $X_6$); the dispersion only ($X_7$ and $X_8$); and in neither the location nor the scale ($X_9$ to $X_{12}$), this latter group being pure noise covariates, i.e., they have no impact on the response. As for the covariate distributions, we include: two skewed covariates, $(X_1, X_{11}) \sim$ Exponential(1); two unbalanced binary covariates $(X_3, X_{10}) \sim$ Bernoulli(0.75); four independent normal covariates $(X_4, X_5, X_7, X_8) \sim$ $\text{N}(0,1)$; and four correlated multivariate normal covariates $(X_2, X_6, X_9, X_{12}) = (Z_1, Z_2, Z_3, Z_4) \sim$ MVN wherein $\text{corr}(Z_j, Z_k)= 0.8^{\lvert j - k \rvert}$. Lastly, three different sample sizes ($n=100$, 500 and 1000) are considered, where each scenario is replicated 1000 times. The Supplementary Material contains some additional simulation studies (not discussed here): scenarios where no covariate enters the dispersion (i.e.,~classical linear regression) and scenarios with only independent normally distributed covariates.

\begin{table}[t!]
\begin{center}
\caption{\label{tab:true_values}True parameter values}%
\begin{tabular}{@{}cccccccccccccc@{}}
  \toprule
    & & \textcolor{blue}{E} & \textcolor{red}{M} & \textcolor{mygreen}{B} & N & N & \textcolor{red}{M} & N & N & \textcolor{red}{M} & \textcolor{mygreen}{B} & \textcolor{blue}{E} & \textcolor{red}{M} \\
    & $X_0$ & \textcolor{blue}{$X_1$} & \textcolor{red}{$X_2$} & \textcolor{mygreen}{$X_3$} & ${X_4}$ & ${X_5}$ & \textcolor{red}{$X_6$} & ${X_7}$ & $X_8$ & \textcolor{red}{${X_9}$} & \textcolor{mygreen}{$X_{10}$} & \textcolor{blue}{$X_{11}$} & \textcolor{red}{$X_{12}$} \\
 \midrule
$\beta$ & 0 & 1 & 0.5 & 0.5 & 1 & 0.5 & 1 & 0 & 0 & 0 & 0 & 0 & 0 \\[0.05cm] 
$\alpha$ & 0 & 0.5 & 1 & 0.5 & 1 & 0 & 0 & 0.5 & 1 & 0 & 0 & 0 & 0 \\
\bottomrule
 \multicolumn{14}{l}{\footnotesize \textcolor{blue}{E = Exponential}, \textcolor{mygreen}{B = Bernoulli}, N = independent normal,} \\
  \multicolumn{14}{l}{\footnotesize \textcolor{red}{M = multivariate normal (correlated)}.}
\end{tabular}
\end{center}
\end{table}

For each scenario, we perform our proposed procedures, MPR-SIC and SPR-SIC; the latter is the SIC implemented for a single-parameter regression model, i.e., penalized linear regression. We compare the performance of our method to the ``\texttt{bamlss}'' package \citep{umlauf18bamlsspaper}, a package which implements penalized distributional regression (among other things). More specifically, we assign LASSO penalties to the location and dispersion regression parameters of a normal distribution, where the associated tuning parameters are selected by minimizing the BIC using a two-dimensional grid search with $50 \times 50$ grid points; hereafter, we refer to this as BAMLSS. We note that the BAMLSS method does not always necessarily bring parameters very close to zero, and, therefore, our interpretation of a zero effect in BAMLSS is based on the associated 95\% credible interval containing zero.

We also apply the adaptive LASSO (ALASSO) method from the ``\texttt{glmnet}'' package \citep{friedman10}, which corresponds to penalized linear regression, where we select the value of the tuning parameter by minimizing the BIC (ALASSO-IC). Note that only cross-validation-minimization is available in the \texttt{glmnet} package, and, therefore, we compute the BIC by evaluating the normal likelihood at the ALASSO estimates, $\hat\beta$, and with the variance estimator
\begin{equation}
    \hat{\sigma}^2=\sum^{n}_{i=1}\frac{(y_i-{x_{i}}^{T}{\hat{\beta}})^2}{n-{k}}
    \label{eq:variance_estimator}
\end{equation}
as suggested in \cite{reid16}, where $k$ is the number of non-zero elements in $\hat\beta$. Since \texttt{glmnet} does not provide parameter inference, we compute standard errors (and hence confidence intervals) using the general sandwich formula provided in \eqref{eq:sandwich}. (The form of the matrix $I(\hat{\theta})$ is slightly different due to the presence of the $L_1$ penalty.) The ALASSO method is misspecified in the scenarios we consider here, since it does not cover dispersion effects. However, we include it as a very commonly used method in practice, and it is useful to see how variable selection in the location is impacted when one fails to model the dispersion. The SPR-SIC method is similarly misspecified and is included as an SIC-based alternative to the ALASSO. Note that the Supplementary Material includes scenarios (not discussed here) where these are not misspecified, i.e., the true model only contains location effects.

\subsection{Simulation Results} \label{sec:sim_results}

Before we consider performance in terms of variable selection and parameter inference, we first briefly review the computational expense. To this end, average computation times for each of the methods are given in Table~\ref{tab:timing}. We can see that our MPR-SIC procedure is 40-50 times faster than BAMLSS, in large part due to the two-dimensional grid required by the latter. Even though the SPR-SIC approach is misspecified here, it is still useful to note that it is 4-8 times faster than the MPR-SIC approach. Thus, as expected, the distributional MPR approach is slightly slower than the SPR approach due to the fact that the former specifies (correctly) a dispersion model, and, hence, has twice the number of parameters to estimate (ignoring intercepts). However, the difference is not as dramatic as MPR-SIC versus BAMLSS since the SIC approaches both have the same penalty with unidimensional $\epsilon$-telescoping. The ALASSO-IC approach is the fastest overall, but it should be noted that the core of its implementation is compiled \texttt{C} code. Even so, the SPR-SIC is still relatively competitive computationally at approximately 10 times slower using only \texttt{R} code.

Turning now to variable selection performance, metrics including the average number of true zero coefficients correctly set to zero (C) and the average number of true non-zero coefficients incorrectly set to zero (IC) are investigated. The probability of choosing the true model (PT) is examined by looking at the proportion of times the true model is selected. The mean squared error (MSE) is computed for each simulation replicate in order to assess in-sample prediction accuracy, and is calculated by $\text{MSE}(\hat{\theta})=(\hat{\theta} - \theta)^T{X^TX}(\hat{\theta} - \theta)/n$ \citep{tibshirani97}. These metrics, averaged over simulation replicates, are presented in Table~\ref{tab:var_selection}.

\begin{table}[b!]

\begin{center}
\caption{\label{tab:timing}Simulation results: average computation time per simulation replicate (in seconds)}%
\begin{tabular}{ccccc}
  \toprule
$n$ & MPR-SIC & BAMLSS & SPR-SIC & ALASSO-IC \\ 
  \midrule
100 & 6.8 & 384.8 & 0.8 & 0.1 \\ 
  500 & 4.1 & 165.4 & 1.0 & 0.1 \\ 
  1000 & 4.8 & 186.5 & 1.3 & 0.1 \\ 
   \bottomrule
   \multicolumn{5}{p{0.6\textwidth}}{\footnotesize Intel(R) Core(TM) i7-10610U CPU @ 1.80GHz   2.30 GHz}\\
\end{tabular}
\end{center}
\end{table}

\begin{table}[ht!]

\begin{center}
\caption{\label{tab:var_selection}Simulation results: model selection metrics}%
    \begin{tabular}{@{}cc   c@{~~}c@{~~}c@{~~}c@{~~}   c@{~~~}   c@{~~}c@{~~}c@{~~}c@{~~}} 
    \toprule
    {} & {} & \multicolumn{4}{c}{MPR-SIC} && \multicolumn{4}{c}{BAMLSS} \\
    \cmidrule(lr){3-6} \cmidrule(){8-11}
    {}     & $n$   & C(6)  & IC(0) & PT & MSE && C(6) & IC(0) & PT & MSE \\
    \midrule
    $\beta$ & 100  &  5.25 & 0.15 & 0.44 & 0.14 && 5.55 & 0.24 & 0.52 & 0.17 \\
    {}      & 500  &  5.88 & 0.00 & 0.88 & 0.01 && 5.67 & 0.00 & 0.73 & 0.02 \\
    {}      & 1000 &  5.95 & 0.00 & 0.95 & 0.00 && 5.70 & 0.00 & 0.74 & 0.01 \\[0.1cm]
    $\alpha$ & 100     & 5.52 & 0.80 & 0.30 & 0.62 && 5.60 & 1.08 & 0.20 & 0.46 \\
    {}       & 500     & 5.92 & 0.00 & 0.93 & 0.04 && 5.35 & 0.00 & 0.67 & 0.08 \\
    {}       & 1000    & 5.95 & 0.00 & 0.95 & 0.02 && 5.08 & 0.00 & 0.63 & 0.06 \\
    \midrule
    {} & {} & \multicolumn{4}{c}{SPR-SIC} && \multicolumn{4}{c}{ALASSO-IC} \\
    \cmidrule(lr){3-6} \cmidrule(){8-11}
    {}     & $n$   & C(6)  & IC(0) & PT & MSE && C(6) & IC(0) & PT & MSE \\
    \midrule
    $\beta$     & 100  & 5.59 & 3.20 & 0.00 & 2.37 && 5.43 & 2.99 & 0.01 & 2.02 \\
    {}          & 500  & 5.84 & 1.57 & 0.11 & 0.64 && 5.58 & 1.13 & 0.22 & 0.61 \\
    {}          & 1000 & 5.88 & 0.70 & 0.37 & 0.27 && 5.70 & 0.45 & 0.45 & 0.27 \\[0.1cm]
    $\alpha$ & 100     & 6.00 & 6.00 & 0.00 & 6.43 && 6.00 & 6.00 & 0.00 & 6.62 \\
    {}       & 500     & 6.00 & 6.00 & 0.00 & 7.12 && 6.00 & 6.00 & 0.00 & 7.16 \\
    {}       & 1000    & 6.00 & 6.00 & 0.00 & 7.15 && 6.00 & 6.00 & 0.00 & 7.17 \\
    \bottomrule
    \multicolumn{11}{p{0.65\textwidth}}{\footnotesize C, average correct zeros; IC, average incorrect zeros; PT, the probability of choosing the true model; MSE, the average mean squared error.}\\
    \end{tabular}
\end{center}
\end{table}

For all of the methods, the C values are close to six (true value) and improve as the sample size increases. For the MPR-SIC method, the IC values are zero in most cases, apart from $n = 100$. This is due to the three smaller-valued weak effects being set to zero incorrectly in some simulation replicates ($\beta_2$, $\beta_3$, $\beta_5$ in the location and $\alpha_1$, $\alpha_3$, $\alpha_7$ in the dispersion). This behaviour is also conveyed by the probability of choosing the true model (PT). The PT values are low for $n=100$, which is due to both $\beta$ and $\alpha$ sometimes having a zero coefficient not set to zero, and, sometimes having a non-zero coefficient incorrectly set to zero. The sample size of $n=100$ is a challenging scenario as the MPR-SIC method is fitting a total of $2(p+1)$ parameters, which in this case is 26 parameters for a relatively small sample size. Taking this into account, we suggest that the performance of the method in this setting is reasonable. The PT values are also high for $n = 500$ and 1000, and appear to be converging to one. The BAMLSS procedure has higher IC values than the MPR-SIC method for $n = 100$ and lower C values for the larger sample sizes. The net effect of this is that the PT values are generally lower than for the MPR-SIC method (except for the location parameter at $n = 100$). For the SPR-SIC and ALASSO-IC methods, it only makes sense to consider their performance in the location, since there is no dispersion model. We can see that the IC values for these approaches are quite large, which means that they are setting some of the non-zero parameters to zero (albeit this is reducing with the sample size). The corresponding PT values are also quite low compared to the MPR-SIC approach. Ultimately, this indicates that erroneously ignoring the dispersion has an impact on the estimation of the location, even though the location and dispersion parameters are orthogonal for the normal distribution.

The estimation and inferential performance of our proposed MPR-SIC method is investigated in Table~\ref{tab:parameter_inference_mprbic}. The average estimate over simulation replicates is shown along with the true standard error (SE), which is the standard deviation of the estimates over simulation replicates, and the average estimated standard error (SEE) over simulation replicates, where the SEE in a given replicate is computed using \eqref{eq:sandwich}; also shown is the empirical coverage probability (CP) for a nominal 95\% confidence interval. We can see that, in all cases, the estimated parameter is close to the true value, albeit there is some bias in the larger $\alpha$ coefficients at $n=100$. The standard errors for both the $\beta$ and $\alpha$ parameters are underestimated at $n=100$, leading to CPs below 0.95. However, at $n \ge 500$ the standard errors are well estimated and the coverage is very close to the desired 0.95 level. The equivalent results for the BAMLSS, SPR-SIC and ALASSO-IC methods are deferred to the Supplementary Material, but we briefly outline them here: although BAMLSS appears to be better at the smallest sample size, the dispersion parameter results are not as good as MPR-SIC for the larger sample sizes (with higher SE values and CP values generally below 0.9); both the SPR-IC and ALASSO-IC methods perform poorly in the location parameter in all respects (biased estimates, large SEs that are underestimated by the SEEs, and quite low CP values).

\begin{table}[b!]

\begin{center}
\caption{\label{tab:parameter_inference_mprbic}Simulation results: estimation and inference metrics}%
\begin{tabular}{@{}c@{~~}c@{~~~} c@{~~}c@{~~}c@{~~}c@{~~} c@{~~~} c@{~~}c@{~~}c@{~~}c@{~~} c@{~~~} c@{~~}c@{~~}c@{~~}c@{}}
\toprule
\multicolumn{16}{l}{MPR-SIC}\\
{} & {} & \multicolumn{4}{c}{$n = 100$} && \multicolumn{4}{c}{$n = 500$} && \multicolumn{4}{c}{$n = 1000$} \\
\cmidrule(r){3-6} \cmidrule(r){8-11} \cmidrule(){13-16}
{} & $\theta$ & $\hat{\theta}$ & SE & SEE & CP && $\hat{\theta}$ & SE & SEE & CP && $\hat{\theta}$ & SE & SEE & CP \\
\midrule
  $\beta_{0}$ & 0.0 &  -0.01 & 0.22 & 0.13 & 0.76 && -0.00 & 0.06 & 0.06 & 0.93 && -0.00 & 0.04 & 0.04 & 0.94 \\
  $\beta_{1}$ & 1.0 &  1.00  & 0.15 & 0.09 & 0.78 && 1.00  & 0.04 & 0.04 & 0.93 && 1.00  & 0.03 & 0.03 & 0.94 \\ 
  $\beta_{2}$ & 0.5 &  0.46  & 0.23 & 0.10 & 0.73 && 0.50  & 0.05 & 0.05 & 0.94 && 0.50  & 0.03 & 0.03 & 0.96 \\ 
  $\beta_{3}$ & 0.5 &  0.50  & 0.11 & 0.07 & 0.82 && 0.50  & 0.03 & 0.03 & 0.93 && 0.50  & 0.02 & 0.02 & 0.95 \\ 
  $\beta_{4}$ & 1.0 &  1.00  & 0.12 & 0.07 & 0.78 && 1.00  & 0.03 & 0.03 & 0.94 && 1.00  & 0.02 & 0.02 & 0.95 \\ 
  $\beta_{5}$ & 0.5 &  0.49  & 0.11 & 0.07 & 0.80 && 0.50  & 0.03 & 0.03 & 0.93 && 0.50  & 0.02 & 0.02 & 0.94 \\ 
  $\beta_{6}$ & 1.0 &  1.03  & 0.23 & 0.11 & 0.70 && 1.00  & 0.05 & 0.05 & 0.94 && 1.00  & 0.03 & 0.03 & 0.95 \\[0.1cm]
  $\alpha_{0}$ & 0.0 &  -0.13 & 0.45 & 0.23 & 0.66 && -0.04 & 0.10 & 0.10 & 0.92 && -0.02 & 0.07 & 0.07 & 0.96 \\
  $\alpha_{1}$ & 0.5 &  0.45  & 0.29 & 0.12 & 0.70 && 0.50  & 0.07 & 0.07 & 0.95 && 0.50  & 0.05 & 0.05 & 0.96 \\ 
  $\alpha_{2}$ & 1.0 &  1.10  & 0.38 & 0.17 & 0.73 && 1.01  & 0.07 & 0.07 & 0.93 && 1.01  & 0.05 & 0.05 & 0.92 \\ 
  $\alpha_{3}$ & 0.5 &  0.49  & 0.37 & 0.13 & 0.58 && 0.51  & 0.08 & 0.08 & 0.95 && 0.50  & 0.05 & 0.05 & 0.94 \\ 
  $\alpha_{4}$ & 1.0 &  1.12  & 0.24 & 0.17 & 0.82 && 1.01  & 0.07 & 0.07 & 0.95 && 1.01  & 0.05 & 0.05 & 0.94 \\ 
  $\alpha_{7}$ & 0.5 &  0.51  & 0.30 & 0.13 & 0.69 && 0.51  & 0.07 & 0.07 & 0.93 && 0.50  & 0.04 & 0.05 & 0.96 \\ 
  $\alpha_{8}$ & 1.0 &  1.11  & 0.24 & 0.17 & 0.81 && 1.02  & 0.07 & 0.07 & 0.93 && 1.01  & 0.05 & 0.05 & 0.94 \\
  \bottomrule
  \multicolumn{16}{p{.8\textwidth}}{\footnotesize SE, standard deviation of estimates over 1000 replications; SEE, average of estimated standard errors over 1000 replications; CP, the empirical coverage probability of a nominal 95\% confidence interval.}\\
\end{tabular}
\end{center}
\end{table}

\begin{table}[b!]
\begin{center}
\caption{\label{tab:sigma_table}Simulation results: out-of-sample prediction coverage probabilities}%
\begin{tabular}{@{}l@{~}c@{~~} c@{~}c@{~}c@{~} c@{~~} c@{~}c@{~}c@{~} c@{~~} c@{~}c@{~}c@{~} c@{~~} c@{~}c@{~}c@{}} 
\toprule
{} && \multicolumn{3}{c}{MPR-SIC} && \multicolumn{3}{c}{BAMLSS} && \multicolumn{3}{c}{SPR-SIC} && \multicolumn{3}{c}{ALASSO-IC} \\
\cmidrule(lr){3-5} \cmidrule(r){7-9} \cmidrule(){11-13} \cmidrule(){15-17}
$n$         && 100  & 500  & 1000 && 100  & 500  & 1000 && 100  & 500  & 1000 && 100  & 500  & 1000 \\
\midrule
Low         && 0.77 & 0.93 & 0.94 && 0.82 & 0.93 & 0.95 && 1.00 & 1.00 & 1.00 && 1.00 & 1.00 & 1.00 \\ 
  Medium    && 0.90 & 0.94 & 0.95 && 0.91 & 0.95 & 0.95 && 0.98 & 1.00 & 1.00 && 0.99 & 1.00 & 1.00 \\ 
  High      && 0.95 & 0.95 & 0.95 && 0.94 & 0.94 & 0.94 && 0.79 & 0.85 & 0.86 && 0.81 & 0.85 & 0.86 \\[0.1cm] 
  Overall   && 0.86 & 0.94 & 0.95 && 0.88 & 0.94 & 0.95 && 0.93 & 0.95 & 0.95 && 0.94 & 0.95 & 0.95 \\
  \bottomrule
  \multicolumn{17}{p{.8\textwidth}}{\footnotesize Variability categorized as low ($\sigma_i \leq 1$), medium ($\sigma_i \in (1, 2.2]$) and high ($\sigma_i > 2.2$). Out-of-sample coverage is calculated for a sample 20\% the size of the original data.}\\
\end{tabular}
\end{center}
\end{table}

\begin{figure}[b!]
\centering
\makebox{\includegraphics[width = \textwidth]{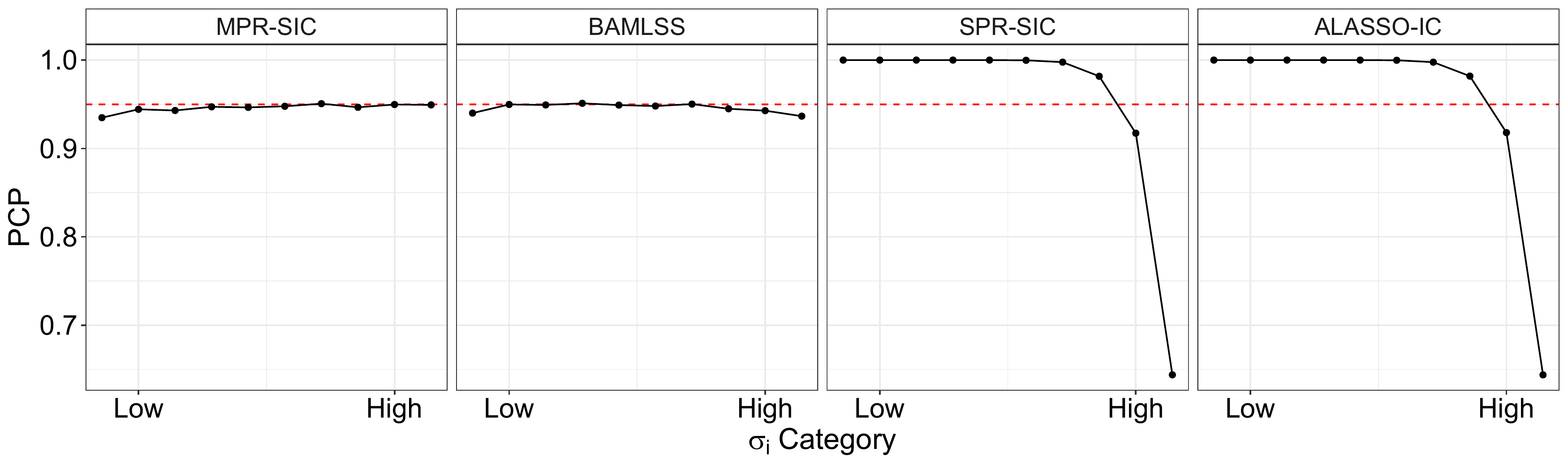}}
\caption{\label{fig:sigma_cp}Prediction coverage probabilities (PCPs) of observations for different dispersion levels, $\sigma_i$. Solid black line indicates the coverage and the red dashed line is a reference line at 0.95.}
\end{figure}

Out-of-sample prediction coverage probabilities (PCPs) for the methods are calculated for a sample 20\% the size of the original data and are shown in Table~\ref{tab:sigma_table}. This is calculated as the proportion of times the true response value lies in a nominal 95\% prediction interval (PI) in each replicate. The average is then taken over the 1000 replicates. The 95\% PIs are calculated as $x_i^T\hat{\beta} \pm 1.96 \sqrt{\exp(x_i^T\hat{\alpha})}$. Note that for the SPR-SIC and ALASSO-IC, the dispersion parameter is held constant across observations and so $\alpha$ is a vector of zeros except for its first element, which is the intercept. The methods appear to perform similarly when examining the overall PCP, although both the MPR-SIC and BAMLSS methods are somewhat poorer than the others in the $n=100$ case. However, note that the observations can be categorized by their variability $\sigma_i$ and, therefore, we split them into groups with low, medium and high variability using the thresholds $\sigma_i \leq 0.7$, $\sigma_i \in (0.7, 1.5]$ and $\sigma_i > 1.5$, respectively, where these thresholds are the tertiles computed numerically from the true underlying distribution of $\sigma_i$. Doing so reveals that none of the methods perform particularly well at $n=100$ when viewed in terms of these three levels of variability. For $n=500$ and 1000, the coverage for the MPR-SIC and BAMLSS procedures remain at approximately 95\%, which is the desired nominal value. In contrast, the PIs are too wide for the low and medium variability cases (leading to 100\% coverage) and too narrow for the large variability cases (leading to approximately 85\% coverage). This is unsurprising since these methods assume a constant $\sigma$, and, therefore, cannot adapt to heterogeneity in the data. This can also be visualized in Figure~\ref{fig:sigma_cp}, which shows the coverage for ten $\sigma_i$ categories in the case where $n=1000$; again we see that the MPR-SIC and BAMLSS methods lie close to 95\%, whereas the other methods are either too high or too low.

In order to examine the generalizability of our simulation study, we have also considered several additional simulation scenarios, whose results can be found in the Supplementary Material. In particular, we have changed the effect sizes and cardinality of the active sets (i.e.,~the sets of covariates with non-zero effects) so they differ across the location and dispersion parameters; we also consider settings where we have doubled the number of covariates (to 24). In general, the performance is comparable to the results presented here, but two noteworthy differences are as follows: (i) when the dispersion effects are much larger than the location effects, the selection performance in the location reduces considerably (albeit inferential performance remains good); and (ii) when the number of covariates is increased to 24, the problem becomes unstable for $n = 100$ and the larger-sample PT values are reduced (by about 10-15 percentage points).

\section{Real Data Analyses} \label{sec:real_data}

\subsection{Overview} \label{sec:real_data_overview}
We consider three real data analyses to illustrate our proposed MPR-SIC method, which is implemented using the $\epsilon$-telescope (Algorithm~\ref{algo:tele}). For each dataset, the resulting MPR-SIC, BAMLSS, SPR-SIC and ALASSO-IC estimates ($\hat\beta, \hat\alpha$) are presented, and note that, for the SPR-SIC and ALASSO-IC methods, $\hat\alpha$ is a vector of zeros except for its first element (the intercept). We also compare these methods in terms of out-of-sample PCP values. Additionally, for the proposed MPR-SIC, we provide the associated standard errors for each non-zero coefficient and the change in BIC, denoted $\Delta$BIC, that arises upon setting that coefficient to zero.  The $\Delta$BIC value provides a measure of the impact of dropping a variable from the location ($\beta$ coefficient) or the dispersion ($\alpha$ coefficient), and, therefore, indicates its importance in these model components. For the other methods, these additional metrics are deferred to Supplementary Material, but we indicate statistical significance by emboldening coefficient values for all methods in the main text. (Note that, for BAMLSS, ``statistical significance'' is based on the credible intervals excluding zero.)

\subsection{Prostate Cancer Data}\label{sec:real_data_pcancer}
We examine the prostate cancer data, which come from a study by \citet{stamey89pcancer} and which appear in \citet{tibshirani96} and \citet{zou05}. The correlation between the level of prostate-specific antigen (PSA) and various clinical measures in 97 men who were about to receive a radical prostatectomy is examined. The predictors consist of eight clinical measures: log(cancer volume ($\text{cm}^3$)) (\texttt{lcavol}), log(prostate weight (g)) (\texttt{lweight}), presence of seminal vesicle invasion (SVI) (\texttt{svi}), age of the patient (\texttt{age}), log(amount of benign prostatic hyperplasia ($\text{cm}^2$)) (\texttt{lbph}), log(capsular penetration (cm)) (\texttt{lcp}), Gleason score (\texttt{gleason}) and percentage of Gleason scores four of five (\texttt{pgg45}). The logarithm of PSA (ng/mL) is the response variable. The presence of SVI (\texttt{svi}) is a binary variable (1=yes, 0=no) and \texttt{gleason} is a discrete numerical variable with four values. The Gleason score relates to prostate cancer grades and the \texttt{pgg45} predictor provides information on the history of the patient. This is the percentage of Gleason scores they received before their final Gleason score in \texttt{gleason}. PSA is a protein that is produced by normal and malignant prostate cells, and is useful as a preoperative marker, as prostate cancer causes PSA to be discharged into the blood.

\begin{figure}[t!]
\centering
\makebox{
\begin{subfigure}[h]{\textwidth}
\centering
\includegraphics[width = 0.95\textwidth]{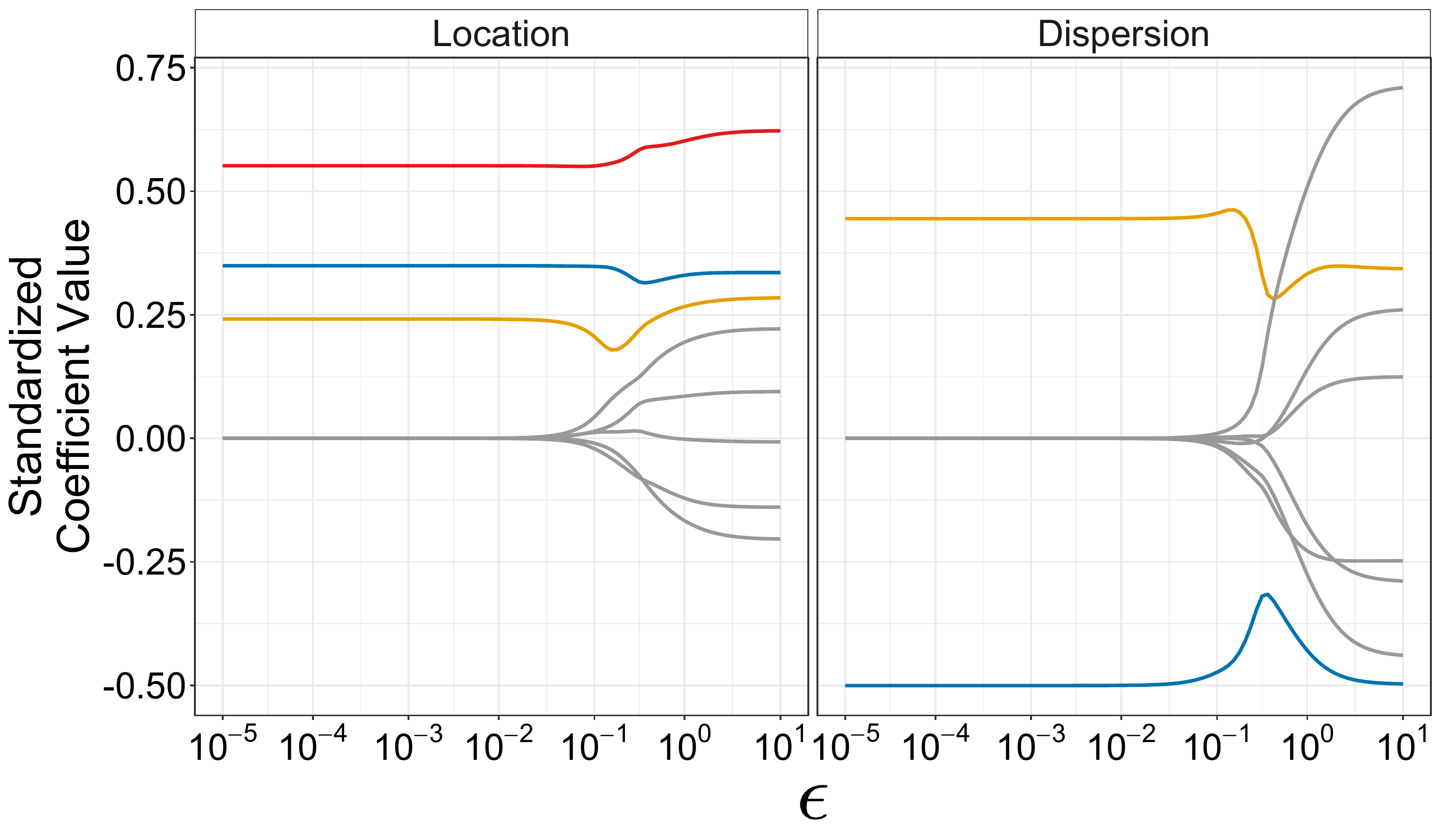}%
\end{subfigure}}
\makebox{
\begin{subfigure}[h]{\textwidth}
\centering
\includegraphics[trim = {0.2cm 0 0.2cm 0.1cm}, clip, scale = 0.5]{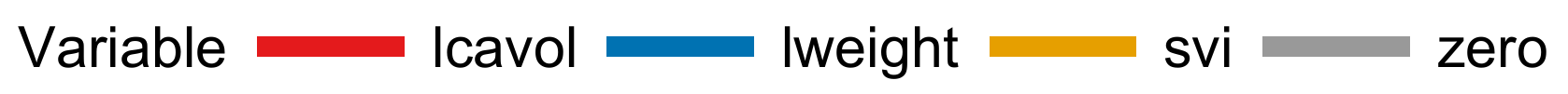}%
\end{subfigure}
}
\caption{\label{fig:pcancer_path}Prostate Cancer Data: standardized coefficient values through the $\epsilon$-telescope for the location and dispersion components. Lines are coloured corresponding to the selected variables, where grey indicates that the variable is set to zero and not selected in the final model.}
\vspace{-0.7cm}
\end{figure}

Figure~\ref{fig:pcancer_path} plots the standardized coefficient values with respect to the MPR-SIC $\epsilon$-telescope, which shows how the method works as $\epsilon$ moves towards zero. We note that the coefficients are essentially unpenalized at the largest $\epsilon = 10^1$ value where there is no variable selection; this is because the penalty in (\ref{eq:smooth_L0_norm}) is close to zero for large $\epsilon$ values. Then, decreasing $\epsilon$ moves the problem towards $L_0$ penalization such that variable selection occurs. In particular, \texttt{lcavol} is selected only in the location component while \texttt{lweight} and \texttt{svi} are selected in both the location and dispersion components. Interestingly, although we decrease to $\epsilon=10^{-5}$, the results here do not change appreciably beyond $\epsilon=10^{-2}$.

\begin{table}[b!]
\begin{center}
\caption{\label{tab:dataset_estimates_pcancer}Prostate Cancer Data: estimation metrics}%
\begin{tabular}{@{}l@{~~} r@{~}c@{~}r@{~} r@{~}c@{~}r@{~} c@{~~} >{}r@{~}>{}r@{~} c@{~~} r@{~}r@{~}  c@{~~} r@{~}r@{}}
\toprule
{} & \multicolumn{6}{c}{MPR-SIC} && \multicolumn{2}{c}{{BAMLSS}} && \multicolumn{2}{c}{SPR-SIC} && \multicolumn{2}{c}{\begin{tabular}[b]{@{}c@{}}ALASSO \\ -IC\end{tabular}}\\
BIC & \multicolumn{6}{c}{224} && \multicolumn{2}{c}{{222}} && \multicolumn{2}{c}{227} && \multicolumn{2}{c}{228}\\
\cmidrule(){2-7} \cmidrule(){9-10} \cmidrule(){12-13} \cmidrule(){15-16}
{} & \multicolumn{2}{c}{$\hat\beta_j$} & $\Delta\text{BIC}$ & \multicolumn{2}{c}{$\hat\alpha_j$} & $\Delta\text{BIC}$ && $\hat\beta_j$ & $\hat\alpha_j$ && $\hat\beta_j$ & $\hat\alpha_j$ && $\hat\beta_j$ & $\hat\alpha_j$ \\
  \midrule
  \texttt{inter}   & -1.26          & (0.53) &       & 3.15           & (1.36)  &       && -0.79          & 2.66            && -0.78          & -0.73 && -0.27          & -0.68 \\ 
  \texttt{lcavol}  & \textbf{0.47}  & (0.06) & 40.39 &                &         &       && \textbf{0.52}  & -0.19           && \textbf{0.53}  &       && \textbf{0.54}  &   \\ 
  \texttt{lweight} & \textbf{0.82}  & (0.14) & 19.79 & \textbf{-1.17} & (0.38)  & 4.78  && \textbf{0.81}  & \textbf{-0.93}  && \textbf{0.66}  &       && \textbf{0.52}  &  \\ 
  \texttt{svi}     & \textbf{0.58}  & (0.22) & 1.64  & \textbf{1.07}  & (0.38)  & 4.09  && \textbf{0.73}  & 0.77            && \textbf{0.67}  &       && \textbf{0.53}  &  \\
  \texttt{age}     &                &        &       &                &         &       && -0.01          & 0.02            &&                &       &&                &  \\
  \texttt{lbph}    &                &        &       &                &         &       && 0.06           & 0.05            &&                &       &&                &  \\
  \texttt{lcp}     &                &        &       &                &         &       && -0.16          & 0.45            &&                &       &&                &  \\
  \texttt{gleason} &                &        &       &                &         &       && 0.02           & -0.12           &&                &       &&                &  \\
  \texttt{pgg45}   &                &        &       &                &         &       && 0.01           & -0.01           &&                &       &&                &  \\
  \bottomrule
  \multicolumn{16}{p{0.8\textwidth}}{\footnotesize Significant effects indicated in bold.}\\
\end{tabular}
\end{center}
\end{table}

\begin{figure}[b!]
\centering
\makebox{\includegraphics[width = 0.75\textwidth]{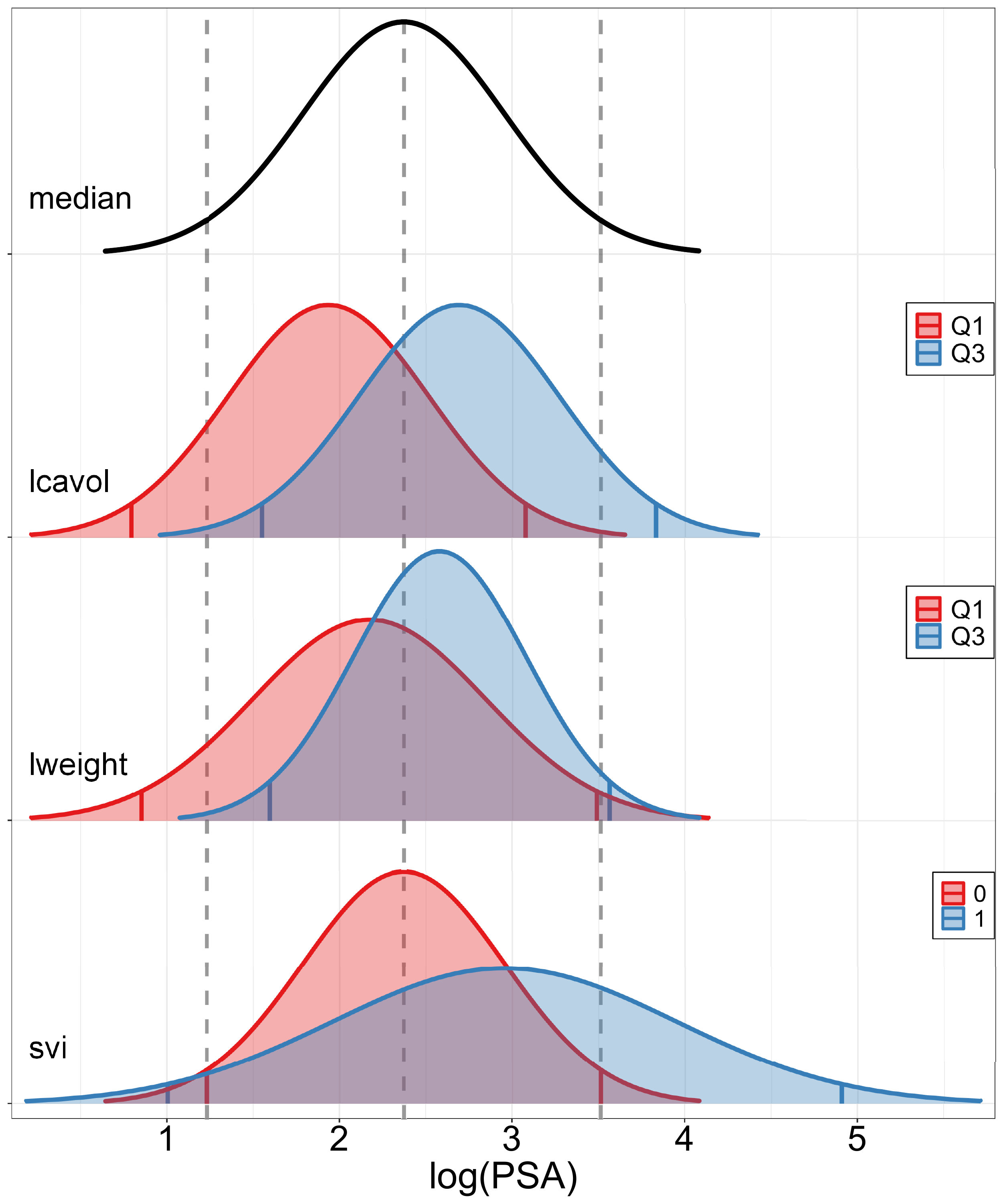}}
\caption{\label{fig:pcancer_densities}Prostate Cancer Data: MPR-SIC model-based conditional density curves. The black curve corresponds to an individual whose covariates (\texttt{lcavol}, \texttt{lweight}, \texttt{svi}) are all equal to median values (serving as a ``baseline'' or ``average'' individual); dashed grey lines mark the 2.5th, 50th and 97.5th quantiles of this density. Keeping two of the covariates fixed at the median values, the red and blue densities correspond to the modification of the third covariate as: ``low'' (Q1, the first quartile) and ``high'' (Q3, the third quartile) for the continuous covariates, \texttt{lcavol} and \texttt{lweight}; and ``absence'' ($=0$) and ``presence'' ($=1$) for the binary covariate, \texttt{svi}; red and blue vertical lines mark the 2.5th and 97.5th quantile values of each density.}
\end{figure}

From Table~\ref{tab:dataset_estimates_pcancer}, we can see that, like the MPR-SIC method, the other three methods also select \texttt{lcavol}, \texttt{lweight}, and \texttt{svi} in the location component; in all cases, their location coefficients are positive. Thus, increased values of log(cancer volume) and log(prostate weight), and the presence of SVI are associated with increased log(PSA) values, and therefore may be indicative of prostate cancer. As for the dispersion component, while the MPR-SIC method selects both \texttt{lweight} and \texttt{svi}, BAMLSS only identifies \texttt{lweight} as being important. Both of these methods have similar BIC values (224 and 222 units respectively), which are lower than the models with only location regression components (227 and 228 units, respectively, for SPR-SIC and ALASSO-IC). Distributional regression approaches improve on classical single parameter regression approaches since they can capture more complex covariate effects, e.g., \texttt{lweight} and \texttt{svi} appear in both the location and dispersion within the MPR-SIC model. Given this additional complexity, it can be helpful to visualize the effects. To this end, inspired by \citet{stadlmann21interactively}, we provide a series of model-based (MPR-SIC) conditional density curves for different covariate combinations in Figure~\ref{fig:pcancer_densities}.

As mentioned, within the MPR-SIC model (and all models considered), increased \texttt{lcavol} values are associated with increased log(PSA). Moreover, the large $\Delta$BIC value of 40.39 identifies the \texttt{lcavol} location effect as being the most important (across all location and dispersion effects) --- and there is a clear location shift in the associated conditional density plots in Figure~\ref{fig:pcancer_densities}. Similarly, increased \texttt{lweight} values are also associated with increased log(PSA), but to a lesser extent than with \texttt{lcavol}. In line with this, there is more overlap between the conditional densities for high and low \texttt{lweight} values and the $\Delta$BIC value is smaller (19.79). In other words, the cancer volume (\texttt{lcavol}) is more strongly associated with PSA values than the prostate weight (\texttt{lweight}). However, the nature of the \texttt{lweight} effect differs from that of \texttt{lcavol} since it impacts the dispersion: increased \texttt{lweight} values are associated with reduced dispersion. As for \texttt{svi}, its presence primarily increases the dispersion. This can be seen both visually from the density plots and confirmed by the fact that removing \texttt{svi} from the dispersion increases the BIC 4.09, whereas, its removal from the location increases the BIC by just 1.64 units.

\subsection{Sniffer Data}\label{sec:real_data_sniffer}
When gasoline is pumped into a tank, hydrocarbon vapours are forced out and emitted into the atmosphere. This is a source of air pollution and in order to reduce this, devices that capture the vapour are set up. Testing these vapour recovery systems involves a ``sniffer" device to measure the amount of vapour that is recovered. A method of estimating the total amount released is required to estimate the efficiency of the system. A laboratory experiment was carried out to discover factors that impact the amount of hydrocarbon vapour released when gasoline is pumped into a tank. Four factors are varied --- vapour pressure (psi) of the dispensed gasoline (\texttt{gaspres}), temperature (\textdegree F) of the dispensed gasoline (\texttt{gastemp}), initial tank temperature (\textdegree F) (\texttt{tanktemp}) and initial vapour pressure (psi) in the tank (\texttt{tankpres}). The quantity of emitted hydrocarbon (g) is the response variable. There are 125 runs in the data. These data have previously been considered by \citet{weisberg13sniffer} who noted that the dispersion may depend on the predictors but did not apply a heteroscedastic model, and \citet{bedrick00sniffer} who used a model with all four predictors in the location along with \texttt{gastemp} and \texttt{gaspres} in the dispersion.

The MPR-SIC, BAMLSS and SPR-SIC methods each select a different combination of variables for the location parameter (see Table~\ref{tab:dataset_estimates_sniffer}). In terms of the selected statistically significant effects, the two location regression models (SPR-SIC and ALASSO-IC) select \texttt{gaspres}, \texttt{gastemp}, and \texttt{tankpres}. However, these models have higher BIC values than the distributional regression models (MPR-SIC and BAMLSS), with the latter models choosing \texttt{tanktemp} rather than \texttt{tankpres} as being important in the location. In any case, the location effects of \texttt{gaspres} and \texttt{gastemp} are positive across all models (albeit \texttt{gaspres} is not statistically significant in BAMLSS), indicating that increased gasoline pressure and temperature values are related to increased amounts of emitted hydrocarbon; moreover, the MPR-SIC model identifies these as the most important effects with $\Delta$BIC values of 68.41 and 56.85, respectively. The initial tank temperature (\texttt{tanktemp}) appears to be less important ($\Delta$BIC $= 5.60$), but its negative location coefficient in the MPR-SIC and BAMLSS models indicates that higher temperatures reduce the emitted hydrocarbon. In addition to the location effect of \texttt{gastemp}, the MPR-SIC model also finds this variable to increase the dispersion. With a $\Delta$BIC value of 17.62, the \texttt{gastemp} \emph{dispersion} effect is far greater than the \texttt{tanktemp} \emph{location} effect; this demonstrates the fact that modelling only the location --- as is most often done in practice --- can miss important features of the process under study. We note that the BAMLSS model is somewhat more complex than the MPR-SIC model, in that there are more coefficients that are far from zero. Overall, the MPR-SIC model achieves the lowest BIC of 616 units.

\begin{table}[h!]
    \begin{center}
    \caption{\label{tab:dataset_estimates_sniffer}Sniffer Data: estimation metrics}%
\resizebox{\textwidth}{!}{
\begin{tabular}{@{}l@{~~} r@{~}c@{~}r@{~} r@{~}c@{~}r@{~} c@{~~} >{}r@{~}>{}r@{~} c@{~~} r@{~}r@{~}  c@{~~} r@{~}r@{}}
\toprule
{} & \multicolumn{6}{c}{MPR-SIC} && \multicolumn{2}{c}{{BAMLSS}} && \multicolumn{2}{c}{SPR-SIC} && \multicolumn{2}{c}{\begin{tabular}[b]{@{}c@{}}ALASSO \\ -IC\end{tabular}}\\
BIC & \multicolumn{6}{c}{616} && \multicolumn{2}{c}{{624}} && \multicolumn{2}{c}{630} && \multicolumn{2}{c}{632}\\
\cmidrule(){2-7} \cmidrule(){9-10} \cmidrule(){12-13} \cmidrule(){15-16}{} & \multicolumn{2}{c}{$\hat\beta_j$} & $\Delta\text{BIC}$ & \multicolumn{2}{c}{$\hat\alpha_j$} & $\Delta\text{BIC}$ && $\hat\beta_j$ & $\hat\alpha_j$ && $\hat\beta_j$ & $\hat\alpha_j$ && $\hat\beta_j$ & $\hat\alpha_j$ \\
\midrule
    \texttt{inter}      & 0.76           & (0.85) &        & -1.35         & (0.64) &        && -1.20          & -0.96          && 0.45           & 2.01 && 0.21            & 2.03 \\ 
    \texttt{gaspres}    & \textbf{5.19}  & (0.51) & 68.41  &               &        &        && 3.34           & \textbf{-3.46} && \textbf{10.84} &      && \textbf{9.79}   &  \\ 
    \texttt{gastemp}    & \textbf{0.23}  & (0.03) & 56.85  & \textbf{0.06} & (0.01) & 17.62  && \textbf{0.26}  & \textbf{0.09}  && \textbf{0.15}  &      && \textbf{0.19}   &  \\ 
    \texttt{tanktemp}   & \textbf{-0.09} & (0.03) & 5.60   &               &        &        && \textbf{-0.15} & 0.01           &&                &      && -0.07           &  \\ 
    \texttt{tankpres}   &                &        &        &               &        &        && 2.69           & \textbf{2.72}  && \textbf{-5.73} &      && \textbf{-4.08}  &  \\
    \bottomrule
\multicolumn{16}{p{0.8\textwidth}}{\footnotesize Significant effects indicated in bold.}\\
\end{tabular}}
\end{center}
\vspace{-1.1cm}
\end{table}

\begin{figure}[b!]
\centering
\begin{subfigure}{.5\textwidth}
  \centering
  \includegraphics[width=\linewidth]{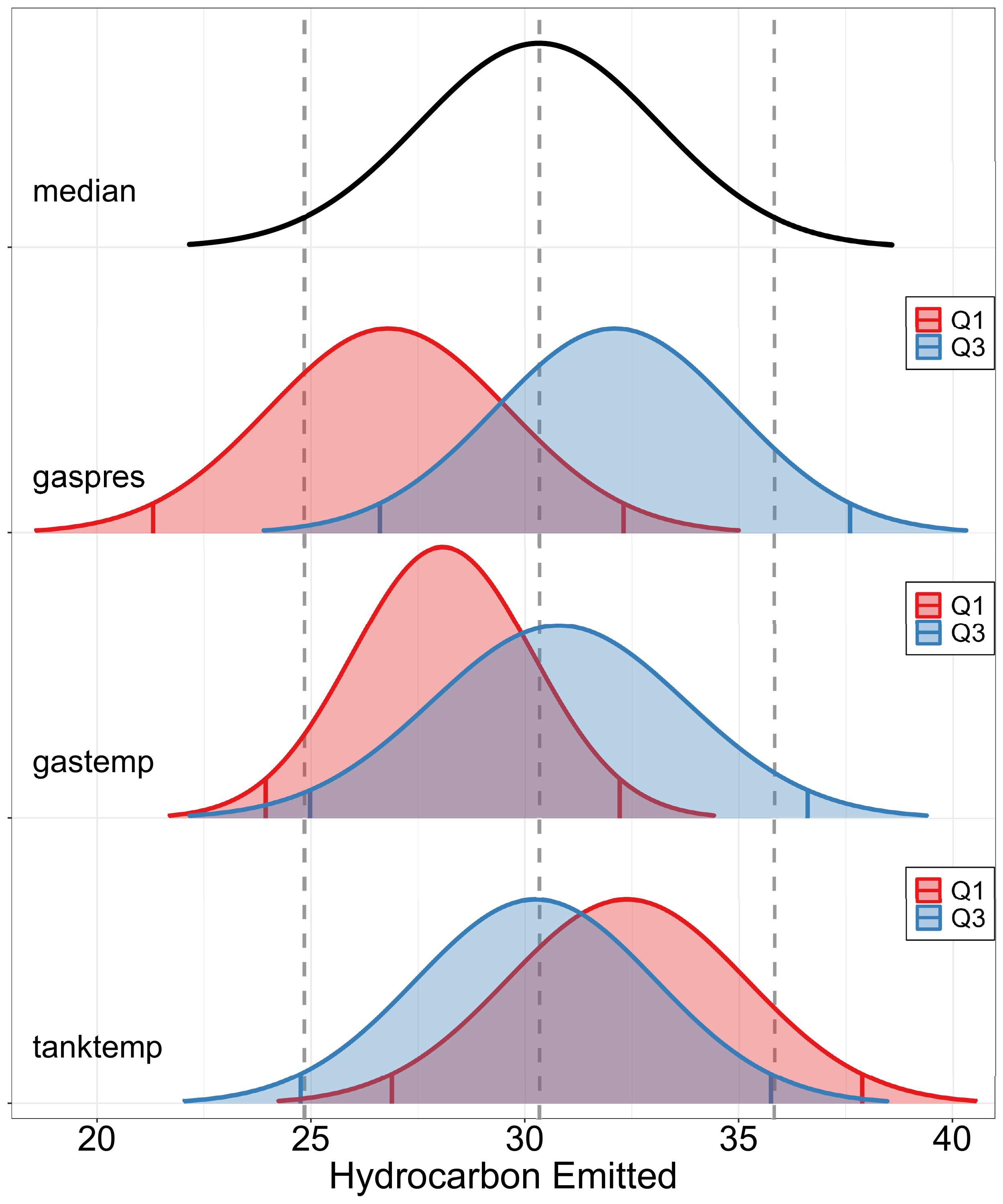}
  \caption{}
  \label{fig:sniffer_densities}
\end{subfigure}%
\begin{subfigure}{.5\textwidth}
  \centering
  \includegraphics[width=\linewidth]{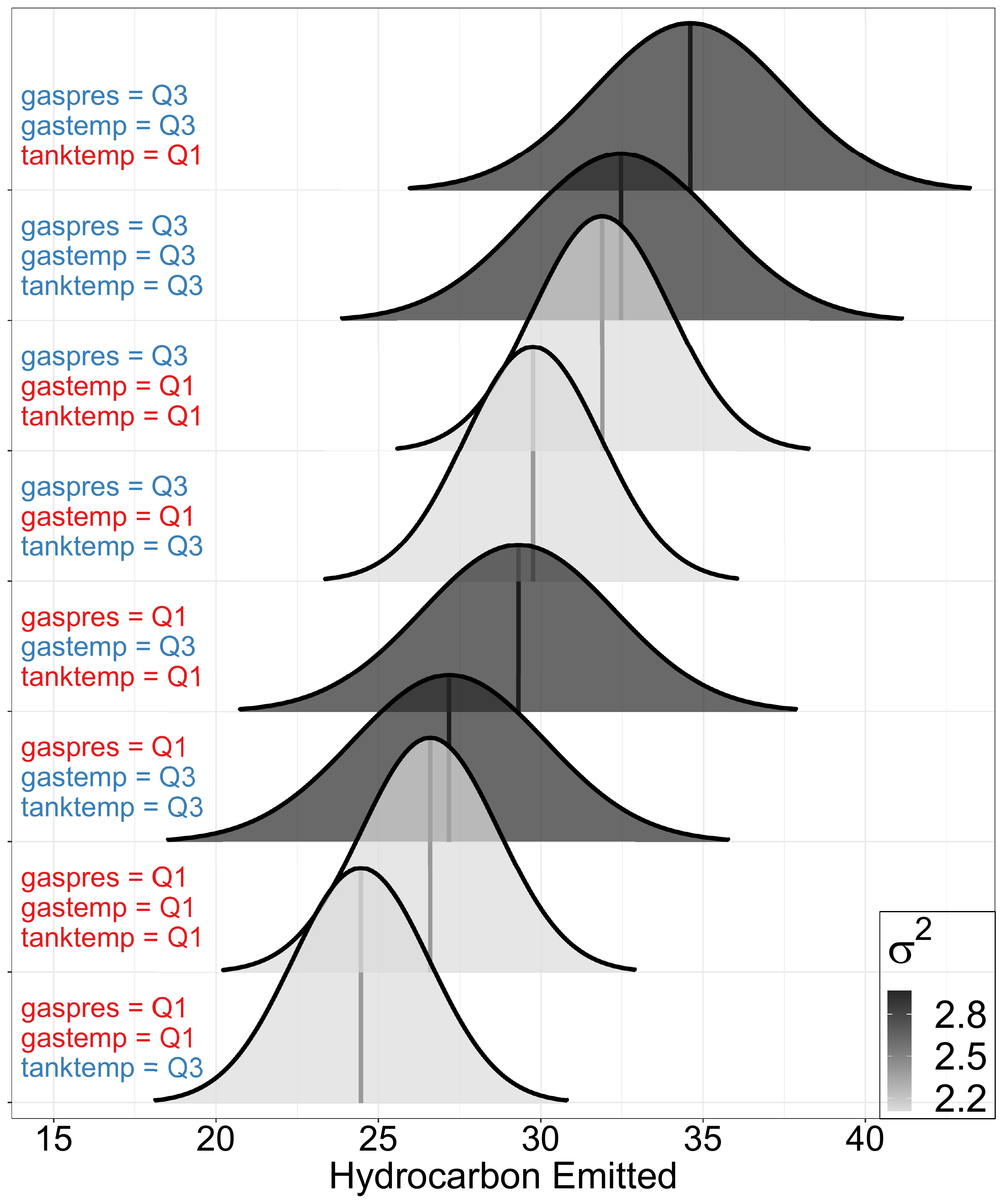}
  \caption{}
  \label{fig:sniffer_combos}
\end{subfigure}
\caption{Sniffer Data: (a) MPR-SIC model-based conditional density curves for each of the selected variables (see Figure~\ref{fig:pcancer_densities} for more details); (b) all eight conditional density curves obtained from the combinations of ``low'' (Q1, the first quartile) and ``high'' (Q3, the third quartile) for each of the three covariates \texttt{gaspres}, \texttt{gastemp}, and \texttt{tankpres}; these are ordered based on mean and coloured based on variance.}
\label{fig:sniffer_all}
\end{figure}
\FloatBarrier
Conditional density plots display the various effects in Figure~\ref{fig:sniffer_densities} (and these are analogous to those shown in Figure~\ref{fig:pcancer_densities} for the prostate cancer data). Here we can clearly see: the large impact of \texttt{gaspres} on the location; the lesser impact of \texttt{gastemp} on the location and its impact on the dispersion; and the weak impact of \text{tanktemp} on the location. Moreover, since there are only three selected variables, and this is an industrial setup each one can be altered in practice, Figure~\ref{fig:sniffer_combos} also displays all of the eight combinations of conditional densities that arise from varying each covariates at either ``low'' or ``high'' values. There is a clear optimal configuration here, which yields both the lowest and least variable hydrocarbon emissions: reduce the gasoline pressure (\texttt{gaspres}) and temperature (\texttt{gastemp}) and increase the initial tank temperature (\texttt{tanktemp}). This setting will yield emissions approximately between 20 and 30, with a mean of 25. In contrast, for the worst setting (high \texttt{gaspres} and \texttt{gastemp} and low \texttt{tanktemp}), emissions will generally be between 27 and 42, with a mean of 35. Thus, air pollution can be reduced considerably by optimizing the setup.

\subsection{Boston House Price Data}\label{sec:real_data_hprice}
Data from a cross-sectional study of 506 communities in the Boston area carried out in 1970 \citep{harrison78hprice2} are available in \citet{wooldridge15hprice2}. The association between median house prices in a particular community with various community characteristics is examined. There are eight explanatory variables: average number of rooms per house (\texttt{rooms}), percentage of the population that are ``lower status" (\texttt{lowstat}), average student-teacher ratio of schools in the community (\texttt{stratio}), log(property tax per \$1000) (\texttt{lproptax}), log(weighted distances to five employment centres in the Boston region) (\texttt{ldist}), crimes committed per capita (\texttt{crime}), log(annual average nitrogen oxide concentration (pphm)) (\texttt{lnox}) and index of accessibility to radial highways (\texttt{radial}). The log(median house price (\$)) is the dependent variable. \citet{diciccio19hprice2} applied a weighted least squares approach to these data (which accounts for heterogeneity but does not model the dispersion), where they only considered the \texttt{rooms}, \texttt{stratio}, \texttt{ldist} and \texttt{lnox} variables.

\begin{table}[b!]
    \begin{center}
    \caption{\label{tab:dataset_estimates_hprice}Boston House Price Data: estimation metrics}%
    \resizebox{\textwidth}{!}{
    \begin{tabular}{@{}l@{~~} r@{~}c@{~}r@{~} r@{~}c@{~}r@{~} c@{~~} >{}r@{~}>{}r@{~} c@{~~} r@{~}r@{~}  c@{~~} r@{~}r@{}}
        \toprule
{} & \multicolumn{6}{c}{MPR-SIC} && \multicolumn{2}{c}{{BAMLSS}} && \multicolumn{2}{c}{SPR-SIC} && \multicolumn{2}{c}{\begin{tabular}[b]{@{}c@{}}ALASSO \\ -IC\end{tabular}}\\
BIC & \multicolumn{6}{c}{-360} && \multicolumn{2}{c}{{-337}} && \multicolumn{2}{c}{-169} && \multicolumn{2}{c}{-169}\\
{} & \multicolumn{2}{c}{$\hat\beta_j$} & $\Delta\text{BIC}$ & \multicolumn{2}{c}{$\hat\alpha_j$} & $\Delta\text{BIC}$ && $\hat\beta_j$ & $\hat\alpha_j$ && $\hat\beta_j$ & $\hat\alpha_j$ && $\hat\beta_j$ & $\hat\alpha_j$ \\
\midrule
    \texttt{inter}      & 11.16              & (0.28) &        & -3.53          & (0.30) &        && 10.51          & -2.39          && 13.26          & -3.30 && 13.18          & -3.28 \\ 
    \texttt{rooms}      & \textbf{0.24}      & (0.01) & 178.27 &                &        &        && \textbf{0.26}  & \textbf{-0.20} && \textbf{0.10}  &       && \textbf{0.10}  &  \\ 
    \texttt{lowstat}    & \textbf{-0.02}     & (0.00) & 95.03  & \textbf{0.03}  & (0.01) & 5.55   && \textbf{-0.02} & \textbf{0.03}  && \textbf{-0.03} &       && \textbf{-0.03} &  \\ 
    \texttt{stratio}    & \textbf{-0.03}     & (0.00) & 51.66  &                &        &        && \textbf{-0.02} & -0.01          && \textbf{-0.04} &       && \textbf{-0.04} &  \\ 
    \texttt{lproptax}   & \textbf{-0.20}     & (0.03) & 39.81  &                &        &        && \textbf{-0.16} & 0.45           && \textbf{-0.26} &       && \textbf{-0.25} &  \\ 
    \texttt{ldist}      & \textbf{-0.16}     & (0.02) & 38.50  & \textbf{-0.92} & (0.16) & 26.80  && \textbf{-0.11} & -1.21          && \textbf{-0.28} &       && \textbf{-0.27} &  \\ 
    \texttt{crime}      & \textbf{-0.01}     & (0.00) & 25.87  &                &        &        && \textbf{-0.01} & -0.01          && \textbf{-0.01} &       && \textbf{-0.01} &  \\ 
    \texttt{lnox}       & \textbf{-0.39}     & (0.08) & 19.18  &                &        &        && \textbf{-0.28} & -1.19          && \textbf{-0.62} &       && \textbf{-0.60} &  \\ 
    \texttt{radial}     & \textbf{0.01}      & (0.00) & 16.74  & \textbf{0.05}  & (0.01) & 20.54  && \textbf{0.00}  & \textbf{0.05}  && \textbf{0.01}  &       && \textbf{0.01}  &  \\ 
  \bottomrule
  \multicolumn{16}{p{0.8\textwidth}}{\footnotesize Significant effects indicated in bold.}\\
\end{tabular}}
\end{center}
\end{table}

\begin{figure}[b!]
\centering
\makebox{\includegraphics[width = 0.85\textwidth]{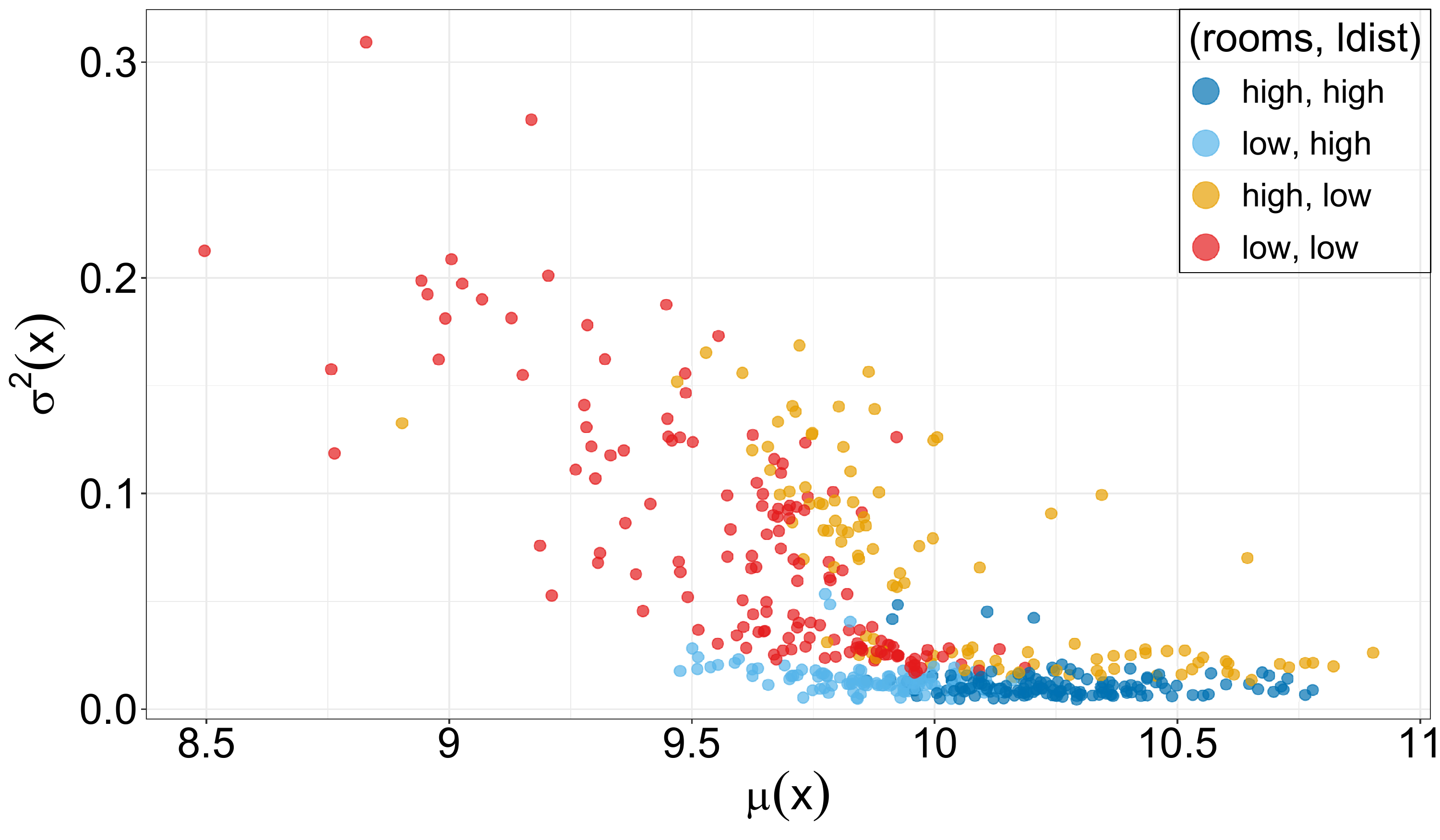}}
\caption{\label{fig:hprice_scatter}Boston House Price Data: mean-variance pairs for each of the 506 communities. The colour of the points relate to whether a community has higher or lower values for \texttt{rooms} and \texttt{ldist} than the medians for each, e.g., the dark blue points correspond to communities where both the \texttt{rooms} and \texttt{ldist} values are higher than the median values.}
\end{figure}

Table~\ref{tab:dataset_estimates_hprice} shows that all eight covariates are included in the location component across the MPR-SIC, {BAMLSS}, SPR-SIC and ALASSO-IC methods. All the coefficients are statistically significant, and the signs of the estimated location coefficients are alike across the methods. Only two covariates (\texttt{rooms} and \texttt{radial}) have a positive effect on house prices. Houses with a greater number of rooms and access to radial highways are generally desirable, which results in increased house prices. The remaining variables may be considered to be undesirable, thus reducing the median house prices. In particular, the percentage of the population in the community that are ``low status" and the student-teacher ratio of schools in the community both have a sizeable impact on the BIC when they are removed from the location parameter. The MPR-SIC method selects three covariates in the dispersion parameter, \texttt{lowstat}, \texttt{ldist} and \texttt{radial}, of which \texttt{ldist} has a negative coefficient while the other two have positive coefficients. We note that \texttt{ldist}, when dropped from the dispersion, leads to a greater $\Delta$BIC value than three of the eight variables in the location component, again highlighting that modelling only the location ignores important effects. Interestingly, the BAMLSS approach finds \texttt{rooms} rather than \texttt{ldist} to be statistically significant, but is otherwise quite comparable to the MPR-SIC approach in the values of the (statistically significant) model coefficients. Moreover, we can see that the BIC values for these two models are much lower than those of the location-regression models (SPR-SIC and ALASSO-IC).

It is useful to estimate the average sales price and the variability for a given community, and, therefore, for each of the 506 communities in the dataset, we compute the mean-variance pairs $(\mu(x_i), \sigma^2(x_i))^T$ where $\mu(x) = x^T\beta$ and $\sigma^2(x) = e^{x^T\alpha}$. These are displayed in Figure~\ref{fig:hprice_scatter} and note also that each point is equivalent to an underlying conditional density (as the normal distribution is fully characterized by its mean and variance). Interestingly, we see that communities with higher prices also tend to have lower variability in these prices. The points are coloured according to the \texttt{rooms} and \texttt{ldist} values, these being the most important location and dispersion variables, respectively. From this, we see that higher values of \texttt{rooms} are associated with increased prices, while higher values of \texttt{ldist} are associated with reduced variability. Thus, from the perspective of the real estate agent, desirable homes are those with a higher number of rooms located a greater distance away from employment centres. That being said, there are certainly other factors influencing house prices and their variability as previously discussed based on Table~\ref{tab:dataset_estimates_hprice}.

\subsection{Prediction Coverage Probabilities}\label{sec:real_data_pcps}

\begin{table}[b!]
\begin{center}
\caption{\label{tab:dataset_PI}Real data analyses results: out-of-sample prediction coverage probabilities}%
\begin{tabular}{@{}l@{~~~}  c@{~~}c@{~~}c@{~~} c@{~~} >{}c@{~~}>{}c@{~~}>{}c@{~~} c@{~~} c@{~~}c@{~~}c@{~~} c@{~~} c@{~~}c@{~~}c@{}} 
\toprule
{} & \multicolumn{3}{c}{MPR-SIC} && \multicolumn{3}{c}{{BAMLSS}} && \multicolumn{3}{c}{SPR-SIC} && \multicolumn{3}{c}{ALASSO-IC} \\
\cmidrule(r){2-4} \cmidrule(r){6-8} \cmidrule(r){10-12} \cmidrule(){14-16}
$n$         & {(a)}  & {(b)}  & {(c)} && {(a)}  & {(b)}  & {(c)} && {(a)}  & {(b)}  & {(c)} && {(a)}  & {(b)}  & {(c)} \\
\midrule
Low         & 0.88 & 0.86 & 0.91 && 0.95 & 0.89 & 0.94 && 0.95 & 0.94 & 0.99 && 0.95 & 0.96 & 0.99 \\ 
Medium      & 0.89 & 0.90 & 0.94 && 0.92 & 0.87 & 0.96 && 0.89 & 0.99 & 0.98 && 0.90 & 0.99 & 0.98 \\ 
High        & 0.96 & 0.95 & 0.92 && 0.96 & 0.98 & 0.91 && 0.79 & 0.87 & 0.80 && 0.79 & 0.87 & 0.80 \\[0.2cm] 
Overall     & 0.89 & 0.90 & 0.93 && 0.90 & 0.92 & 0.94 && 0.86 & 0.92 & 0.93 && 0.87 & 0.93 & 0.93 \\
  \bottomrule
  \multicolumn{16}{p{0.8\textwidth}}{\footnotesize(a) Prostate cancer data, low: $\sigma_i \leq 0.6$, medium: $\sigma_i \in (0.6,1.2]$, high: $\sigma_i > 1.2$.}\\
    \multicolumn{16}{p{.8\textwidth}}{\footnotesize(b) Sniffer data, low: $\sigma_i \leq 2.5$, medium: $\sigma_i \in (2.5,2.9]$, high: $\sigma_i > 2.9$.}\\
  \multicolumn{16}{p{.8\textwidth}}{\footnotesize(c) Boston house price data, low: $\sigma_i \leq 0.1$, medium: $\sigma_i \in (0.1,0.2]$, high: $\sigma_i > 0.2$.}\\
\end{tabular}
\end{center}
\end{table}

\begin{figure}[t!]
\centering
\makebox{\includegraphics[width = \textwidth]{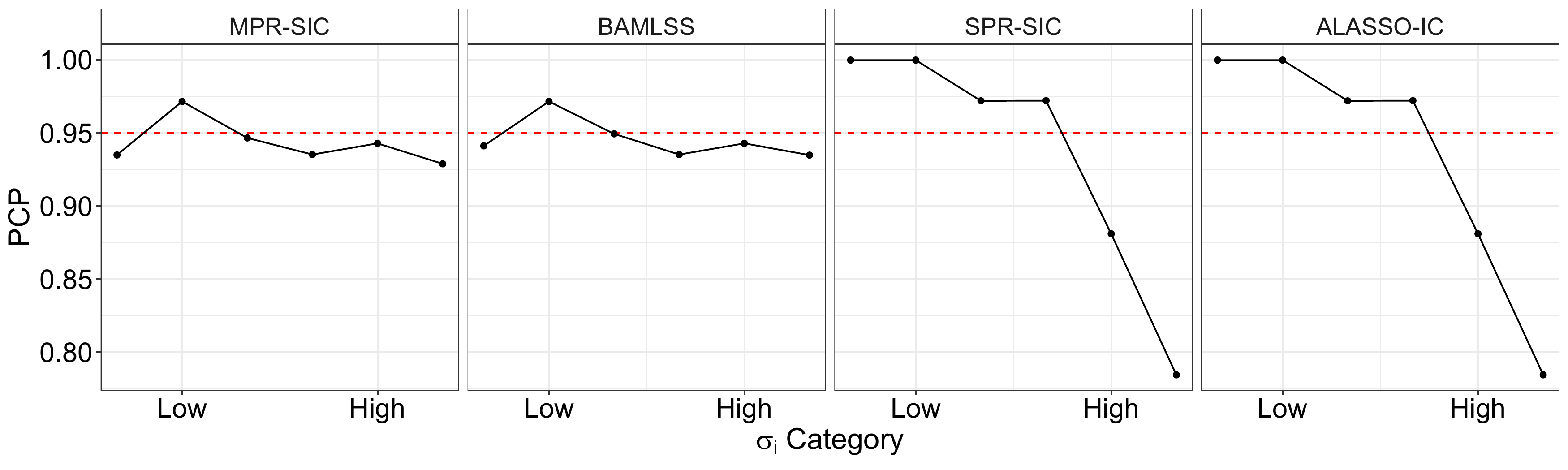}}
\caption{\label{fig:hprice_pcp}Boston House Price Data: prediction coverage probabilities (PCPs) of observations for different dispersion levels, $\sigma_i$. Solid black line indicates the coverage and the red dashed line is a reference line at 0.95.}
\end{figure}

Table~\ref{tab:dataset_PI} contains the out-of-sample PCPs overall and split by category of variability calculated using 10-fold cross-validation for the prostate cancer, sniffer, and Boston house price data. Considering the PCPs from an overall point of view, the MPR-SIC, BAMLSS, SPR-SIC and ALASSO-IC methods perform similarly across all three data analyses. As expected based on the simulation, the coverage improves with respect to sample size, as we compare results from the smaller prostate cancer and sniffer datasets with the larger Boston house price dataset. The overall pattern is that both the SPR-SIC and ALASSO-IC methods tend to produce wider PIs for the low and medium variability categories, and narrower PIs for the high category --- but do okay in terms of coverage for the low variability cases in the two smaller datasets (prostate cancer and sniffer data). The MPR-SIC and BAMLSS approaches are more balanced and tend towards good coverage with increasing sample size --- whereas the other two methods continue to produce overly wide intervals for the low and medium variability categories and overly narrow intervals for the high category. This effect can be seen in Figure~\ref{fig:hprice_pcp} for the Boston house price data where the coverage is displayed for six $\sigma_i$ categories; the coverage for both the MPR-SIC and BAMLSS methods lie close to 95\%, while this is not the case for the other methods (that do not model the dispersion).

\section{Discussion} \label{sec:discussion}
Our proposed variable selection procedure uses a smooth $L_0$ norm to facilitate smooth information criterion optimization, and is extended for application in the developing area of distributional regression. This enables straightforward selection of more complex covariate effects afforded by modelling multiple distributional parameters simultaneously. The smooth objective function means that this is achieved using standard gradient based optimization procedures, i.e., Newton-Raphson. Moreover, because the objective function is an information criterion, the approach circumvents the need for penalty tuning parameter optimization, e.g., this is fixed at $\lambda = \log(n)$ in the BIC case. This is something that can be computationally intensive in LASSO-type problems, especially in the context of distributional regression modelling due to the additional parameters to be estimated and the fact that, in theory, there may be a separate tuning parameter for each distributional parameter. We provide a publicly available package \texttt{smoothic} for the implementation of our proposed methods \citep{smoothic_package}.

Through extensive simulation studies, we have demonstrated that the procedure has very favourable performance in terms of variable selection, parameter inference, and out-of-sample prediction; this is true in both single and multiparameter settings. Results from the real data analyses illustrate the effectiveness of our procedure, and in particular, the advantage of modelling the dispersion is clear from the fact that we have found dispersion effects that are stronger (in BIC terms) than location effects.

The methods proposed in this article are not restricted for use with only the normal distribution. We believe that the techniques presented herein can be extended for use with non-normal models, for example, the generalized linear model family. Moreover, we anticipate that the methods will also extend to the setting of robust statistical modelling, which is particularly important given the ever-growing presence of complex datasets \citep{fan14robustadaptive, avella15robustreview, maronna19robustbook}. Such an extension would be capable of dealing with heteroscedasticity and outliers, while also carrying out variable selection and parameter estimation simultaneously. Additionally, an anonymous reviewer has pointed out that, in combination with our proposal, a fused or group penalty would also be useful in practice for nominal or ordinal covariates. Such extensions will be a focus of our future work.

\section*{Acknowledgements}
This work was carried out within the Confirm Smart Manufacturing Research Centre (https://confirm.ie/) funded by Science Foundation Ireland (grant number: 16/RC/3918). We would also like to thank Prof.~James Gleeson for his support.

\bibliographystyle{apalike}
\bibliography{bibfile}

\clearpage
\vspace*{0.5em}
\begin{spacing}{1.5}
\begin{center}
    {\LARGE Supplementary Material for ``Variable Selection Using a Smooth Information Criterion for Distributional Regression Models"} \par
    \vskip 1.5em
       {\large   Meadhbh O'Neill \qquad \qquad Kevin Burke}
\end{center}
\end{spacing}

\appendix
\counterwithin{figure}{section}
\counterwithin{table}{section}
\renewcommand\thefigure{\thesection\arabic{figure}}
\renewcommand\thetable{\thesection\arabic{table}}

\section{Simulation Results:\\ Fixed Smoothing Parameter $\epsilon$}\label{app:jumpin}
This section contains additional simulation results for the MPR-SIC method. The smoothing parameter $\epsilon$ is fixed at a single value (i.e.,~the $\epsilon$-telescoping procedure is not performed). The smoothing parameter is fixed at $\epsilon = 10^{-1},$ $10^{-2},$ and $10^{-3}$.

\begin{itemize}
    \item Table~\ref{tab:var_selection_jumpin} is analogous to Table 4 of the main paper, but showing the model selection metrics when the smoothing parameter $\epsilon$ is fixed at a single value. The variable selection performance is poor for all values of $\epsilon$ that are tested.
    \item Table~\ref{tab:parameter_inference_jumpin} in analogous to Table 5 of the main paper, but showing the estimation and inference metrics when the smoothing parameter $\epsilon$ is fixed at a single value. The estimation and inferential performance for the location component of the model does not appear to be impacted. However, for $\epsilon = 10^{-2}$ and $10^{-3}$, the estimation and inferential performance for the dispersion component does not perform well --- the estimated values do not move from the initial values that are used in the optimization.
\end{itemize}

\begin{table}[h!]
\begin{center}
\caption{\label{tab:var_selection_jumpin}Simulation results: model selection metrics for MPR-SIC with fixed smoothing parameter $\epsilon$}%
\resizebox{\textwidth}{!}{
\begin{tabular}{@{}ccc@{~~}c@{~~}c@{~~}c@{~~}c@{~~~}c@{~~}c@{~~}c@{~~}c@{~~}c@{~~~}c@{~~}c@{~~}c@{~~}c@{}}
  \toprule
 {} & {} & \multicolumn{4}{c}{$\epsilon = 10^{-1}$} && \multicolumn{4}{c}{$\epsilon = 10^{-2}$} && \multicolumn{4}{c}{$\epsilon = 10^{-3}$} \\
 \cmidrule(lr){3-6} \cmidrule(r){8-11} \cmidrule(){13-16}
 {}     & $n$   & C(6)  & IC(0) & PT & MSE && C(6) & IC(0) & PT & MSE && C(6) & IC(0) & PT & MSE \\ 
 \midrule
 $\beta$ & 100  & 0.00 & 0.00 & 0.00 & 1.59 && 0.00 & 0.00 & 0.00 & 2.48 && 0.00 & 0.00 & 0.00 & 2.52 \\
 {}      & 500  & 0.00 & 0.00 & 0.00 & 0.12 && 0.00 & 0.00 & 0.00 & 0.36 && 0.01 & 0.00 & 0.00 & 0.53 \\
 {}      & 1000 & 0.00 & 0.00 & 0.00 & 0.05 && 0.00 & 0.00 & 0.00 & 0.13 && 0.01 & 0.00 & 0.00 & 0.27 \\[0.2cm]

 $\alpha$ & 100     & 0.31 & 0.31 & 0.00 & 3.42 && 0.47 & 0.47 & 0.00 & 6.21 && 0.08 & 0.07 & 0.00 & 6.21 \\ 
 {}       & 500     & 0.01 & 0.01 & 0.00 & 0.35 && 0.13 & 0.13 & 0.00 & 4.49 && 0.33 & 0.32 & 0.00 & 7.01 \\ 
 {}       & 1000    & 0.03 & 0.03 & 0.00 & 0.16 && 0.02 & 0.02 & 0.00 & 1.41 && 0.46 & 0.45 & 0.00 & 7.09 \\
 \bottomrule
   \multicolumn{16}{p{.9\textwidth}}{\footnotesize C, average correct zeros; IC, average incorrect zeros; PT, the probability of choosing the true model; MSE, the average mean squared error.}\\
\end{tabular}}
\end{center}
\end{table}

\begin{table}[h!]
\begin{center}
\caption{\label{tab:parameter_inference_jumpin}Simulation results: estimation and inference metrics for MPR-SIC with fixed smoothing parameter $\epsilon$}%
\begin{tabular}{@{}c@{~~}c@{~~~} c@{~~}c@{~~}c@{~~}c@{~~} c@{~~~} c@{~~}c@{~~}c@{~~}c@{~~} c@{~~~} c@{~~}c@{~~}c@{~~}c@{}}
\toprule
\multicolumn{2}{l}{$\epsilon = 10^{-1}$} & \multicolumn{4}{c}{$n = 100$} && \multicolumn{4}{c}{$n = 500$} && \multicolumn{4}{c}{$n = 1000$} \\
\cmidrule(r){3-6} \cmidrule(r){8-11} \cmidrule(){13-16}
{} & $\theta$ & $\hat{\theta}$ & SE & SEE & CP && $\hat{\theta}$ & SE & SEE & CP && $\hat{\theta}$ & SE & SEE & CP \\
\midrule
  $\beta_{0}$ & 0.0 & 0.04 & 0.67 & 1.94 & 0.93  && 0.01 & 0.17 & 0.30 & 0.92 && -0.00 & 0.12 & 0.08 & 0.93 \\
  $\beta_{1}$ & 1.0 & 1.00 & 0.53 & 0.87 & 0.87  && 0.99 & 0.10 & 0.13 & 0.93 && 1.00 & 0.07 & 0.04 & 0.94 \\ 
  $\beta_{2}$ & 0.5 & 0.52 & 0.64 & 1.27 & 0.72  && 0.50 & 0.19 & 0.17 & 0.90 && 0.50 & 0.10 & 0.06 & 0.93 \\ 
  $\beta_{3}$ & 0.5 & 0.46 & 0.36 & 1.10 & 0.82  && 0.50 & 0.08 & 0.09 & 0.94 && 0.50 & 0.03 & 0.03 & 0.94 \\ 
  $\beta_{4}$ & 1.0 & 0.99 & 0.41 & 0.68 & 0.90  && 1.00 & 0.06 & 0.08 & 0.93 && 1.00 & 0.03 & 0.03 & 0.93 \\ 
  $\beta_{5}$ & 0.5 & 0.46 & 0.36 & 0.66 & 0.81  && 0.50 & 0.08 & 0.08 & 0.93 && 0.50 & 0.04 & 0.03 & 0.94 \\ 
  $\beta_{6}$ & 1.0 & 0.95 & 0.77 & 1.43 & 0.75  && 0.98 & 0.27 & 0.54 & 0.91 && 1.00 & 0.14 & 0.09 & 0.90 \\[0.1cm]
  
  $\alpha_{0}$ & 0.0 & 1.43 & 0.62 & 0.91 & 0.28  && 0.14 & 0.44 & 0.41 & 0.82 && 0.06 & 0.34 & 0.13 & 0.83 \\ 
  $\alpha_{1}$ & 0.5 & 0.12 & 0.22 & 0.27 & 0.35  && 0.45 & 0.12 & 0.11 & 0.85  && 0.48 & 0.08 & 0.06 & 0.89 \\ 
  $\alpha_{2}$ & 1.0 & 0.16 & 0.30 & 0.41 & 0.26  && 0.88 & 0.32 & 0.43 & 0.83  && 0.95 & 0.20 & 0.28 & 0.90 \\ 
  $\alpha_{3}$ & 0.5 & 0.06 & 0.16 & 0.24 & 0.17  && 0.44 & 0.17 & 0.34 & 0.88  && 0.48 & 0.09 & 0.07 & 0.89 \\ 
  $\alpha_{4}$ & 1.0 & 0.24 & 0.34 & 0.41 & 0.33  && 0.91 & 0.21 & 0.15 & 0.79  && 0.96 & 0.14 & 0.06 & 0.85 \\ 
  $\alpha_{7}$ & 0.5 & 0.09 & 0.20 & 0.28 & 0.27  && 0.45 & 0.13 & 0.16 & 0.89  && 0.48 & 0.08 & 0.06 & 0.89 \\ 
  $\alpha_{8}$ & 1.0 & 0.24 & 0.35 & 0.49 & 0.30  && 0.91 & 0.22 & 0.24 & 0.77  && 0.96 & 0.15 & 0.06 & 0.84 \\
  \bottomrule
\end{tabular}
  \end{center}
  \end{table}

  \begin{table}[h!]
  \ContinuedFloat
\begin{center}
\begin{tabular}{@{}c@{~~}c@{~~~} c@{~~}c@{~~}c@{~~}c@{~~} c@{~~~} c@{~~}c@{~~}c@{~~}c@{~~} c@{~~~} c@{~~}c@{~~}c@{~~}c@{}}
\toprule
\multicolumn{2}{l}{$\epsilon = 10^{-2}$} & \multicolumn{4}{c}{$n = 100$} && \multicolumn{4}{c}{$n = 500$} && \multicolumn{4}{c}{$n = 1000$} \\
\cmidrule(r){3-6} \cmidrule(r){8-11} \cmidrule(){13-16}
{} & $\theta$ & $\hat{\theta}$ & SE & SEE & CP && $\hat{\theta}$ & SE & SEE & CP && $\hat{\theta}$ & SE & SEE & CP \\
\midrule
  $\beta_{0}$ & 0.0 & -0.01 & 0.78 & 2.04 & 0.94 && -0.00 & 0.31 & 0.67 & 0.91 && 0.01 & 0.19 & 0.28 & 0.90 \\ 
  $\beta_{1}$ & 1.0 & 1.05 & 0.69 & 0.52 & 0.79 && 1.00 & 0.24 & 0.28 & 0.78 && 0.99 & 0.11 & 0.09 & 0.90 \\ 
  $\beta_{2}$ & 0.5 & 0.54 & 0.76 & 1.04 & 0.78 && 0.49 & 0.29 & 0.36 & 0.87 && 0.50 & 0.13 & 0.14 & 0.93 \\ 
  $\beta_{3}$ & 0.5 & 0.49 & 0.37 & 1.21 & 0.87 && 0.49 & 0.14 & 0.19 & 0.97 && 0.50 & 0.06 & 0.08 & 0.98 \\ 
  $\beta_{4}$ & 1.0 & 1.01 & 0.53 & 0.42 & 0.84 && 1.00 & 0.18 & 0.22 & 0.86 && 1.00 & 0.08 & 0.07 & 0.92 \\ 
  $\beta_{5}$ & 0.5 & 0.49 & 0.41 & 0.47 & 0.82 && 0.49 & 0.14 & 0.26 & 0.94 && 0.50 & 0.06 & 0.06 & 0.95 \\ 
  $\beta_{6}$ & 1.0 & 0.97 & 0.82 & 1.25 & 0.83 && 1.00 & 0.32 & 1.61 & 0.91 && 1.00 & 0.19 & 0.20 & 0.91 \\[0.1cm]
  
  $\alpha_{0}$ & 0.0 & 2.25 & 0.60 & 0.15 & 0.00 && 1.81 & 0.74 & 0.30 & 0.07 && 0.70 & 0.82 & 0.27 & 0.34 \\ 
  $\alpha_{1}$ & 0.5 & 0.00 & 0.00 & 0.00 & 0.00 && 0.11 & 0.20 & 0.05 & 0.17 && 0.36 & 0.21 & 0.07 & 0.67 \\ 
  $\alpha_{2}$ & 1.0 & 0.00 & 0.00 & 0.00 & 0.00 && 0.24 & 0.41 & 0.08 & 0.20 && 0.75 & 0.41 & 0.20 & 0.69 \\ 
  $\alpha_{3}$ & 0.5 & 0.00 & 0.00 & 0.00 & 0.00 && 0.03 & 0.09 & 0.01 & 0.03 && 0.14 & 0.22 & 0.04 & 0.25 \\ 
  $\alpha_{4}$ & 1.0 & 0.00 & 0.00 & 0.00 & 0.00 && 0.25 & 0.42 & 0.05 & 0.21 && 0.77 & 0.39 & 0.09 & 0.68 \\ 
  $\alpha_{7}$ & 0.5 & 0.00 & 0.00 & 0.00 & 0.00 && 0.06 & 0.16 & 0.02 & 0.09 && 0.29 & 0.24 & 0.08 & 0.56 \\ 
  $\alpha_{8}$ & 1.0 & 0.00 & 0.00 & 0.00 & 0.00 && 0.24 & 0.41 & 0.11 & 0.19 && 0.77 & 0.39 & 0.09 & 0.68 \\
  \bottomrule

&&&&&&&&&&&&&&&\\[-0.15cm]
\multicolumn{2}{l}{$\epsilon = 10^{-3}$} & \multicolumn{4}{c}{$n = 100$} && \multicolumn{4}{c}{$n = 500$} && \multicolumn{4}{c}{$n = 1000$} \\
\cmidrule(r){3-6} \cmidrule(r){8-11} \cmidrule(){13-16}
{} & $\theta$ & $\hat{\theta}$ & SE & SEE & CP && $\hat{\theta}$ & SE & SEE & CP && $\hat{\theta}$ & SE & SEE & CP \\
\midrule
  $\beta_{0}$ & 0.0 & -0.01 & 0.78 & 0.77 & 0.93 && 0.02 & 0.38 & 0.55 & 0.92 && 0.00 & 0.27 & 0.27 & 0.90 \\ 
  $\beta_{1}$ & 1.0 & 1.05 & 0.69 & 0.37 & 0.78 && 1.00 & 0.34 & 0.19 & 0.73 && 1.00 & 0.25 & 0.12 & 0.73 \\ 
  $\beta_{2}$ & 0.5 & 0.54 & 0.76 & 0.60 & 0.85 && 0.50 & 0.33 & 0.33 & 0.89 && 0.50 & 0.23 & 0.20 & 0.90 \\ 
  $\beta_{3}$ & 0.5 & 0.50 & 0.37 & 0.43 & 0.94 && 0.50 & 0.16 & 0.21 & 0.98 && 0.50 & 0.11 & 0.14 & 0.99 \\ 
  $\beta_{4}$ & 1.0 & 1.01 & 0.54 & 0.35 & 0.83 && 1.00 & 0.24 & 0.18 & 0.84 && 1.00 & 0.17 & 0.12 & 0.84 \\ 
  $\beta_{5}$ & 0.5 & 0.49 & 0.42 & 0.35 & 0.87 && 0.49 & 0.17 & 0.18 & 0.94 && 0.50 & 0.12 & 0.12 & 0.95 \\ 
  $\beta_{6}$ & 1.0 & 0.97 & 0.82 & 0.75 & 0.89 && 1.00 & 0.35 & 0.46 & 0.96 && 0.99 & 0.25 & 0.28 & 0.95 \\[0.1cm]
  
  $\alpha_{0}$ & 0.0 & 2.25 & 0.60 & 0.14 & 0.00 && 2.54 & 0.34 & 0.06 & 0.00 && 2.57 & 0.27 & 0.04 & 0.00 \\ 
  $\alpha_{1}$ & 0.5 & 0.00 & 0.00 & 0.00 & 0.00 && 0.00 & 0.02 & 0.00 & 0.00 && 0.00 & 0.02 & 0.00 & 0.00 \\ 
  $\alpha_{2}$ & 1.0 & 0.00 & 0.00 & 0.00 & 0.00 && 0.00 & 0.00 & 0.00 & 0.00 && 0.00 & 0.03 & 0.00 & 0.00 \\ 
  $\alpha_{3}$ & 0.5 & 0.00 & 0.00 & 0.00 & 0.00 && 0.00 & 0.00 & 0.00 & 0.00 && 0.00 & 0.02 & 0.00 & 0.00 \\ 
  $\alpha_{4}$ & 1.0 & 0.00 & 0.00 & 0.00 & 0.00 && 0.00 & 0.00 & 0.00 & 0.00 && 0.00 & 0.03 & 0.00 & 0.00 \\ 
  $\alpha_{7}$ & 0.5 & 0.00 & 0.00 & 0.00 & 0.00 && 0.00 & 0.00 & 0.00 & 0.00 && 0.00 & 0.02 & 0.00 & 0.00 \\ 
  $\alpha_{8}$ & 1.0 & 0.00 & 0.00 & 0.00 & 0.00 && 0.00 & 0.00 & 0.00 & 0.00 && 0.00 & 0.00 & 0.00 & 0.00 \\
  \bottomrule
  \multicolumn{16}{p{.9\textwidth}}{\footnotesize SE, standard deviation of estimates over 1000 replications; SEE, average of estimated standard errors over 1000 replications; CP, the empirical coverage probability of a nominal 95\% confidence interval.}\\
\end{tabular}
\end{center}
\end{table}

\FloatBarrier

\section{Simulation Results:\\ Fewer steps $T$ in the $\epsilon$-telescope}\label{app:lesssteps}
This section contains additional simulation results for the MPR-SIC method, where fewer steps $T$ are implemented in the $\epsilon$-telescope. Results for sequences of $T=50$ and $T=10$ steps.
\begin{itemize}
    \item Table~\ref{tab:var_selection_lesssteps} is analogous to Table 4 of the main paper, but showing the model selection metrics for fewer steps $T$ in the $\epsilon$-telescope. For $T=50$, the performance is comparable with the performance when $T=100$. When $T=10$, the model selection metrics are poor as the zero coefficients are not being driven close enough to zero.
    \item Table~\ref{tab:parameter_inference_lesssteps} is analogous to Table 5 of the main paper, but gives the estimation and inference metrics when fewer steps $T$ are used in the $\epsilon$-telescope. Estimation and inferential results are comparable with the case when $T=100$.
\end{itemize}

\begin{table}[h!]
\begin{center}
\caption{\label{tab:var_selection_lesssteps}Simulation results: model selection metrics for MPR-SIC with fewer steps $T$}%
\begin{tabular}{@{}cc  c@{~~}c@{~~}c@{~~}c@{~~}   c@{~~~}   c@{~~}c@{~~}c@{~~}c@{}}
  \toprule
 {} & {} & \multicolumn{4}{c}{$T=50$} && \multicolumn{4}{c}{$T = 10$} \\
 \cmidrule(lr){3-6} \cmidrule(r){8-11}
 {}     & $n$   & C(6)  & IC(0) & PT & MSE && C(6) & IC(0) & PT & MSE \\ 
 \midrule
 $\beta$ & 100      & 5.14 & 0.14 & 0.43 & 0.14 && 0.49 & 0.00 & 0.00 & 0.20  \\
 {}      & 500      & 5.86 & 0.00 & 0.88 & 0.01 && 0.32 & 0.00 & 0.00 & 0.02  \\
 {}      & 1000     & 5.92 & 0.00 & 0.94 & 0.00 && 0.34 & 0.00 & 0.01 & 0.01  \\[0.2cm]

 $\alpha$ & 100     & 5.42 & 0.71 & 0.32 & 0.63 && 0.43 & 0.02 & 0.00 & 0.97 \\ 
 {}       & 500     & 5.89 & 0.00 & 0.91 & 0.04 && 0.24 & 0.00 & 0.00 & 0.06 \\ 
 {}       & 1000    & 5.92 & 0.00 & 0.94 & 0.02 && 0.27 & 0.00 & 0.00 & 0.03 \\
 \bottomrule
   \multicolumn{11}{p{.5\textwidth}}{\footnotesize C, average correct zeros; IC, average incorrect zeros; PT, the probability of choosing the true model; MSE, the average mean squared error.}\\
\end{tabular}
\end{center}
\end{table}

\begin{table}[h!]

\begin{center}
\caption{\label{tab:parameter_inference_lesssteps}Simulation results: estimation and inference metrics for MPR-SIC with fewer steps $T$}%
\begin{tabular}{@{}c@{~~}c@{~~~} c@{~~}c@{~~}c@{~~}c@{~~} c@{~~~} c@{~~}c@{~~}c@{~~}c@{~~} c@{~~~} c@{~~}c@{~~}c@{~~}c@{}}
\toprule
\multicolumn{16}{l}{$T=50$}\\
{} & {} & \multicolumn{4}{c}{$n = 100$} && \multicolumn{4}{c}{$n = 500$} && \multicolumn{4}{c}{$n = 1000$} \\
\cmidrule(r){3-6} \cmidrule(r){8-11} \cmidrule(){13-16}
{} & $\theta$ & $\hat{\theta}$ & SE & SEE & CP && $\hat{\theta}$ & SE & SEE & CP && $\hat{\theta}$ & SE & SEE & CP \\
\midrule
  $\beta_{0}$ & 0.0 & 0.01 & 0.22 & 0.13 & 0.77 && -0.00 & 0.06 & 0.05 & 0.93 && -0.00 & 0.04 & 0.04 & 0.94 \\
  $\beta_{1}$ & 1.0 & 1.00 & 0.15 & 0.10 & 0.77 && 1.00 & 0.04 & 0.04 & 0.93 && 1.00 & 0.03 & 0.03 & 0.96 \\ 
  $\beta_{2}$ & 0.5 & 0.48 & 0.23 & 0.10 & 0.72 && 0.50 & 0.05 & 0.05 & 0.92 && 0.50 & 0.03 & 0.03 & 0.94 \\ 
  $\beta_{3}$ & 0.5 & 0.49 & 0.11 & 0.07 & 0.79 && 0.50 & 0.03 & 0.03 & 0.94 && 0.50 & 0.02 & 0.02 & 0.94 \\ 
  $\beta_{4}$ & 1.0 & 1.00 & 0.11 & 0.07 & 0.80 && 1.00 & 0.03 & 0.03 & 0.94 && 1.00 & 0.02 & 0.02 & 0.93 \\ 
  $\beta_{5}$ & 0.5 & 0.50 & 0.11 & 0.07 & 0.77 && 0.50 & 0.03 & 0.03 & 0.93 && 0.50 & 0.02 & 0.02 & 0.95 \\ 
  $\beta_{6}$ & 1.0 & 1.01 & 0.24 & 0.11 & 0.71 && 1.00 & 0.05 & 0.05 & 0.91 && 1.00 & 0.04 & 0.03 & 0.94 \\[0.2cm]
  
  $\alpha_{0}$ & 0.0 & -0.18 & 0.44 & 0.23 & 0.66 && -0.03 & 0.11 & 0.10 & 0.90 && -0.02 & 0.07 & 0.07 & 0.91 \\
  $\alpha_{1}$ & 0.5 & 0.45 & 0.28 & 0.13 & 0.73 && 0.50 & 0.07 & 0.07 & 0.93 && 0.50 & 0.05 & 0.05 & 0.94 \\ 
  $\alpha_{2}$ & 1.0 & 1.07 & 0.37 & 0.17 & 0.77 && 1.01 & 0.08 & 0.07 & 0.92 && 1.01 & 0.05 & 0.05 & 0.93 \\ 
  $\alpha_{3}$ & 0.5 & 0.52 & 0.36 & 0.14 & 0.59 && 0.51 & 0.08 & 0.08 & 0.95 && 0.51 & 0.05 & 0.05 & 0.94 \\ 
  $\alpha_{4}$ & 1.0 & 1.11 & 0.24 & 0.17 & 0.82 && 1.01 & 0.07 & 0.07 & 0.95 && 1.01 & 0.05 & 0.05 & 0.94 \\ 
  $\alpha_{7}$ & 0.5 & 0.51 & 0.28 & 0.14 & 0.73 && 0.51 & 0.06 & 0.07 & 0.96 && 0.50 & 0.05 & 0.05 & 0.94 \\ 
  $\alpha_{8}$ & 1.0 & 1.11 & 0.24 & 0.17 & 0.81 && 1.01 & 0.07 & 0.07 & 0.93 && 1.01 & 0.05 & 0.05 & 0.93 \\
  \bottomrule
  &&&&&&&&&&&&&&&\\[-0.15cm]
  \multicolumn{16}{l}{$T=10$}\\
{} & {} & \multicolumn{4}{c}{$n = 100$} && \multicolumn{4}{c}{$n = 500$} && \multicolumn{4}{c}{$n = 1000$} \\
\cmidrule(r){3-6} \cmidrule(r){8-11} \cmidrule(){13-16}
{} & $\theta$ & $\hat{\theta}$ & SE & SEE & CP && $\hat{\theta}$ & SE & SEE & CP && $\hat{\theta}$ & SE & SEE & CP \\
\midrule
  $\beta_{0}$ & 0.0 & 0.01 & 0.26 & 0.18 & 0.79 && -0.00 & 0.08 & 0.08 & 0.93 && -0.00 & 0.05 & 0.05 & 0.94 \\
  $\beta_{1}$ & 1.0 & 1.00 & 0.17 & 0.11 & 0.78 && 1.00 & 0.05 & 0.05 & 0.93 && 1.00 & 0.03 & 0.03 & 0.95 \\ 
  $\beta_{2}$ & 0.5 & 0.50 & 0.20 & 0.13 & 0.79 && 0.50 & 0.06 & 0.05 & 0.92 && 0.50 & 0.03 & 0.03 & 0.95 \\ 
  $\beta_{3}$ & 0.5 & 0.49 & 0.12 & 0.08 & 0.80 && 0.50 & 0.03 & 0.03 & 0.93 && 0.50 & 0.02 & 0.02 & 0.94 \\ 
  $\beta_{4}$ & 1.0 & 1.00 & 0.13 & 0.08 & 0.81 && 1.00 & 0.03 & 0.03 & 0.95 && 1.00 & 0.02 & 0.02 & 0.93 \\ 
  $\beta_{5}$ & 0.5 & 0.51 & 0.11 & 0.08 & 0.78 && 0.50 & 0.03 & 0.03 & 0.93 && 0.50 & 0.02 & 0.02 & 0.94 \\ 
  $\beta_{6}$ & 1.0 & 1.00 & 0.25 & 0.16 & 0.79 && 1.00 & 0.07 & 0.07 & 0.93 && 1.00 & 0.05 & 0.04 & 0.93 \\[0.2cm]
  
  $\alpha_{0}$ & 0.0 & -0.43 & 0.48 & 0.35 & 0.69 && -0.06 & 0.13 & 0.13 & 0.89 && -0.03 & 0.09 & 0.10 & 0.92 \\
  $\alpha_{1}$ & 0.5 & 0.54 & 0.25 & 0.19 & 0.88 && 0.50 & 0.07 & 0.07 & 0.93 && 0.50 & 0.05 & 0.05 & 0.95 \\ 
  $\alpha_{2}$ & 1.0 & 1.17 & 0.44 & 0.32 & 0.83 && 1.03 & 0.11 & 0.12 & 0.93 && 1.01 & 0.08 & 0.08 & 0.94 \\ 
  $\alpha_{3}$ & 0.5 & 0.64 & 0.32 & 0.22 & 0.78 && 0.51 & 0.08 & 0.08 & 0.94 && 0.51 & 0.05 & 0.05 & 0.94 \\ 
  $\alpha_{4}$ & 1.0 & 1.17 & 0.27 & 0.20 & 0.79 && 1.02 & 0.07 & 0.07 & 0.94 && 1.01 & 0.05 & 0.05 & 0.94 \\ 
  $\alpha_{7}$ & 0.5 & 0.59 & 0.26 & 0.20 & 0.85 && 0.51 & 0.07 & 0.07 & 0.96 && 0.50 & 0.05 & 0.05 & 0.94 \\ 
  $\alpha_{8}$ & 1.0 & 1.18 & 0.27 & 0.20 & 0.79 && 1.02 & 0.07 & 0.07 & 0.92 && 1.01 & 0.05 & 0.05 & 0.94 \\
  \bottomrule
  \multicolumn{16}{p{.8\textwidth}}{\footnotesize SE, standard deviation of estimates over 1000 replications; SEE, average of estimated standard errors over 1000 replications; CP, the empirical coverage probability of a nominal 95\% confidence interval.}\\
\end{tabular}
\end{center}
\end{table}

\clearpage

\section{Simulation Results:\\ Single Parameter Setting}\label{app:spr_setting}
This section contains the results of a simulation study carried out in a single parameter setting. Note that fewer replicates for the BAMLSS procedure are used (200 replicates) due to the computational intensity of the method. The other approaches are averaged over 1000 replicates.
\begin{itemize}
    \item Table~\ref{tab:true_values_spr} is analogous to Table 2 of the main paper, but showing the true values that are used to simulate homoscedastic data.
    \item Table~\ref{tab:pcp_spr_setting} is analogous to Table 6 of the main paper, but showing the out-of-sample prediction coverage probabilities from the single parameter setting.
    \item Table~\ref{tab:var_selection_spr_setting} is analogous to Table 4 of the main paper, but showing the model selection metrics from the single parameter setting.
    \item Table~\ref{tab:parameter_inference_mprsic_sprsic_alassoic_spr_setting} is analogous to Table 5 of the main paper, but showing the estimation and inference metrics from the single parameter setting.
\end{itemize}

\begin{table}[h!]
\begin{center}
\caption{\label{tab:true_values_spr}True parameter values}%
\begin{tabular}{@{}cccccccccccccc@{}}
  \toprule
    & & \textcolor{blue}{E} & \textcolor{red}{M} & \textcolor{mygreen}{B} & N & N & \textcolor{red}{M} & N & N & \textcolor{red}{M} & \textcolor{mygreen}{B} & \textcolor{blue}{E} & \textcolor{red}{M} \\
       & $X_0$ & \textcolor{blue}{$X_1$} & \textcolor{red}{$X_2$} & \textcolor{mygreen}{$X_3$} & ${X_4}$ & ${X_5}$ & \textcolor{red}{$X_6$} & ${X_7}$ & $X_8$ & \textcolor{red}{${X_9}$} & \textcolor{mygreen}{$X_{10}$} & \textcolor{blue}{$X_{11}$} & \textcolor{red}{$X_{12}$} \\
    \midrule
$\beta$ & 0 & 1 & 0.5 & 0.5 & 1 & 0.5 & 1 & 0 & 0 & 0 & 0 & 0 & 0 \\[0.05cm] 
$\alpha$ & 0 & 0 & 0 & 0 & 0 & 0 & 0 & 0 & 0 & 0 & 0 & 0 & 0 \\
\bottomrule
   \multicolumn{14}{l}{\footnotesize \textcolor{blue}{E = Exponential}, \textcolor{mygreen}{B = Bernoulli}, N = independent normal,} \\
  \multicolumn{14}{l}{\footnotesize \textcolor{red}{M = multivariate normal (correlated)}.}
\end{tabular}
\end{center}
\end{table}

\begin{table}[h!]

\begin{center}
\caption{\label{tab:pcp_spr_setting}Out-of-sample prediction coverage probabilities}%
\begin{tabular}{@{}l@{~}c@{~~} c@{~}c@{~}c@{~} c@{~~} c@{~}c@{~}c@{~} c@{~~} c@{~}c@{~}c@{~} c@{~~} c@{~}c@{~}c@{}} 
\toprule
{} && \multicolumn{3}{c}{MPR-SIC} && \multicolumn{3}{c}{BAMLSS} && \multicolumn{3}{c}{SPR-SIC} && \multicolumn{3}{c}{ALASSO-IC} \\
\cmidrule(lr){3-5} \cmidrule(r){7-9} \cmidrule(){11-13} \cmidrule(){15-17}
$n$         && 100  & 500  & 1000 && 100  & 500  & 1000 && 100  & 500  & 1000 && 100  & 500  & 1000 \\
\midrule
  Overall   && 0.90 & 0.94 & 0.95 && 0.89 & 0.94 & 0.95 && 0.92 & 0.95 & 0.95 && 0.93 & 0.95 & 0.95 \\
  \bottomrule
  \multicolumn{17}{p{0.8\textwidth}}{\footnotesize Out-of-sample coverage is calculated for a sample 20\% the size of the original data.}
\end{tabular}
\end{center}
\end{table}

\begin{table}[h!]

\begin{center}
\caption{\label{tab:var_selection_spr_setting}Model selection metrics}%
   \begin{tabular}{@{}cc   c@{~~}c@{~~}c@{~~}c@{~~}   c@{~~~}   c@{~~}c@{~~}c@{~~}c@{~~}} 
    \toprule
    {} & {} & \multicolumn{4}{c}{MPR-SIC} && \multicolumn{4}{c}{BAMLSS} \\
    \cmidrule(lr){3-6} \cmidrule(){8-11}
    {}     & $n$   & C(6)  & IC(0) & PT & MSE && C(6) & IC(0) & PT & MSE \\
    \midrule
    $\beta$ & 100  &  5.64  & 0.27 & 0.55 & 0.12  && 5.41  & 0.34 & 0.40 & 0.17 \\
    {}      & 500  &  5.93  & 0.00 & 0.94 & 0.02  && 5.63  & 0.00 & 0.68 & 0.03 \\
    {}      & 1000 &  5.95  & 0.00 & 0.95 & 0.01  && 5.64  & 0.00 & 0.71 & 0.01 \\
    \midrule
    {}     & $n$   & C(12)  & IC(0) & PT & MSE && C(12) & IC(0) & PT & MSE \\
    \midrule
    $\alpha$ & 100     & 11.26 & 0.00 & 0.51 & 0.24   && 11.01 & 0.00 & 0.46 & 0.49 \\
    {}       & 500     & 11.87 & 0.00 & 0.88 & 0.01   && 11.28 & 0.00 & 0.52 & 0.06 \\
    {}       & 1000    & 11.91 & 0.00 & 0.92 & 0.00   && 11.39 & 0.00 & 0.62 & 0.03 \\
    \midrule
    {} & {} & \multicolumn{4}{c}{SPR-SIC} && \multicolumn{4}{c}{ALASSO-IC} \\
    \cmidrule(lr){3-6} \cmidrule(){8-11}
    {}     & $n$   & C(6)  & IC(0) & PT & MSE && C(6) & IC(0) & PT & MSE \\
    \midrule
    $\beta$     & 100  & 5.75  & 0.25 & 0.60 & 0.10  && 5.57  & 0.18 & 0.55 & 0.11 \\
    {}          & 500  & 5.94  & 0.00 & 0.94 & 0.02  && 5.91  & 0.00 & 0.92 & 0.02 \\
    {}          & 1000 & 5.95  & 0.00 & 0.95 & 0.01  && 5.95  & 0.00 & 0.95 & 0.01 \\
    \midrule
    {}     & $n$   & C(12)  & IC(0) & PT & MSE && C(12) & IC(0) & PT & MSE \\
    \midrule
    $\alpha$ & 100     & 12.00 & 0.00 & 1.00 & 0.03  && 12.00 & 0.00 & 1.00 & 0.03 \\
    {}       & 500     & 12.00 & 0.00 & 1.00 & 0.00  && 12.00 & 0.00 & 1.00 & 0.00 \\
    {}       & 1000    & 12.00 & 0.00 & 1.00 & 0.00  && 12.00 & 0.00 & 1.00 & 0.00 \\
    \bottomrule
    \multicolumn{11}{p{0.65\textwidth}}{\footnotesize C, average correct zeros; IC, average incorrect zeros; PT, the probability of choosing the true model; MSE, the average mean squared error.}\\
    \end{tabular}
\end{center}
\end{table}

\begin{table}[t!]

\begin{center}
\caption{\label{tab:parameter_inference_mprsic_sprsic_alassoic_spr_setting}Estimation and inference metrics for MPR-SIC, BAMLSS, SPR-SIC and ALASSO-IC methods}%
\begin{subtable}{\textwidth}
    \centering
    \vspace{-0.1cm}
\begin{tabular}{@{}c@{~~}c@{~~~} c@{~~}c@{~~}c@{~~}c@{~~} c@{~~~} c@{~~}c@{~~}c@{~~}c@{~~} c@{~~~} c@{~~}c@{~~}c@{~~}c@{}}
\toprule
\multicolumn{16}{l}{MPR-SIC}\\
{} & {} & \multicolumn{4}{c}{$n = 100$} && \multicolumn{4}{c}{$n = 500$} && \multicolumn{4}{c}{$n = 1000$} \\
\cmidrule(r){3-6} \cmidrule(r){8-11} \cmidrule(){13-16}
{} & $\theta$ & $\hat{\theta}$ & SE & SEE & CP && $\hat{\theta}$ & SE & SEE & CP && $\hat{\theta}$ & SE & SEE & CP \\
\midrule
  $\beta_{0}$ & 0.0 & 0.01  & 0.19 & 0.15 & 0.87 && -0.00 & 0.07 & 0.07 & 0.93 && -0.00 & 0.05 & 0.05 & 0.94 \\ 
  $\beta_{1}$ & 1.0 & 1.00  & 0.12 & 0.10 & 0.88 && 1.00  & 0.05 & 0.04 & 0.94 && 1.00  & 0.03 & 0.03 & 0.95 \\ 
  $\beta_{2}$ & 0.5 & 0.44  & 0.27 & 0.12 & 0.72 && 0.50  & 0.08 & 0.07 & 0.94 && 0.50  & 0.05 & 0.05 & 0.95 \\ 
  $\beta_{3}$ & 0.5 & 0.49  & 0.15 & 0.11 & 0.88 && 0.50  & 0.05 & 0.05 & 0.94 && 0.50  & 0.04 & 0.04 & 0.96 \\ 
  $\beta_{4}$ & 1.0 & 1.00  & 0.11 & 0.09 & 0.90 && 1.00  & 0.04 & 0.04 & 0.95 && 1.00  & 0.03 & 0.03 & 0.95 \\ 
  $\beta_{5}$ & 0.5 & 0.49  & 0.13 & 0.09 & 0.88 && 0.50  & 0.05 & 0.04 & 0.94 && 0.50  & 0.03 & 0.03 & 0.95 \\ 
  $\beta_{6}$ & 1.0 & 1.05  & 0.27 & 0.15 & 0.71 && 1.00  & 0.08 & 0.07 & 0.94 && 1.00  & 0.05 & 0.05 & 0.93 \\[0.1cm]
  $\alpha_{0}$ & 0.0 & -0.09 & 0.32 & 0.16 & 0.76 && -0.02 & 0.08 & 0.06 & 0.91 && -0.01 & 0.05 & 0.05 & 0.92 \\  
  \bottomrule
  \end{tabular}
\end{subtable}
\medskip

\begin{subtable}{\textwidth}
\centering
\vspace{-0.1cm}
\begin{tabular}{@{}c@{~~}c@{~~~} c@{~~}c@{~~}c@{~~}c@{~~} c@{~~~} c@{~~}c@{~~}c@{~~}c@{~~} c@{~~~} c@{~~}c@{~~}c@{~~}c@{}}
  \multicolumn{16}{l}{BAMLSS}\\
{} & {} & \multicolumn{4}{c}{$n = 100$} && \multicolumn{4}{c}{$n = 500$} && \multicolumn{4}{c}{$n = 1000$} \\
\cmidrule(r){3-6} \cmidrule(r){8-11} \cmidrule(){13-16}
{} & $\theta$ & $\hat{\theta}$ & SE & SEE & CP && $\hat{\theta}$ & SE & SEE & CP && $\hat{\theta}$ & SE & SEE & CP \\
\midrule
  $\beta_{0}$ & 0.0 & 0.01 & 0.24 & - & 0.91 && -0.01 & 0.09 & - & 0.95 && 0.01 & 0.06 & - & 0.96 \\ 
  $\beta_{1}$ & 1.0 & 1.00 & 0.13 & - & 0.90 && 1.01  & 0.05 & - & 0.94 && 1.00 & 0.03 & - & 0.92 \\ 
  $\beta_{2}$ & 0.5 & 0.49 & 0.22 & - & 0.89 && 0.50  & 0.08 & - & 0.94 && 0.50 & 0.05 & - & 0.94 \\ 
  $\beta_{3}$ & 0.5 & 0.50 & 0.13 & - & 0.92 && 0.50  & 0.05 & - & 0.95 && 0.50 & 0.04 & - & 0.95 \\ 
  $\beta_{4}$ & 1.0 & 0.99 & 0.11 & - & 0.95 && 1.00  & 0.04 & - & 0.97 && 1.00 & 0.03 & - & 0.93 \\ 
  $\beta_{5}$ & 0.5 & 0.50 & 0.13 & - & 0.91 && 0.50  & 0.05 & - & 0.90 && 0.50 & 0.03 & - & 0.92 \\ 
  $\beta_{6}$ & 1.0 & 1.00 & 0.30 & - & 0.89 && 1.00  & 0.10 & - & 0.91 && 0.99 & 0.06 & - & 0.96 \\[0.1cm]
  $\alpha_{0}$ & 0.0 &  0.01 & 0.41 & - & 0.92 && -0.01 & 0.13 & - & 0.94 && 0.01 & 0.09 & - & 0.95 \\ 
  \bottomrule
    \end{tabular}
    \end{subtable}
     \end{center}
     \end{table}

\begin{table}[h!]
\vspace{-5cm}
\ContinuedFloat
\begin{center}
\begin{subtable}{\textwidth}
    \centering
    \vspace{-0.1cm}    
\begin{tabular}{@{}c@{~~}c@{~~~} c@{~~}c@{~~}c@{~~}c@{~~} c@{~~~} c@{~~}c@{~~}c@{~~}c@{~~} c@{~~~} c@{~~}c@{~~}c@{~~}c@{}}
\toprule  
  \multicolumn{16}{l}{SPR-SIC}\\
{} & {} & \multicolumn{4}{c}{$n = 100$} && \multicolumn{4}{c}{$n = 500$} && \multicolumn{4}{c}{$n = 1000$} \\
\cmidrule(r){3-6} \cmidrule(r){8-11} \cmidrule(){13-16}
{} & $\theta$ & $\hat{\theta}$ & SE & SEE & CP && $\hat{\theta}$ & SE & SEE & CP && $\hat{\theta}$ & SE & SEE & CP \\
\midrule
  $\beta_{0}$ & 0.0     & 0.06 & 0.18 & 0.16 & 0.92 && 0.01 & 0.07 & 0.07 & 0.93 && 0.01 & 0.05 & 0.05 & 0.94 \\ 
  $\beta_{1}$ & 1.0     & 0.98 & 0.11 & 0.11 & 0.92 && 1.00 & 0.05 & 0.05 & 0.95 && 1.00 & 0.03 & 0.03 & 0.95 \\ 
  $\beta_{2}$ & 0.5     & 0.41 & 0.24 & 0.15 & 0.81 && 0.48 & 0.08 & 0.08 & 0.91 && 0.48 & 0.05 & 0.05 & 0.93 \\ 
  $\beta_{3}$ & 0.5     & 0.43 & 0.15 & 0.12 & 0.85 && 0.48 & 0.05 & 0.05 & 0.94 && 0.49 & 0.04 & 0.04 & 0.94 \\ 
  $\beta_{4}$ & 1.0     & 0.98 & 0.11 & 0.10 & 0.94 && 0.99 & 0.04 & 0.05 & 0.95 && 1.00 & 0.03 & 0.03 & 0.95 \\ 
  $\beta_{5}$ & 0.5     & 0.45 & 0.13 & 0.10 & 0.88 && 0.49 & 0.05 & 0.05 & 0.92 && 0.49 & 0.03 & 0.03 & 0.94 \\ 
  $\beta_{6}$ & 1.0     & 1.04 & 0.24 & 0.17 & 0.80 && 1.01 & 0.08 & 0.08 & 0.92 && 1.01 & 0.06 & 0.05 & 0.93 \\
  \bottomrule
  \end{tabular}
\end{subtable}

\medskip

\begin{subtable}{\textwidth}
    \centering
    \vspace{-0.1cm}
\begin{tabular}{@{}c@{~~}c@{~~~} c@{~~}c@{~~}c@{~~}c@{~~} c@{~~~} c@{~~}c@{~~}c@{~~}c@{~~} c@{~~~} c@{~~}c@{~~}c@{~~}c@{}}
  \multicolumn{16}{l}{ALASSO-IC}\\
{} & {} & \multicolumn{4}{c}{$n = 100$} && \multicolumn{4}{c}{$n = 500$} && \multicolumn{4}{c}{$n = 1000$} \\
\cmidrule(r){3-6} \cmidrule(r){8-11} \cmidrule(){13-16}
{} & $\theta$ & $\hat{\theta}$ & SE & SEE & CP && $\hat{\theta}$ & SE & SEE & CP && $\hat{\theta}$ & SE & SEE & CP \\
\midrule
  $\beta_{0}$ & 0.0     &  -0.00 & 0.25 & 0.24 & 0.93 && 0.00 & 0.11 & 0.11 & 0.94 && 0.00 & 0.08 & 0.08 & 0.96\\
  $\beta_{1}$ & 1.0     &  0.87  & 0.36 & 0.22 & 0.83 && 0.97 & 0.13 & 0.11 & 0.92 && 0.99 & 0.09 & 0.08 & 0.91 \\ 
  $\beta_{2}$ & 0.5     &  0.33  & 0.33 & 0.13 & 0.56 && 0.44 & 0.18 & 0.10 & 0.79 && 0.47 & 0.12 & 0.08 & 0.81 \\ 
  $\beta_{3}$ & 0.5     &  0.31  & 0.29 & 0.14 & 0.60 && 0.45 & 0.15 & 0.11 & 0.85 && 0.47 & 0.10 & 0.08 & 0.88 \\ 
  $\beta_{4}$ & 1.0     &  0.85  & 0.33 & 0.22 & 0.84 && 0.97 & 0.12 & 0.11 & 0.93 && 0.99 & 0.08 & 0.08 & 0.95 \\ 
  $\beta_{5}$ & 0.5     &  0.32  & 0.28 & 0.14 & 0.64 && 0.43 & 0.13 & 0.11 & 0.88 && 0.46 & 0.09 & 0.08 & 0.90 \\ 
  $\beta_{6}$ & 1.0     &  0.85  & 0.33 & 0.22 & 0.84 && 0.97 & 0.12 & 0.11 & 0.94 && 0.99 & 0.08 & 0.08 & 0.95 \\
  \bottomrule
  \multicolumn{16}{p{0.85\textwidth}}{\footnotesize SE, standard deviation of estimates over 1000 replications; SEE, average of estimated standard errors over 1000 replications (200 replications for BAMLSS); CP, the empirical coverage probability of a nominal 95\% confidence interval.}\\
\end{tabular}
\end{subtable}
\end{center}
\end{table}

\clearpage

\section{Simulation Results:\\ Normal Setting}\label{app:mpr_normal}
This section contains additional simulation results for the MPR-SIC, BAMLSS, SPR-SIC and ALASSO-IC methods in a normal setting. Data are simulated from the normal MPR model, where $X_4$, $X_5$, $X_7$ and $X_9$ are Bernoulli(0.5) and the remainder are $N(0,1)$. Note that fewer replicates for the BAMLSS procedure are used (200 replicates) due to the computational intensity of the method. The other approaches are averaged over 1000 replicates.
\begin{itemize}
    \item Table~\ref{tab:app_normal_true_values} is analogous to Table 2 of the main paper, but showing the true values that are used to simulate data in a normal setting.
    \item Table~\ref{tab:app_normal_sigma_table} is analogous to Table 6 of the main paper, but showing the out-of-sample prediction coverage probabilities for the normal setting.
    \item Table~\ref{tab:app_normal_var_selection} is analogous to Table 4 of the main paper, but showing the model selection metrics for the normal setting.
    \item Table~\ref{tab:app_normal_parameter_inference} is analogous to Table 5 of the main paper, but showing the estimation and inference metrics for the normal setting.
\end{itemize}

\begin{table}[h!]
\begin{center}
\caption{\label{tab:app_normal_true_values}True parameter values}%
\begin{tabular}{@{}cccccccccccccc@{}}
  \toprule
    & $X_0$ & $X_1$ & $X_2$ & $X_3$ & $\boldsymbol{X_4}$ & $\boldsymbol{X_5}$ & $X_6$ & $\boldsymbol{X_7}$ & $X_8$ & $\boldsymbol{X_9}$ & $X_{10}$ & $X_{11}$ & $X_{12}$ \\ 
 \midrule
$\beta$ & 0 & 1 & 0.5 & 0.5 & 1 & 0.5 & 1 & 0 & 0 & 0 & 0 & 0 & 0 \\[0.05cm] 
$\alpha$ & 0 & 0.5 & 1 & 0.5 & 1 & 0 & 0 & 0.5 & 1 & 0 & 0 & 0 & 0 \\
\bottomrule
\multicolumn{14}{p{.5\textwidth}}{\footnotesize Binary covariates indicated in bold.}\\
\end{tabular}
\end{center}
\end{table}

\begin{table}[h!]
\begin{center}
\caption{\label{tab:app_normal_sigma_table}Simulation results: out-of-sample prediction coverage probabilities}%
\begin{tabular}{@{}l@{~}c@{~~} c@{~}c@{~}c@{~} c@{~~} >{}c@{~}>{}c@{~}>{}c@{~} c@{~~} c@{~}c@{~}c@{~} c@{~~} c@{~}c@{~}c@{}} 
\toprule
{} && \multicolumn{3}{c}{MPR-SIC} && \multicolumn{3}{c}{{BAMLSS}} && \multicolumn{3}{c}{SPR-SIC} && \multicolumn{3}{c}{ALASSO-IC} \\
\cmidrule(lr){3-5} \cmidrule(r){7-9} \cmidrule(){11-13} \cmidrule(){15-17}
$n$         && 100  & 500  & 1000 && 100  & 500  & 1000 && 100  & 500  & 1000 && 100  & 500  & 1000 \\
\midrule
Low         && 0.78 & 0.93 & 0.94 && 0.80 & 0.93 & 0.94 && 1.00 & 1.00 & 1.00 && 1.00 & 1.00 & 1.00 \\ 
  Medium    && 0.89 & 0.94 & 0.95 && 0.90 & 0.94 & 0.95 && 0.97 & 1.00 & 1.00 && 0.98 & 1.00 & 1.00 \\ 
  High      && 0.94 & 0.95 & 0.95 && 0.94 & 0.95 & 0.95 && 0.79 & 0.84 & 0.85 && 0.81 & 0.85 & 0.85 \\[0.1cm] 
  Overall   && 0.86 & 0.94 & 0.95 && 0.87 & 0.94 & 0.95 && 0.93 & 0.95 & 0.95 && 0.93 & 0.95 & 0.95 \\
  \bottomrule
  \multicolumn{17}{p{.8\textwidth}}{\footnotesize Variability categorized as low ($\sigma_i \leq 0.7$), medium ($\sigma_i \in (0.7, 1.5]$) and high ($\sigma_i > 1.5$). Out-of-sample coverage is calculated for a sample 20\% the size of the original data.}\\
\end{tabular}
\end{center}
\end{table}

\begin{figure}[h!]

\centering
\makebox{\includegraphics[width = \textwidth]{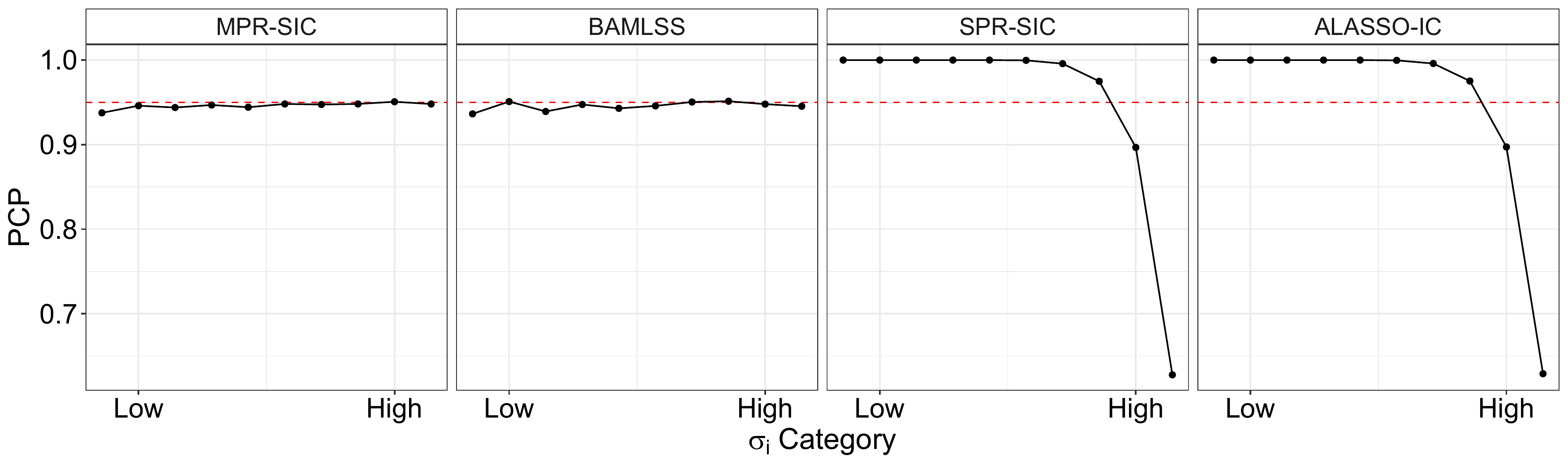}}
\caption{\label{fig:app_normal_sigma_cp}Prediction coverage probabilities (PCPs) of observations split by variability $\sigma_i$. Solid black line indicates the coverage and the red dashed line is a reference line at 0.95.}
\end{figure}

\begin{table}[h!]
\begin{center}
\caption{\label{tab:app_normal_var_selection}Simulation results: model selection metrics}%
    \begin{tabular}{@{}cc   c@{~~}c@{~~}c@{~~}c@{~~}   c@{~~~}   c@{~~}c@{~~}c@{~~}c@{~~}} 
    \toprule
    {} & {} & \multicolumn{4}{c}{MPR-SIC} && \multicolumn{4}{c}{{BAMLSS}} \\
    \cmidrule(lr){3-6} \cmidrule(){8-11}
    {}     & $n$   & C(6)  & IC(0) & PT & MSE && {C(6)} & {IC(0)} & {PT} & {MSE} \\
    \midrule
    $\beta$ & 100  &  5.27 & 0.00 & 0.50 & 0.05 && {5.63} & {0.00} & {0.71} & {0.08} \\
    {}      & 500  &  5.90 & 0.00 & 0.90 & 0.00 && {5.61} & {0.00} & {0.69} & {0.01} \\
    {}      & 1000 &  5.93 & 0.00 & 0.93 & 0.00 && {5.72} & {0.00} & {0.76} & {0.00} \\[0.1cm]
    $\alpha$ & 100     & 5.48 & 0.68 & 0.31 & 0.61 && {5.58} & {0.83} & {0.27} & {0.47} \\
    {}       & 500     & 5.93 & 0.00 & 0.93 & 0.04 && {5.70} & {0.00} & {0.74} & {0.06} \\
    {}       & 1000    & 5.94 & 0.00 & 0.94 & 0.02 && {5.71} & {0.00} & {0.73} & {0.03} \\
    \midrule
    {} & {} & \multicolumn{4}{c}{SPR-SIC} && \multicolumn{4}{c}{ALASSO-IC} \\
    \cmidrule(lr){3-6} \cmidrule(){8-11}
    {}     & $n$   & C(6)  & IC(0) & PT & MSE && C(6) & IC(0) & PT & MSE \\
    \midrule
    $\beta$     & 100  & 5.65 & 1.49 & 0.16 & 0.91 && 5.38 & 1.31 & 0.14 & 0.86 \\
    {}          & 500  & 5.84 & 0.15 & 0.75 & 0.15 && 5.68 & 0.10 & 0.68 & 0.15 \\
    {}          & 1000 & 5.91 & 0.01 & 0.90 & 0.06 && 5.80 & 0.01 & 0.82 & 0.07 \\[0.1cm]
    $\alpha$ & 100     & 6.00 & 6.00 & 0.00 & 6.34 && 6.00 & 6.00 & 0.00 & 6.58 \\
    {}       & 500     & 6.00 & 6.00 & 0.00 & 6.83 && 6.00 & 6.00 & 0.00 & 6.89 \\
    {}       & 1000    & 6.00 & 6.00 & 0.00 & 6.90 && 6.00 & 6.00 & 0.00 & 6.93 \\
    \bottomrule
    \multicolumn{11}{p{0.65\textwidth}}{\footnotesize C, average correct zeros; IC, average incorrect zeros; PT, the probability of choosing the true model; MSE, the average mean squared error.}\\
    \end{tabular}
\end{center}
\end{table}

\begin{table}[h!]
\begin{center}
\caption{\label{tab:app_normal_parameter_inference}Estimation and inference metrics for MPR-SIC, BAMLSS, SPR-SIC and ALASSO-IC methods}%
\begin{subtable}{\textwidth}
    \centering
    \vspace{-0.1cm}
\begin{tabular}{@{}c@{~~}c@{~~~} c@{~~}c@{~~}c@{~~}c@{~~} c@{~~~} c@{~~}c@{~~}c@{~~}c@{~~} c@{~~~} c@{~~}c@{~~}c@{~~}c@{}}
\toprule
\multicolumn{16}{l}{MPR-SIC}\\
{} & {} & \multicolumn{4}{c}{$n = 100$} && \multicolumn{4}{c}{$n = 500$} && \multicolumn{4}{c}{$n = 1000$} \\
\cmidrule(r){3-6} \cmidrule(r){8-11} \cmidrule(){13-16}
{} & $\theta$ & $\hat{\theta}$ & SE & SEE & CP && $\hat{\theta}$ & SE & SEE & CP && $\hat{\theta}$ & SE & SEE & CP \\
\midrule
  $\beta_{0}$ & 0.0 & -0.00  & 0.12 & 0.08 & 0.84  && 0.00  & 0.04 & 0.04 & 0.93  && 0.00  & 0.03 & 0.03 & 0.95 \\ 
  $\beta_{1}$ & 1.0 & 1.00   & 0.07 & 0.05 & 0.82  && 1.00  & 0.02 & 0.02 & 0.93  && 1.00  & 0.01 & 0.01 & 0.95 \\ 
  $\beta_{2}$ & 0.5 & 0.49   & 0.08 & 0.05 & 0.81  && 0.50  & 0.02 & 0.02 & 0.93  && 0.50  & 0.01 & 0.01 & 0.96 \\ 
  $\beta_{3}$ & 0.5 & 0.50   & 0.07 & 0.05 & 0.80  && 0.50  & 0.02 & 0.02 & 0.94  && 0.50  & 0.01 & 0.01 & 0.95 \\ 
  $\beta_{4}$ & 1.0 & 1.00   & 0.09 & 0.06 & 0.82  && 1.00  & 0.03 & 0.03 & 0.94  && 1.00  & 0.02 & 0.02 & 0.95 \\ 
  $\beta_{5}$ & 0.5 & 0.50   & 0.07 & 0.04 & 0.80  && 0.50  & 0.02 & 0.02 & 0.94  && 0.50  & 0.01 & 0.01 & 0.93 \\ 
  $\beta_{6}$ & 1.0 & 1.00   & 0.07 & 0.05 & 0.79  && 1.00  & 0.02 & 0.02 & 0.94  && 1.00  & 0.01 & 0.01 & 0.94 \\[0.1cm]
  $\alpha_{0}$ & 0.0 & -0.22  & 0.18 & 0.15 & 0.66  && -0.03 & 0.07 & 0.06 & 0.91  && -0.02 & 0.05 & 0.04 & 0.94 \\ 
  $\alpha_{1}$ & 0.5 & 0.49   & 0.31 & 0.13 & 0.68  && 0.51  & 0.07 & 0.07 & 0.94  && 0.50  & 0.05 & 0.05 & 0.95 \\ 
  $\alpha_{2}$ & 1.0 & 1.12   & 0.25 & 0.17 & 0.81  && 1.01  & 0.07 & 0.07 & 0.95  && 1.00  & 0.05 & 0.05 & 0.95 \\ 
  $\alpha_{3}$ & 0.5 & 0.49   & 0.31 & 0.13 & 0.67  && 0.51  & 0.07 & 0.07 & 0.94  && 0.50  & 0.04 & 0.05 & 0.95 \\ 
  $\alpha_{4}$ & 1.0 & 1.09   & 0.21 & 0.16 & 0.84  && 1.01  & 0.07 & 0.06 & 0.93  && 1.00  & 0.05 & 0.05 & 0.95 \\ 
  $\alpha_{7}$ & 0.5 & 0.49   & 0.30 & 0.13 & 0.67  && 0.51  & 0.07 & 0.06 & 0.94  && 0.50  & 0.05 & 0.05 & 0.94 \\ 
  $\alpha_{8}$ & 1.0 & 1.13   & 0.24 & 0.17 & 0.81  && 1.02  & 0.07 & 0.07 & 0.94  && 1.01  & 0.05 & 0.05 & 0.95 \\  
  \bottomrule
  \end{tabular}
  \end{subtable}
  
  \medskip

\begin{subtable}{\textwidth}

    \centering
    \vspace{-0.1cm}
\begin{tabular}{@{}c@{~~}c@{~~~} c@{~~}c@{~~}c@{~~}c@{~~} c@{~~~} c@{~~}c@{~~}c@{~~}c@{~~} c@{~~~} c@{~~}c@{~~}c@{~~}c@{}}
  \multicolumn{16}{l}{BAMLSS}\\
{} & {} & \multicolumn{4}{c}{$n = 100$} && \multicolumn{4}{c}{$n = 500$} && \multicolumn{4}{c}{$n = 1000$} \\
\cmidrule(r){3-6} \cmidrule(r){8-11} \cmidrule(){13-16}
{} & $\theta$ & $\hat{\theta}$ & SE & SEE & CP && $\hat{\theta}$ & SE & SEE & CP && $\hat{\theta}$ & SE & SEE & CP \\
\midrule
  $\beta_{0}$ & 0.0 & -0.01 & 0.13 & - & 0.90  && 0.00  & 0.05 & - & 0.94 && 0.00  & 0.03 & - & 0.95 \\ 
  $\beta_{1}$ & 1.0 & 0.99  & 0.07 & - & 0.93  && 1.00  & 0.02 & - & 0.93 && 1.00  & 0.01 & - & 0.95 \\ 
  $\beta_{2}$ & 0.5 & 0.49  & 0.08 & - & 0.92  && 0.50  & 0.02 & - & 0.92 && 0.50  & 0.01 & - & 0.94 \\ 
  $\beta_{3}$ & 0.5 & 0.50  & 0.07 & - & 0.92  && 0.50  & 0.02 & - & 0.94 && 0.50  & 0.01 & - & 0.96 \\ 
  $\beta_{4}$ & 1.0 & 1.00  & 0.09 & - & 0.90  && 1.00  & 0.03 & - & 0.91 && 1.00  & 0.02 & - & 0.94 \\ 
  $\beta_{5}$ & 0.5 & 0.51  & 0.07 & - & 0.91  && 0.50  & 0.02 & - & 0.88 && 0.50  & 0.01 & - & 0.95 \\ 
  $\beta_{6}$ & 1.0 & 1.00  & 0.07 & - & 0.92  && 1.00  & 0.02 & - & 0.95 && 1.00  & 0.01 & - & 0.94 \\[0.1cm]
  $\alpha_{0}$ & 0.0 &  -0.03 & 0.17 & - & 0.94  && -0.00 & 0.07 & - & 0.95 && -0.00 & 0.04 & - & 0.96 \\ 
  $\alpha_{1}$ & 0.5 &  0.50  & 0.22 & - & 0.92  && 0.51  & 0.07 & - & 0.92 && 0.50  & 0.05 & - & 0.92 \\ 
  $\alpha_{2}$ & 1.0 &  1.03  & 0.21 & - & 0.95  && 1.00  & 0.07 & - & 0.94 && 1.00  & 0.04 & - & 0.95 \\ 
  $\alpha_{3}$ & 0.5 &  0.51  & 0.19 & - & 0.97  && 0.50  & 0.07 & - & 0.92 && 0.50  & 0.05 & - & 0.94 \\ 
  $\alpha_{4}$ & 1.0 &  1.00  & 0.20 & - & 0.94  && 1.00  & 0.06 & - & 0.96 && 1.00  & 0.05 & - & 0.95 \\ 
  $\alpha_{7}$ & 0.5 &  0.51  & 0.21 & - & 0.90  && 0.51  & 0.07 & - & 0.93 && 0.50  & 0.04 & - & 0.96 \\ 
  $\alpha_{8}$ & 1.0 &  1.00  & 0.20 & - & 0.94  && 1.00  & 0.06 & - & 0.96 && 1.00  & 0.05 & - & 0.96 \\  
  \bottomrule
    \end{tabular}
    \end{subtable}
    \end{center}%
\end{table}%

\begin{table}[t!]
\vspace{-9cm}
\ContinuedFloat
\begin{center}

\begin{subtable}{\textwidth}
    \centering
    \vspace{-0.1cm}    
\begin{tabular}{@{}c@{~~}c@{~~~} c@{~~}c@{~~}c@{~~}c@{~~} c@{~~~~} c@{~~}c@{~~}c@{~~}c@{~~} c@{~~~~} c@{~~}c@{~~}c@{~~}c@{}}
  \toprule 
  \multicolumn{16}{l}{SPR-SIC}\\ 
{} & {} & \multicolumn{4}{c}{$n = 100$} && \multicolumn{4}{c}{$n = 500$} && \multicolumn{4}{c}{$n = 1000$} \\
\cmidrule(r){3-6} \cmidrule(r){8-11} \cmidrule(){13-16}
{} & $\theta$ & $\hat{\theta}$ & SE & SEE & CP && $\hat{\theta}$ & SE & SEE & CP && $\hat{\theta}$ & SE & SEE & CP \\
\midrule
  $\beta_{0}$ & 0.0     & -0.01 & 0.25 & 0.23 & 0.92 && 0.00 & 0.11 & 0.11 & 0.94 && 0.00 & 0.08 & 0.08 & 0.96 \\ 
  $\beta_{1}$ & 1.0     & 0.99  & 0.36 & 0.21 & 0.87 && 1.00 & 0.12 & 0.11 & 0.92 && 1.00 & 0.09 & 0.08 & 0.92 \\ 
  $\beta_{2}$ & 0.5     & 0.39  & 0.40 & 0.12 & 0.47 && 0.49 & 0.18 & 0.10 & 0.83 && 0.51 & 0.11 & 0.08 & 0.84 \\ 
  $\beta_{3}$ & 0.5     & 0.36  & 0.36 & 0.11 & 0.51 && 0.50 & 0.16 & 0.10 & 0.89 && 0.50 & 0.09 & 0.08 & 0.93 \\ 
  $\beta_{4}$ & 1.0     & 0.97  & 0.31 & 0.21 & 0.89 && 1.00 & 0.11 & 0.11 & 0.94 && 1.00 & 0.07 & 0.08 & 0.95 \\ 
  $\beta_{5}$ & 0.5     & 0.39  & 0.35 & 0.12 & 0.56 && 0.49 & 0.14 & 0.10 & 0.93 && 0.50 & 0.08 & 0.08 & 0.95 \\ 
  $\beta_{6}$ & 1.0     & 0.96  & 0.34 & 0.21 & 0.88 && 1.00 & 0.11 & 0.11 & 0.94 && 1.00 & 0.08 & 0.08 & 0.97 \\
  \end{tabular}
\end{subtable}%


\medskip
\begin{subtable}{\textwidth}
\centering
\vspace{-0.1cm}
\begin{tabular}{@{}c@{~~}c@{~~~} c@{~~}c@{~~}c@{~~}c@{~~} c@{~~~~} c@{~~}c@{~~}c@{~~}c@{~~} c@{~~~~} c@{~~}c@{~~}c@{~~}c@{}}
\toprule  
  \multicolumn{16}{l}{ALASSO-IC}\\
{} & {} & \multicolumn{4}{c}{$n = 100$} && \multicolumn{4}{c}{$n = 500$} && \multicolumn{4}{c}{$n = 1000$} \\
\cmidrule(r){3-6} \cmidrule(r){8-11} \cmidrule(){13-16}
{} & $\theta$ & $\hat{\theta}$ & SE & SEE & CP && $\hat{\theta}$ & SE & SEE & CP && $\hat{\theta}$ & SE & SEE & CP \\
\midrule
  $\beta_{0}$ & 0.0     &  -0.00 & 0.25 & 0.24 & 0.93 && 0.00 & 0.11 & 0.11 & 0.94 && 0.00 & 0.08 & 0.08 & 0.96\\
  $\beta_{1}$ & 1.0     &  0.87  & 0.36 & 0.22 & 0.83 && 0.97 & 0.13 & 0.11 & 0.92 && 0.99 & 0.09 & 0.08 & 0.91 \\ 
  $\beta_{2}$ & 0.5     &  0.33  & 0.33 & 0.13 & 0.56 && 0.44 & 0.18 & 0.10 & 0.79 && 0.47 & 0.12 & 0.08 & 0.81 \\ 
  $\beta_{3}$ & 0.5     &  0.31  & 0.29 & 0.14 & 0.60 && 0.45 & 0.15 & 0.11 & 0.85 && 0.47 & 0.10 & 0.08 & 0.88 \\ 
  $\beta_{4}$ & 1.0     &  0.85  & 0.33 & 0.22 & 0.84 && 0.97 & 0.12 & 0.11 & 0.93 && 0.99 & 0.08 & 0.08 & 0.95 \\ 
  $\beta_{5}$ & 0.5     &  0.32  & 0.28 & 0.14 & 0.64 && 0.43 & 0.13 & 0.11 & 0.88 && 0.46 & 0.09 & 0.08 & 0.90 \\ 
  $\beta_{6}$ & 1.0     &  0.85  & 0.33 & 0.22 & 0.84 && 0.97 & 0.12 & 0.11 & 0.94 && 0.99 & 0.08 & 0.08 & 0.95 \\
  \bottomrule
  \multicolumn{16}{p{0.8\textwidth}}{\footnotesize SE, standard deviation of estimates over 1000 replications; SEE, average of estimated standard errors over 1000 replications (200 replications for BAMLSS); CP, the empirical coverage probability of a nominal 95\% confidence interval.}
\end{tabular}%
\end{subtable}%
\end{center}%
\end{table}%

\FloatBarrier%
\section{Simulation Results:\\ Multiparameter Setting} \label{app:mpr_setting}
This section displays additional simulation results for data simulated using the normal MPR model. Table~\ref{tab:parameter_inference_sprsic_alassoic} is analogous to Table 5 of the main paper, but showing estimation and inference metrics for the BAMLSS, SPR-SIC and ALASSO-IC methods.

\begin{table}[h!]
   \begin{center}
    \caption{\label{tab:parameter_inference_sprsic_alassoic}Estimation and inference metrics for BAMLSS, SPR-SIC and ALASSO-IC methods}%
    \begin{tabular}{@{}c@{~~}c@{~~~} c@{~~}c@{~~}c@{~~}c@{~~} c@{~~~} c@{~~}c@{~~}c@{~~}c@{~~} c@{~~~} c@{~~}c@{~~}c@{~~}c@{}}
    \toprule
      \multicolumn{16}{l}{BAMLSS}\\
    {} & {} & \multicolumn{4}{c}{$n = 100$} && \multicolumn{4}{c}{$n = 500$} && \multicolumn{4}{c}{$n = 1000$} \\
    \cmidrule(r){3-6} \cmidrule(r){8-11} \cmidrule(){13-16}
    {} & $\theta$ & $\hat{\theta}$ & SE & SEE & CP && $\hat{\theta}$ & SE & SEE & CP && $\hat{\theta}$ & SE & SEE & CP \\
    \midrule
      $\beta_{0}$ & 0.0 & -0.00 & 0.25 & - & 0.90    && 0.00 & 0.08 & - & 0.92     && -0.00 & 0.05 & - & 0.94 \\ 
      $\beta_{1}$ & 1.0 & 1.01 & 0.15  & - & 0.92    && 1.00 & 0.05 & - & 0.94     && 1.00 & 0.03  & - & 0.95 \\ 
      $\beta_{2}$ & 0.5 & 0.49 & 0.18  & - & 0.93    && 0.50 & 0.05 & - & 0.95     && 0.50 & 0.03  & - & 0.96 \\ 
      $\beta_{3}$ & 0.5 & 0.50 & 0.11  & - & 0.93    && 0.50 & 0.03 & - & 0.94     && 0.50 & 0.02  & - & 0.95 \\ 
      $\beta_{4}$ & 1.0 & 1.01 & 0.12  & - & 0.91    && 1.00 & 0.03 & - & 0.95     && 1.00 & 0.02  & - & 0.95 \\ 
      $\beta_{5}$ & 0.5 & 0.50 & 0.11  & - & 0.93    && 0.50 & 0.03 & - & 0.95     && 0.50 & 0.02  & - & 0.95 \\ 
      $\beta_{6}$ & 1.0 & 1.01 & 0.23  & - & 0.94    && 1.00 & 0.07 & - & 0.94     && 1.00 & 0.04  & - & 0.96 \\[0.1cm]
      $\alpha_{0}$ & 0.0 &  -0.03 & 0.40 & - & 0.92    && 0.02 & 0.17 & - & 0.90     && 0.04 & 0.15  & - & 0.86 \\ 
      $\alpha_{1}$ & 0.5 &  0.50 & 0.21  & - & 0.92    && 0.49 & 0.08 & - & 0.89     && 0.49 & 0.06  & - & 0.86 \\ 
      $\alpha_{2}$ & 1.0 &  1.04 & 0.35  & - & 0.92    && 0.97 & 0.15 & - & 0.90     && 0.95 & 0.15  & - & 0.84 \\ 
      $\alpha_{3}$ & 0.5 &  0.50 & 0.25  & - & 0.93    && 0.49 & 0.09 & - & 0.90     && 0.48 & 0.07  & - & 0.85 \\ 
      $\alpha_{4}$ & 1.0 &  1.01 & 0.20  & - & 0.94    && 0.98 & 0.10 & - & 0.91     && 0.97 & 0.10  & - & 0.85 \\ 
      $\alpha_{7}$ & 0.5 &  0.52 & 0.21  & - & 0.94    && 0.49 & 0.08 & - & 0.88     && 0.49 & 0.06  & - & 0.86 \\ 
      $\alpha_{8}$ & 1.0 &  1.01 & 0.21  & - & 0.94    && 0.99 & 0.10 & - & 0.91     && 0.97 & 0.10  & - & 0.86 \\  
      \bottomrule
      &&&&&&&&&&&&&&&\\[-0.15cm]
      \multicolumn{16}{l}{SPR-SIC}\\
    {} & {} & \multicolumn{4}{c}{$n = 100$} && \multicolumn{4}{c}{$n = 500$} && \multicolumn{4}{c}{$n = 1000$} \\
    \cmidrule(r){3-6} \cmidrule(r){8-11} \cmidrule(){13-16}
    {} & $\theta$ & $\hat{\theta}$ & SE & SEE & CP && $\hat{\theta}$ & SE & SEE & CP && $\hat{\theta}$ & SE & SEE & CP \\
    \midrule
      $\beta_{0}$ & 0.0     & 0.28 & 0.78 & 0.45 & 0.72  && 0.10 & 0.37 & 0.24 & 0.83  && 0.01 & 0.26 & 0.18 & 0.88 \\ 
      $\beta_{1}$ & 1.0     & 0.90 & 0.76 & 0.23 & 0.59  && 0.98 & 0.38 & 0.16 & 0.69  && 1.01 & 0.27 & 0.12 & 0.69 \\ 
      $\beta_{2}$ & 0.5     & 0.51 & 0.78 & 0.12 & 0.09  && 0.36 & 0.52 & 0.08 & 0.21  && 0.35 & 0.37 & 0.09 & 0.44 \\ 
      $\beta_{3}$ & 0.5     & 0.18 & 0.38 & 0.06 & 0.17  && 0.34 & 0.30 & 0.10 & 0.57  && 0.46 & 0.19 & 0.12 & 0.88 \\ 
      $\beta_{4}$ & 1.0     & 0.90 & 0.63 & 0.24 & 0.68  && 0.99 & 0.27 & 0.16 & 0.84  && 1.00 & 0.16 & 0.12 & 0.85 \\ 
      $\beta_{5}$ & 0.5     & 0.25 & 0.41 & 0.09 & 0.28  && 0.41 & 0.28 & 0.11 & 0.68  && 0.48 & 0.15 & 0.11 & 0.92 \\ 
      $\beta_{6}$ & 1.0     & 0.76 & 0.82 & 0.16 & 0.33  && 1.06 & 0.53 & 0.16 & 0.46  && 1.11 & 0.33 & 0.15 & 0.54 \\
      \bottomrule
      \end{tabular}
      \end{center}
      \end{table}
      
      \begin{table}[h!]
      \ContinuedFloat
   \begin{center}
   \begin{tabular}{@{}c@{~~}c@{~~~} c@{~~}c@{~~}c@{~~}c@{~~} c@{~~~} c@{~~}c@{~~}c@{~~}c@{~~} c@{~~~} c@{~~}c@{~~}c@{~~}c@{}}
    \toprule
      \multicolumn{16}{l}{ALASSO-IC}\\
    {} & {} & \multicolumn{4}{c}{$n = 100$} && \multicolumn{4}{c}{$n = 500$} && \multicolumn{4}{c}{$n = 1000$} \\
    \cmidrule(r){3-6} \cmidrule(r){8-11} \cmidrule(){13-16}
    {} & $\theta$ & $\hat{\theta}$ & SE & SEE & CP && $\hat{\theta}$ & SE & SEE & CP && $\hat{\theta}$ & SE & SEE & CP \\
    \midrule
      $\beta_{0}$ & 0.0  & 0.45 & 0.70 & 0.47 & 0.68  && 0.20 & 0.37 & 0.25 & 0.79  && 0.09 & 0.26 & 0.18 & 0.85\\
      $\beta_{1}$ & 1.0  & 0.74 & 0.67 & 0.24 & 0.59  && 0.91 & 0.38 & 0.16 & 0.66  && 0.97 & 0.28 & 0.12 & 0.66 \\ 
      $\beta_{2}$ & 0.5  & 0.38 & 0.60 & 0.16 & 0.26  && 0.38 & 0.41 & 0.15 & 0.49  && 0.37 & 0.31 & 0.13 & 0.64 \\ 
      $\beta_{3}$ & 0.5  & 0.14 & 0.29 & 0.09 & 0.23  && 0.29 & 0.23 & 0.12 & 0.66  && 0.38 & 0.16 & 0.12 & 0.85 \\ 
      $\beta_{4}$ & 1.0  & 0.72 & 0.56 & 0.25 & 0.67  && 0.90 & 0.29 & 0.16 & 0.78  && 0.96 & 0.17 & 0.12 & 0.81 \\ 
      $\beta_{5}$ & 0.5  & 0.18 & 0.30 & 0.10 & 0.32  && 0.34 & 0.23 & 0.12 & 0.72  && 0.41 & 0.15 & 0.11 & 0.85 \\ 
      $\beta_{6}$ & 1.0  & 0.71 & 0.66 & 0.25 & 0.54  && 0.97 & 0.42 & 0.21 & 0.71  && 1.05 & 0.27 & 0.17 & 0.73 \\
      \bottomrule
      \multicolumn{16}{p{0.8\textwidth}}{\footnotesize SE, standard deviation of estimates over 1000 replications; SEE, average of estimated standard errors over 1000 replications; CP, the empirical coverage probability of a nominal 95\% confidence interval.}\\
    \end{tabular}
    \end{center}
    \end{table}

\FloatBarrier

\section{Simulation Results:\\ Different Effect Sizes}\label{app:effectsizes}
This section contains additional simulation results for the MPR-SIC method, with different effect sizes.

\subsection{Greater Location Effects}
\begin{itemize}
    \item Table~\ref{tab:true_values_effectsloc} is analogous to Table 2 of the main paper, but showing the new true values that are used to simulate data. The location component has greater effect sizes, resulting in a mean-driven problem.
    \item Table~\ref{tab:var_selection_effectsloc} is analogous to Table 4 of the main paper, but showing the model selection metrics for the setting with greater location effects. The results are comparable to the setting where the location and dispersion effects are on the same scale.
    \item Table~\ref{tab:parameter_inference_effectsloc} is analogous to Table 5 of the main paper, but gives the estimation and inference metrics for the setting with greater location effects. The results are comparable to the setting where the location and dispersion effects are on the same scale. 
\end{itemize}

\begin{table}[h!]
\begin{center}
\caption{\label{tab:true_values_effectsloc}True parameter values for the setting with greater location effects}%
\begin{tabular}{@{}cccccccccccccc@{}}
  \toprule
  & & \textcolor{blue}{E} & \textcolor{red}{M} & \textcolor{mygreen}{B} & N & N & \textcolor{red}{M} & N & N & \textcolor{red}{M} & \textcolor{mygreen}{B} & \textcolor{blue}{E} & \textcolor{red}{M} \\
    & $X_0$ & \textcolor{blue}{$X_1$} & \textcolor{red}{$X_2$} & \textcolor{mygreen}{$X_3$} & ${X_4}$ & ${X_5}$ & \textcolor{red}{$X_6$} & ${X_7}$ & $X_8$ & \textcolor{red}{${X_9}$} & \textcolor{mygreen}{$X_{10}$} & \textcolor{blue}{$X_{11}$} & \textcolor{red}{$X_{12}$} \\
 \midrule
$\beta$ & 0 & 10 & 5 & 5 & 10 & 5 & 10 & 0 & 0 & 0 & 0 & 0 & 0 \\[0.05cm] 
$\alpha$ & 0 & 0.5 & 1 & 0.5 & 1 & 0 & 0 & 0.5 & 1 & 0 & 0 & 0 & 0 \\
\bottomrule
\multicolumn{14}{l}{\footnotesize \textcolor{blue}{E = Exponential}, \textcolor{mygreen}{B = Bernoulli}, N = independent normal,} \\
  \multicolumn{14}{l}{\footnotesize \textcolor{red}{M = multivariate normal (correlated)}.}
\end{tabular}
\end{center}
\end{table}

\begin{table}[h!]

\begin{center}
\caption{\label{tab:var_selection_effectsloc}Simulation results: model selection metrics for the setting with greater location effects}%
\begin{tabular}{@{}ccc@{~~}c@{~~}c@{~~}c@{}}
  \toprule
 {} & {} & \multicolumn{4}{c}{MPR-SIC}\\
 \cmidrule(l){3-6} 
 {}     & $n$   & C(6)  & IC(0) & PT & MSE \\ 
 \midrule
 $\beta$ & 100  & 5.28 & 0.00 & 0.53 & 0.13  \\
 {}      & 500  & 5.91 & 0.00 & 0.91 & 0.01  \\
 {}      & 1000 & 5.95 & 0.00 & 0.95 & 0.00  \\[0.2cm]

 $\alpha$ & 100     & 5.54 & 0.75 & 0.34 & 0.61  \\ 
 {}       & 500     & 5.92 & 0.00 & 0.93 & 0.03  \\ 
 {}       & 1000    & 5.95 & 0.00 & 0.95 & 0.02  \\
 \bottomrule
   \multicolumn{6}{p{.35\textwidth}}{\footnotesize C, average correct zeros; IC, average incorrect zeros; PT, the probability of choosing the true model; MSE, the average mean squared error.}\\
\end{tabular}
\end{center}
\end{table}

\begin{table}[h!]

\begin{center}
\caption{\label{tab:parameter_inference_effectsloc}Simulation results: estimation and inference metrics for the setting with greater location effects}%
\begin{tabular}{@{}c@{~~}c@{~~~} c@{~~}c@{~~}c@{~~}c@{~~} c@{~~~} c@{~~}c@{~~}c@{~~}c@{~~} c@{~~~} c@{~~}c@{~~}c@{~~}c@{}}
\toprule
\multicolumn{16}{l}{MPR-SIC}\\
{} & {} & \multicolumn{4}{c}{$n = 100$} && \multicolumn{4}{c}{$n = 500$} && \multicolumn{4}{c}{$n = 1000$} \\
\cmidrule(r){3-6} \cmidrule(r){8-11} \cmidrule(){13-16}
{} & $\theta$ & $\hat{\theta}$ & SE & SEE & CP && $\hat{\theta}$ & SE & SEE & CP && $\hat{\theta}$ & SE & SEE & CP \\
\midrule
  $\beta_{0}$ & 0.00 & 0.01 & 0.21 & 0.13 & 0.79 && -0.00 & 0.06 & 0.05 & 0.94 && -0.00 & 0.04 & 0.04 & 0.94 \\ 
  $\beta_{1}$ & 10.0 & 10.0 & 0.15 & 0.10 & 0.78 && 10.0 & 0.04 & 0.04 & 0.94 && 10.0 & 0.03 & 0.03 & 0.96 \\ 
  $\beta_{2}$ & 5.00 & 5.01 & 0.18 & 0.12 & 0.80 && 5.00 & 0.05 & 0.05 & 0.92 && 5.00 & 0.03 & 0.03 & 0.94 \\ 
  $\beta_{3}$ & 5.00 & 4.99 & 0.10 & 0.07 & 0.80 && 5.00 & 0.03 & 0.03 & 0.94 && 5.00 & 0.02 & 0.02 & 0.94 \\ 
  $\beta_{4}$ & 10.0 & 10.0 & 0.11 & 0.07 & 0.81 && 10.0 & 0.03 & 0.03 & 0.94 && 10.0 & 0.02 & 0.02 & 0.93 \\ 
  $\beta_{5}$ & 5.00 & 5.01 & 0.10 & 0.07 & 0.79 && 5.00 & 0.03 & 0.03 & 0.93 && 5.00 & 0.02 & 0.02 & 0.95 \\ 
  $\beta_{6}$ & 10.0 & 9.99 & 0.20 & 0.12 & 0.77 && 10.0 & 0.05 & 0.05 & 0.92 && 10.0 & 0.03 & 0.03 & 0.94 \\[0.2cm]
  
  $\alpha_{0}$ & 0.00 &-0.17 & 0.43 & 0.23 & 0.66 && -0.03 & 0.11 & 0.10 & 0.90 && -0.02 & 0.07 & 0.07 & 0.92 \\
  $\alpha_{1}$ & 0.50 & 0.45 & 0.28 & 0.13 & 0.72 && 0.50 & 0.07 & 0.07 & 0.93 && 0.50 & 0.05 & 0.05 & 0.94 \\ 
  $\alpha_{2}$ & 1.00 & 1.07 & 0.36 & 0.17 & 0.78 && 1.01 & 0.07 & 0.07 & 0.92 && 1.01 & 0.05 & 0.05 & 0.93 \\ 
  $\alpha_{3}$ & 0.50 & 0.51 & 0.36 & 0.14 & 0.59 && 0.51 & 0.08 & 0.08 & 0.95 && 0.51 & 0.05 & 0.05 & 0.94 \\ 
  $\alpha_{4}$ & 1.00 & 1.11 & 0.24 & 0.17 & 0.82 && 1.01 & 0.06 & 0.07 & 0.95 && 1.01 & 0.05 & 0.05 & 0.94 \\ 
  $\alpha_{7}$ & 0.50 & 0.51 & 0.29 & 0.14 & 0.72 && 0.51 & 0.06 & 0.07 & 0.96 && 0.50 & 0.05 & 0.05 & 0.94 \\ 
  $\alpha_{8}$ & 1.00 & 1.11 & 0.24 & 0.17 & 0.82 && 1.01 & 0.07 & 0.07 & 0.93 && 1.01 & 0.05 & 0.05 & 0.94 \\
  \bottomrule
  \multicolumn{16}{p{.8\textwidth}}{\footnotesize SE, standard deviation of estimates over 1000 replications; SEE, average of estimated standard errors over 1000 replications; CP, the empirical coverage probability of a nominal 95\% confidence interval.}\\
\end{tabular}
\end{center}
\end{table}

\FloatBarrier
\subsection{Greater Dispersion Effects}
\begin{itemize}
    \item Table~\ref{tab:true_values_effectsdisp} is analogous to Table 2 of the main paper, but showing the new true values that are used to simulate data. The dispersion component has greater effect sizes, resulting in a dispersion-driven problem.
    \item Table~\ref{tab:var_selection_effectsdisp} is analogous to Table 4 of the main paper, but showing the model selection metrics for the setting with greater location effects. This dispersion-driven problem results in poor variable selection metrics for the location component of the model.
    \item Table~\ref{tab:parameter_inference_effectsdisp} is analogous to Table 5 of the main paper, but gives the estimation and inference metrics for the setting with greater location effects. For $n=100$, the coverage of the location parameters is poor, but improves as the sample size increases.
\end{itemize}

\begin{table}[h!]
\begin{center}
\caption{\label{tab:true_values_effectsdisp}True parameter values for the setting with greater dispersion effects}%
\begin{tabular}{@{}cccccccccccccc@{}}
  \toprule
    & & \textcolor{blue}{E} & \textcolor{red}{M} & \textcolor{mygreen}{B} & N & N & \textcolor{red}{M} & N & N & \textcolor{red}{M} & \textcolor{mygreen}{B} & \textcolor{blue}{E} & \textcolor{red}{M} \\
    & $X_0$ & \textcolor{blue}{$X_1$} & \textcolor{red}{$X_2$} & \textcolor{mygreen}{$X_3$} & ${X_4}$ & ${X_5}$ & \textcolor{red}{$X_6$} & ${X_7}$ & $X_8$ & \textcolor{red}{${X_9}$} & \textcolor{mygreen}{$X_{10}$} & \textcolor{blue}{$X_{11}$} & \textcolor{red}{$X_{12}$} \\
 \midrule
$\beta$ & 0 & 1 & 0.5 & 0.5 & 1 & 0.5 & 1 & 0 & 0 & 0 & 0 & 0 & 0 \\[0.05cm]
$\alpha$ & 0 & 1.5 & 3 & 1.5 & 3 & 0 & 0 & 1.5 & 3 & 0 & 0 & 0 & 0 \\
\bottomrule
\multicolumn{14}{l}{\footnotesize \textcolor{blue}{E = Exponential}, \textcolor{mygreen}{B = Bernoulli}, N = independent normal,} \\
\multicolumn{14}{l}{\footnotesize \textcolor{red}{M = multivariate normal (correlated)}.}
\end{tabular}
\end{center}
\end{table}

\begin{table}[h!]

\begin{center}
\caption{\label{tab:var_selection_effectsdisp}Simulation results: model selection metrics for the setting with greater dispersion effects}%
\begin{tabular}{@{}ccc@{~~}c@{~~}c@{~~}c@{}}
  \toprule
 {} & {} & \multicolumn{4}{c}{MPR-SIC}\\
 \cmidrule(l){3-6} 
 {}     & $n$   & C(6)  & IC(0) & PT & MSE \\ 
 \midrule
 $\beta$ & 100  & 5.15 & 0.00 & 0.52 & 0.02  \\
 {}      & 500  & 4.42 & 0.00 & 0.26 & 0.00  \\
 {}      & 1000 & 2.97 & 0.00 & 0.04 & 0.00  \\[0.2cm]

 $\alpha$ & 100     & 5.60 & 0.00 & 0.69 & 0.50  \\ 
 {}       & 500     & 5.93 & 0.00 & 0.93 & 0.04  \\ 
 {}       & 1000    & 5.94 & 0.00 & 0.94 & 0.02  \\
 \bottomrule
   \multicolumn{6}{p{.35\textwidth}}{\footnotesize C, average correct zeros; IC, average incorrect zeros; PT, the probability of choosing the true model; MSE, the average mean squared error.}\\
\end{tabular}
\end{center}
\end{table}

\begin{table}[h!]

\begin{center}
\caption{\label{tab:parameter_inference_effectsdisp}Simulation results: estimation and inference metrics for the setting with greater dispersion effects}%
\begin{tabular}{@{}c@{~~}c@{~~~} c@{~~}c@{~~}c@{~~}c@{~~} c@{~~~} c@{~~}c@{~~}c@{~~}c@{~~} c@{~~~} c@{~~}c@{~~}c@{~~}c@{}}
\toprule
\multicolumn{16}{l}{MPR-SIC}\\
{} & {} & \multicolumn{4}{c}{$n = 100$} && \multicolumn{4}{c}{$n = 500$} && \multicolumn{4}{c}{$n = 1000$} \\
\cmidrule(r){3-6} \cmidrule(r){8-11} \cmidrule(){13-16}
{} & $\theta$ & $\hat{\theta}$ & SE & SEE & CP && $\hat{\theta}$ & SE & SEE & CP && $\hat{\theta}$ & SE & SEE & CP \\
\midrule
  $\beta_{0}$ & 0.0 & 0.00 & 0.07 & 0.03 & 0.70 && -0.00 & 0.01 & 0.00 & 0.92 && 0.00 & 0.00 & 0.00 & 0.92 \\ 
  $\beta_{1}$ & 1.0 & 1.00 & 0.05 & 0.02 & 0.70 && 1.00 & 0.00 & 0.00 & 0.92 && 1.00 & 0.00 & 0.00 & 0.94 \\ 
  $\beta_{2}$ & 0.5 & 0.50 & 0.05 & 0.02 & 0.75 && 0.50 & 0.00 & 0.00 & 0.92 && 0.50 & 0.00 & 0.00 & 0.94 \\ 
  $\beta_{3}$ & 0.5 & 0.50 & 0.03 & 0.01 & 0.72 && 0.50 & 0.00 & 0.00 & 0.92 && 0.50 & 0.00 & 0.00 & 0.93 \\ 
  $\beta_{4}$ & 1.0 & 1.00 & 0.04 & 0.02 & 0.70 && 1.00 & 0.00 & 0.00 & 0.93 && 1.00 & 0.00 & 0.00 & 0.93 \\ 
  $\beta_{5}$ & 0.5 & 0.50 & 0.03 & 0.01 & 0.73 && 0.50 & 0.00 & 0.00 & 0.92 && 0.50 & 0.00 & 0.00 & 0.93 \\ 
  $\beta_{6}$ & 1.0 & 1.00 & 0.06 & 0.02 & 0.70 && 1.00 & 0.00 & 0.00 & 0.90 && 1.00 & 0.00 & 0.00 & 0.94 \\[0.2cm]
  
  $\alpha_{0}$ & 0.0 & -0.35 & 0.33 & 0.25 & 0.67 && -0.04 & 0.11 & 0.10 & 0.91 && -0.02 & 0.07 & 0.07 & 0.92 \\
  $\alpha_{1}$ & 1.5 & 1.54 & 0.20 & 0.17 & 0.91 && 1.50 & 0.07 & 0.07 & 0.93 && 1.50 & 0.05 & 0.05 & 0.94 \\ 
  $\alpha_{2}$ & 3.0 & 3.16 & 0.27 & 0.18 & 0.79 && 3.02 & 0.07 & 0.07 & 0.92 && 3.01 & 0.05 & 0.05 & 0.93 \\ 
  $\alpha_{3}$ & 1.5 & 1.63 & 0.25 & 0.19 & 0.83 && 1.51 & 0.08 & 0.07 & 0.94 && 1.51 & 0.05 & 0.05 & 0.94 \\ 
  $\alpha_{4}$ & 3.0 & 3.16 & 0.22 & 0.17 & 0.82 && 3.02 & 0.06 & 0.07 & 0.94 && 3.01 & 0.05 & 0.05 & 0.94 \\ 
  $\alpha_{7}$ & 1.5 & 1.59 & 0.20 & 0.16 & 0.88 && 1.51 & 0.06 & 0.07 & 0.95 && 1.50 & 0.05 & 0.05 & 0.94 \\ 
  $\alpha_{8}$ & 3.0 & 3.16 & 0.22 & 0.17 & 0.80 && 3.02 & 0.07 & 0.07 & 0.92 && 3.01 & 0.05 & 0.05 & 0.94 \\
  \bottomrule
  \multicolumn{16}{p{.8\textwidth}}{\footnotesize SE, standard deviation of estimates over 1000 replications; SEE, average of estimated standard errors over 1000 replications; CP, the empirical coverage probability of a nominal 95\% confidence interval.}\\
\end{tabular}
\end{center}
\end{table}

\FloatBarrier
\section{Simulation Results:\\ Imbalance of Active Sets}\label{app:cardinalty}
This section contains additional simulation results for the MPR-SIC method, with changes in the cardinality of the active sets.

\subsection{Location Component}
\begin{itemize}
    \item Table~\ref{tab:true_values_cardloc} is analogous to Table 2 of the main paper, but showing the new true values that are used to simulate data. The cardinality of the active set for the location component is greater than the dispersion component.
    \item Table~\ref{tab:var_selection_cardloc} is analogous to Table 4 of the main paper, but showing the model selection metrics for the setting where the active set of the location component contains more parameters than the dispersion component. Performance is comparable with the performance when the cardinality of the active sets are identical.
    \item Table~\ref{tab:parameter_inference_cardloc} is analogous to Table 5 of the main paper, but gives the estimation and inference metrics for the setting where the cardinality of the active set for the location component is greater than the dispersion component. Performance is comparable with the performance when the cardinality of the active sets are identical.
\end{itemize}

\begin{table}[h!]
\begin{center}
\caption{\label{tab:true_values_cardloc}True parameter values for the setting where the cardinality of the active set is greater for the location component}%
\begin{tabular}{@{}cccccccccccccc@{}}
  \toprule
    & & \textcolor{blue}{E} & \textcolor{red}{M} & \textcolor{mygreen}{B} & N & N & \textcolor{red}{M} & N & N & \textcolor{red}{M} & \textcolor{mygreen}{B} & \textcolor{blue}{E} & \textcolor{red}{M} \\
    & $X_0$ & \textcolor{blue}{$X_1$} & \textcolor{red}{$X_2$} & \textcolor{mygreen}{$X_3$} & ${X_4}$ & ${X_5}$ & \textcolor{red}{$X_6$} & ${X_7}$ & $X_8$ & \textcolor{red}{${X_9}$} & \textcolor{mygreen}{$X_{10}$} & \textcolor{blue}{$X_{11}$} & \textcolor{red}{$X_{12}$} \\
 \midrule
$\beta$ & 0 & 1 & 0.5 & 0.5 & 1 & 0.5 & 1 & 0 & 0 & 0 & 0 & 0 & 0 \\[0.05cm] 
$\alpha$ & 0 & 0 & 1 & 0 & 1 & 0 & 0 & 0 & 1 & 0 & 0 & 0 & 0 \\
\bottomrule
\multicolumn{14}{l}{\footnotesize \textcolor{blue}{E = Exponential}, \textcolor{mygreen}{B = Bernoulli}, N = independent normal,} \\
\multicolumn{14}{l}{\footnotesize \textcolor{red}{M = multivariate normal (correlated)}.}
\end{tabular}
\end{center}
\end{table}

\begin{table}[h!]

\begin{center}
\caption{\label{tab:var_selection_cardloc}Simulation results: model selection metrics for the setting where the cardinality of the active set is greater for the location component}%
\begin{tabular}{@{}ccc@{~~}c@{~~}c@{~~}c@{}}
  \toprule
 {} & {} & \multicolumn{4}{c}{MPR-SIC}\\
 \cmidrule(l){3-6} 
 {}     & $n$   & C(6)  & IC(0) & PT & MSE \\ 
 \midrule
 $\beta$ & 100  & 5.40 & 0.03 & 0.57 & 0.06  \\
 {}      & 500  & 5.91 & 0.00 & 0.92 & 0.01  \\
 {}      & 1000 & 5.95 & 0.00 & 0.95 & 0.00  \\
 \midrule
 {}     & $n$   & C(9)  & IC(0) & PT & MSE \\
 \midrule
 $\alpha$ & 100     & 8.34 & 0.05 & 0.55 & 0.38  \\ 
 {}       & 500     & 8.89 & 0.00 & 0.90 & 0.02  \\ 
 {}       & 1000    & 8.93 & 0.00 & 0.93 & 0.01  \\
 \bottomrule
   \multicolumn{6}{p{.35\textwidth}}{\footnotesize C, average correct zeros; IC, average incorrect zeros; PT, the probability of choosing the true model; MSE, the average mean squared error.}\\
\end{tabular}
\end{center}
\end{table}

\begin{table}[h!]

\begin{center}
\caption{\label{tab:parameter_inference_cardloc}Simulation results: estimation and inference metrics for the setting where the cardinality of the active set is greater for the location component}%
\begin{tabular}{@{}c@{~~}c@{~~~} c@{~~}c@{~~}c@{~~}c@{~~} c@{~~~} c@{~~}c@{~~}c@{~~}c@{~~} c@{~~~} c@{~~}c@{~~}c@{~~}c@{}}
\toprule
\multicolumn{16}{l}{MPR-SIC}\\
{} & {} & \multicolumn{4}{c}{$n = 100$} && \multicolumn{4}{c}{$n = 500$} && \multicolumn{4}{c}{$n = 1000$} \\
\cmidrule(r){3-6} \cmidrule(r){8-11} \cmidrule(){13-16}
{} & $\theta$ & $\hat{\theta}$ & SE & SEE & CP && $\hat{\theta}$ & SE & SEE & CP && $\hat{\theta}$ & SE & SEE & CP \\
\midrule
  $\beta_{0}$ & 0.0 & 0.01 & 0.16 & 0.11 & 0.81 && -0.00 & 0.05 & 0.05 & 0.94 && 0.00 & 0.03 & 0.03 & 0.94 \\ 
  $\beta_{1}$ & 1.0 & 1.00 & 0.08 & 0.06 & 0.84 && 1.00 & 0.02 & 0.02 & 0.95 && 1.00 & 0.02 & 0.02 & 0.95 \\ 
  $\beta_{2}$ & 0.5 & 0.50 & 0.14 & 0.09 & 0.82 && 0.50 & 0.04 & 0.04 & 0.92 && 0.50 & 0.03 & 0.03 & 0.95 \\ 
  $\beta_{3}$ & 0.5 & 0.50 & 0.09 & 0.06 & 0.84 && 0.50 & 0.03 & 0.03 & 0.95 && 0.50 & 0.02 & 0.02 & 0.94 \\ 
  $\beta_{4}$ & 1.0 & 1.00 & 0.08 & 0.06 & 0.84 && 1.00 & 0.02 & 0.02 & 0.94 && 1.00 & 0.02 & 0.02 & 0.94 \\ 
  $\beta_{5}$ & 0.5 & 0.50 & 0.07 & 0.05 & 0.85 && 0.50 & 0.02 & 0.02 & 0.93 && 0.50 & 0.02 & 0.02 & 0.95 \\ 
  $\beta_{6}$ & 1.0 & 1.00 & 0.15 & 0.09 & 0.80 && 1.00 & 0.04 & 0.04 & 0.93 && 1.00 & 0.03 & 0.03 & 0.93 \\[0.2cm]
  
  $\alpha_{0}$ & 0.0 & -0.18 & 0.28 & 0.16 & 0.68 && -0.02 & 0.08 & 0.06 & 0.90 && -0.01 & 0.05 & 0.05 & 0.93 \\
  $\alpha_{2}$ & 1.0 & 1.06 & 0.34 & 0.17 & 0.81 && 1.01 & 0.07 & 0.07 & 0.93 && 1.01 & 0.05 & 0.05 & 0.93 \\ 
  $\alpha_{4}$ & 1.0 & 1.10 & 0.22 & 0.17 & 0.84 && 1.01 & 0.06 & 0.07 & 0.95 && 1.01 & 0.05 & 0.05 & 0.95 \\ 
  $\alpha_{8}$ & 1.0 & 1.11 & 0.22 & 0.17 & 0.85 && 1.01 & 0.07 & 0.07 & 0.93 && 1.01 & 0.05 & 0.05 & 0.93 \\
  \bottomrule
  \multicolumn{16}{p{.8\textwidth}}{\footnotesize SE, standard deviation of estimates over 1000 replications; SEE, average of estimated standard errors over 1000 replications; CP, the empirical coverage probability of a nominal 95\% confidence interval.}\\
\end{tabular}
\end{center}
\end{table}

\FloatBarrier
\subsection{Dispersion Component}
\begin{itemize}
    \item Table~\ref{tab:true_values_carddisp} is analogous to Table 2 of the main paper, but showing the new true values that are used to simulate data. The cardinality of the active set for the dispersion component is greater than the location component.
    \item Table~\ref{tab:var_selection_carddisp} is analogous to Table 4 of the main paper, but showing the model selection metrics for the setting where the active set of the dispersion component contains more parameters than the location component. Performance is comparable with the performance when the cardinality of the active sets are identical.
    \item Table~\ref{tab:parameter_inference_carddisp} is analogous to Table 5 of the main paper, but gives the estimation and inference metrics for the setting where the cardinality of the active set for the dispersion component is greater than the location component. Performance is comparable with the performance when the cardinality of the active sets are identical.
\end{itemize}

\begin{table}[h!]
\begin{center}
\caption{\label{tab:true_values_carddisp}True parameter values for the setting where the cardinality of the active set is greater for the dispersion component}%
\begin{tabular}{@{}cccccccccccccc@{}}
  \toprule
    & & \textcolor{blue}{E} & \textcolor{red}{M} & \textcolor{mygreen}{B} & N & N & \textcolor{red}{M} & N & N & \textcolor{red}{M} & \textcolor{mygreen}{B} & \textcolor{blue}{E} & \textcolor{red}{M} \\
       & $X_0$ & \textcolor{blue}{$X_1$} & \textcolor{red}{$X_2$} & \textcolor{mygreen}{$X_3$} & ${X_4}$ & ${X_5}$ & \textcolor{red}{$X_6$} & ${X_7}$ & $X_8$ & \textcolor{red}{${X_9}$} & \textcolor{mygreen}{$X_{10}$} & \textcolor{blue}{$X_{11}$} & \textcolor{red}{$X_{12}$} \\
 \midrule
$\beta$ & 0 & 1 & 0 & 0 & 1 & 0 & 1 & 0 & 0 & 0 & 0 & 0 & 0 \\[0.05cm] 
$\alpha$ & 0 & 0.5 & 1 & 0.5 & 1 & 0 & 0 & 0.5 & 1 & 0 & 0 & 0 & 0 \\
\bottomrule
\multicolumn{14}{l}{\footnotesize \textcolor{blue}{E = Exponential}, \textcolor{mygreen}{B = Bernoulli}, N = independent normal,} \\
\multicolumn{14}{l}{\footnotesize \textcolor{red}{M = multivariate normal (correlated)}.}
\end{tabular}
\end{center}
\end{table}

\begin{table}[h!]

\begin{center}
\caption{\label{tab:var_selection_carddisp}Simulation results: model selection metrics for the setting where the cardinality of the active set is greater for the dispersion component}%
\begin{tabular}{@{}ccc@{~~}c@{~~}c@{~~}c@{}}
  \toprule
 {} & {} & \multicolumn{4}{c}{MPR-SIC}\\
 \cmidrule(l){3-6} 
 {}     & $n$   & C(9)  & IC(0) & PT & MSE \\ 
 \midrule
 $\beta$ & 100  & 8.04 & 0.00 & 0.45 & 0.10  \\
 {}      & 500  & 8.87 & 0.00 & 0.88 & 0.01  \\
 {}      & 1000 & 8.93 & 0.00 & 0.93 & 0.00  \\
 \midrule
 {}     & $n$   & C(6)  & IC(0) & PT & MSE \\
 \midrule
 $\alpha$ & 100     & 5.57 & 0.73 & 0.34 & 0.53  \\ 
 {}       & 500     & 5.93 & 0.00 & 0.93 & 0.03  \\ 
 {}       & 1000    & 5.95 & 0.00 & 0.95 & 0.02  \\
 \bottomrule
   \multicolumn{6}{p{.35\textwidth}}{\footnotesize C, average correct zeros; IC, average incorrect zeros; PT, the probability of choosing the true model; MSE, the average mean squared error.}\\
\end{tabular}
\end{center}
\end{table}

\begin{table}[h!]

\begin{center}
\caption{\label{tab:parameter_inference_carddisp}Simulation results: estimation and inference metrics for the setting where the cardinality of the active set is greater for the dispersion component}%
\begin{tabular}{@{}c@{~~}c@{~~~} c@{~~}c@{~~}c@{~~}c@{~~} c@{~~~} c@{~~}c@{~~}c@{~~}c@{~~} c@{~~~} c@{~~}c@{~~}c@{~~}c@{}}
\toprule
\multicolumn{16}{l}{MPR-SIC}\\
{} & {} & \multicolumn{4}{c}{$n = 100$} && \multicolumn{4}{c}{$n = 500$} && \multicolumn{4}{c}{$n = 1000$} \\
\cmidrule(r){3-6} \cmidrule(r){8-11} \cmidrule(){13-16}
{} & $\theta$ & $\hat{\theta}$ & SE & SEE & CP && $\hat{\theta}$ & SE & SEE & CP && $\hat{\theta}$ & SE & SEE & CP \\
\midrule
  $\beta_{0}$ & 0.0 & 0.01 & 0.19 & 0.12 & 0.77 && 0.00 & 0.06 & 0.05 & 0.93 && -0.00 & 0.04 & 0.04 & 0.94 \\ 
  $\beta_{1}$ & 1.0 & 1.00 & 0.14 & 0.09 & 0.79 && 1.00 & 0.04 & 0.04 & 0.94 && 1.00 & 0.03 & 0.03 & 0.96 \\ 
  $\beta_{4}$ & 1.0 & 1.00 & 0.10 & 0.07 & 0.81 && 1.00 & 0.03 & 0.03 & 0.95 && 1.00 & 0.02 & 0.02 & 0.93 \\ 
  $\beta_{6}$ & 1.0 & 0.99 & 0.16 & 0.07 & 0.74 && 1.00 & 0.04 & 0.03 & 0.93 && 1.00 & 0.02 & 0.02 & 0.92 \\[0.2cm]
  
  $\alpha_{0}$ & 0.0 & -0.12 & 0.42 & 0.22 & 0.68 && -0.02 & 0.11 & 0.10 & 0.91 && -0.01 & 0.07 & 0.07 & 0.92 \\
  $\alpha_{1}$ & 0.5 & 0.44 & 0.27 & 0.13 & 0.73 && 0.50 & 0.07 & 0.07 & 0.93 && 0.50 & 0.05 & 0.05 & 0.95 \\ 
  $\alpha_{2}$ & 1.0 & 1.05 & 0.34 & 0.16 & 0.80 && 1.01 & 0.07 & 0.07 & 0.93 && 1.00 & 0.05 & 0.05 & 0.93 \\ 
  $\alpha_{3}$ & 0.5 & 0.49 & 0.35 & 0.13 & 0.61 && 0.51 & 0.08 & 0.07 & 0.95 && 0.50 & 0.05 & 0.05 & 0.94 \\ 
  $\alpha_{4}$ & 1.0 & 1.08 & 0.23 & 0.17 & 0.84 && 1.01 & 0.06 & 0.07 & 0.95 && 1.00 & 0.05 & 0.05 & 0.94 \\ 
  $\alpha_{7}$ & 0.5 & 0.50 & 0.28 & 0.13 & 0.74 && 0.50 & 0.06 & 0.07 & 0.96 && 0.50 & 0.05 & 0.05 & 0.94 \\ 
  $\alpha_{8}$ & 1.0 & 1.08 & 0.22 & 0.17 & 0.83 && 1.01 & 0.07 & 0.07 & 0.93 && 1.00 & 0.05 & 0.05 & 0.94 \\
  \bottomrule
  \multicolumn{16}{p{.8\textwidth}}{\footnotesize SE, standard deviation of estimates over 1000 replications; SEE, average of estimated standard errors over 1000 replications; CP, the empirical coverage probability of a nominal 95\% confidence interval.}\\
\end{tabular}
\end{center}
\end{table}

\FloatBarrier
\section{Simulation Results:\\ Additional Covariates}\label{app:morecovariates}
This section contains additional simulation results for the MPR-SIC method, with additional covariates.

\subsection{Additional Noise}
\begin{itemize}
    \item Table~\ref{tab:true_values_morecovariates_scen1} is analogous to Table 2 of the main paper, but showing the new true values that are used to simulate data with additional covariates.
    \item Table~\ref{tab:var_selection_morecovariates_scen1} is analogous to Table 4 of the main paper, but showing the model selection metrics for the scenario with additional covariates.
    \item Table~\ref{tab:parameter_inference_morecovariates_scen1} is analogous to Table 5 of the main paper, but gives the estimation and inference metrics for the scenario with additional covariates.
\end{itemize}
\begingroup
\setlength{\tabcolsep}{4pt} 
\begin{table}[h!]
    \begin{center}
    \caption{\label{tab:true_values_morecovariates_scen1}True parameter values for the scenario with additional covariates}%
    \begin{tabular}{@{}cccccccccccccc@{}}
      \toprule
      & & \textcolor{blue}{E} & \textcolor{red}{M} & \textcolor{mygreen}{B} & N & N & \textcolor{red}{M} & N & N & \textcolor{red}{M} & \textcolor{mygreen}{B} & \textcolor{blue}{E} & \textcolor{red}{M} \\
     & $X_0$ & \textcolor{blue}{$X_1$} & \textcolor{red}{$X_2$} & \textcolor{mygreen}{$X_3$} & ${X_4}$ & ${X_5}$ & \textcolor{red}{$X_6$} & ${X_7}$ & $X_8$ & \textcolor{red}{${X_9}$} & \textcolor{mygreen}{$X_{10}$} & \textcolor{blue}{$X_{11}$} & \textcolor{red}{$X_{12}$} \\
     \midrule
    $\beta$ & 0 & 1 & 0 & 0 & 1 & 0 & 1 & 0 & 0 & 0 & 0 & 0 & 0 \\[0.05cm] 
    $\alpha$ & 0 & 0.5 & 1 & 0.5 & 1 & 0 & 0 & 0.5 & 1 & 0 & 0 & 0 & 0 \\
    \midrule
    & & N & N & N & N & N & N & N & N & N & N & N & N \\
    & {} & $X_{13}$ & ${X_{14}}$ & ${X_{15}}$ & $X_{16}$ & ${X_{17}}$ & $X_{18}$ & ${X_{19}}$ & $X_{20}$ & $X_{21}$ & $X_{22}$ & $X_{23}$ & $X_{24}$ \\ 
    \midrule
    $\beta$ & {} & 0 & 0 & 0 & 0 & 0 & 0 & 0 & 0 & 0 & 0 & 0 & 0 \\[0.05cm] 
    $\alpha$ & {} & 0 & 0 & 0 & 0 & 0 & 0 & 0 & 0 & 0 & 0 & 0 & 0 \\
    \bottomrule
    \multicolumn{14}{l}{\footnotesize \textcolor{blue}{E = Exponential}, \textcolor{mygreen}{B = Bernoulli}, N = independent normal,} \\
  \multicolumn{14}{l}{\footnotesize \textcolor{red}{M = multivariate normal (correlated)}.}
    \end{tabular}
    \end{center}
    \end{table}
\endgroup
\begin{table}[h!]

\begin{center}
\caption{\label{tab:var_selection_morecovariates_scen1}Simulation results: model selection metrics for the scenario with additional covariates}%
\begin{tabular}{@{}ccc@{~~}c@{~~}c@{~~}c@{}}
  \toprule
 {} & {} & \multicolumn{4}{c}{MPR-SIC}\\
 \cmidrule(l){3-6} 
 {}     & $n$   & C(18)  & IC(0) & PT & MSE \\ 
 \midrule
 $\beta$ & 500  & 17.64 & 0.00 & 0.71 & 0.01  \\
 {}      & 1000 & 17.82 & 0.00 & 0.84 & 0.00  \\
 {}      & 2000 & 17.87 & 0.00 & 0.88 & 0.00  \\[0.2cm]

 $\alpha$ & 500     & 17.72 & 0.00 & 0.76 & 0.04  \\ 
 {}       & 100     & 17.81 & 0.00 & 0.82 & 0.02  \\ 
 {}       & 2000    & 17.90 & 0.00 & 0.91 & 0.01  \\
 \bottomrule
   \multicolumn{6}{p{.35\textwidth}}{\footnotesize C, average correct zeros; IC, average incorrect zeros; PT, the probability of choosing the true model; MSE, the average mean squared error.}\\
\end{tabular}
\end{center}
\end{table}

\begin{table}[h!]

\begin{center}
\caption{\label{tab:parameter_inference_morecovariates_scen1}Simulation results: estimation and inference metrics for the scenario with additional covariates}%
\begin{tabular}{@{}c@{~~}c@{~~~} c@{~~}c@{~~}c@{~~}c@{~~} c@{~~~} c@{~~}c@{~~}c@{~~}c@{~~} c@{~~~} c@{~~}c@{~~}c@{~~}c@{}}
\toprule
\multicolumn{16}{l}{MPR-SIC}\\
{} & {} & \multicolumn{4}{c}{$n = 500$} && \multicolumn{4}{c}{$n = 1000$} && \multicolumn{4}{c}{$n = 2000$} \\
\cmidrule(r){3-6} \cmidrule(r){8-11} \cmidrule(){13-16}
{} & $\theta$ & $\hat{\theta}$ & SE & SEE & CP && $\hat{\theta}$ & SE & SEE & CP && $\hat{\theta}$ & SE & SEE & CP \\
\midrule
  $\beta_{0}$ & 0.0 & -0.00 & 0.06 & 0.06 & 0.92 && 0.00 & 0.04 & 0.04 & 0.94 && -0.00 & 0.03 & 0.03 & 0.95 \\
  $\beta_{1}$ & 1.0 & 1.00 & 0.04 & 0.04 & 0.93 && 1.00 & 0.03 & 0.03 & 0.94 && 1.00 & 0.02 & 0.02 & 0.95 \\ 
  $\beta_{2}$ & 0.5 & 0.50 & 0.05 & 0.05 & 0.92 && 0.50 & 0.03 & 0.03 & 0.95 && 0.50 & 0.02 & 0.02 & 0.94 \\ 
  $\beta_{3}$ & 0.5 & 0.50 & 0.03 & 0.03 & 0.93 && 0.50 & 0.02 & 0.02 & 0.94 && 0.50 & 0.01 & 0.01 & 0.95 \\ 
  $\beta_{4}$ & 1.0 & 1.00 & 0.03 & 0.03 & 0.92 && 1.00 & 0.02 & 0.02 & 0.94 && 1.00 & 0.01 & 0.01 & 0.95 \\ 
  $\beta_{5}$ & 0.5 & 0.50 & 0.03 & 0.03 & 0.93 && 0.50 & 0.02 & 0.02 & 0.94 && 0.50 & 0.01 & 0.01 & 0.95 \\ 
  $\beta_{6}$ & 1.0 & 1.00 & 0.06 & 0.05 & 0.91 && 1.00 & 0.03 & 0.03 & 0.94 && 1.00 & 0.02 & 0.02 & 0.93 \\[0.2cm]
  
  $\alpha_{0}$ & 0.0 & -0.04 & 0.11 & 0.10 & 0.91 && -0.02 & 0.07 & 0.07 & 0.95 && -0.01 & 0.05 & 0.05 & 0.95 \\
  $\alpha_{1}$ & 0.5 & 0.50 & 0.07 & 0.07 & 0.95 && 0.50 & 0.05 & 0.05 & 0.94 && 0.50 & 0.03 & 0.03 & 0.95 \\ 
  $\alpha_{2}$ & 1.0 & 1.02 & 0.08 & 0.07 & 0.92 && 1.01 & 0.05 & 0.05 & 0.93 && 1.00 & 0.03 & 0.03 & 0.94 \\ 
  $\alpha_{3}$ & 0.5 & 0.52 & 0.08 & 0.08 & 0.95 && 0.51 & 0.05 & 0.05 & 0.94 && 0.50 & 0.04 & 0.04 & 0.95 \\ 
  $\alpha_{4}$ & 1.0 & 1.02 & 0.07 & 0.07 & 0.93 && 1.01 & 0.05 & 0.05 & 0.95 && 1.00 & 0.03 & 0.03 & 0.95 \\ 
  $\alpha_{7}$ & 0.5 & 0.51 & 0.07 & 0.07 & 0.95 && 0.50 & 0.05 & 0.05 & 0.94 && 0.50 & 0.03 & 0.03 & 0.96 \\ 
  $\alpha_{8}$ & 1.0 & 1.02 & 0.07 & 0.07 & 0.93 && 1.01 & 0.05 & 0.05 & 0.94 && 1.00 & 0.03 & 0.03 & 0.95 \\
  \bottomrule
  \multicolumn{16}{p{.8\textwidth}}{\footnotesize SE, standard deviation of estimates over 1000 replications; SEE, average of estimated standard errors over 1000 replications; CP, the empirical coverage probability of a nominal 95\% confidence interval.}\\
\end{tabular}
\end{center}
\end{table}

\FloatBarrier
\subsection{Repeated Covariates}
\begin{itemize}
    \item Table~\ref{tab:true_values_morecovariates_scen2} is analogous to Table 2 of the main paper, but showing the new true values that are used to simulate data with additional covariates.
    \item Table~\ref{tab:var_selection_morecovariates_scen2} is analogous to Table 4 of the main paper, but showing the model selection metrics for the scenario with additional covariates.
    \item Table~\ref{tab:parameter_inference_morecovariates_scen2} is analogous to Table 5 of the main paper, but gives the estimation and inference metrics for the scenario with additional covariates.
\end{itemize}
\begingroup
\setlength{\tabcolsep}{4pt} 
\begin{table}[h!]
    \begin{center}
    \caption{\label{tab:true_values_morecovariates_scen2}True parameter values for the scenario with additional covariates}%
    \begin{tabular}{@{}cccccccccccccc@{}}
      \toprule
         & & \textcolor{blue}{E} & \textcolor{red}{$\text{M}_1$} & \textcolor{mygreen}{B} & N & N & \textcolor{red}{$\text{M}_1$} & N & N & \textcolor{red}{$\text{M}_1$} & \textcolor{mygreen}{B} & \textcolor{blue}{E} & \textcolor{red}{$\text{M}_1$} \\
     & $X_0$ & \textcolor{blue}{$X_1$} & \textcolor{red}{$X_2$} & \textcolor{mygreen}{$X_3$} & ${X_4}$ & ${X_5}$ & \textcolor{red}{$X_6$} & ${X_7}$ & $X_8$ & \textcolor{red}{${X_9}$} & \textcolor{mygreen}{$X_{10}$} & \textcolor{blue}{$X_{11}$} & \textcolor{red}{$X_{12}$} \\
     \midrule
    $\beta$ & 0 & 1 & 0 & 0 & 1 & 0 & 1 & 0 & 0 & 0 & 0 & 0 & 0 \\[0.05cm] 
    $\alpha$ & 0 & 0.5 & 1 & 0.5 & 1 & 0 & 0 & 0.5 & 1 & 0 & 0 & 0 & 0 \\
    \midrule
         & & \textcolor{blue}{E} & \textcolor{orange}{$\text{M}_2$} & \textcolor{mygreen}{B} & N & N & \textcolor{orange}{$\text{M}_2$} & N & N & \textcolor{orange}{$\text{M}_2$} & \textcolor{mygreen}{B} & \textcolor{blue}{E} & \textcolor{orange}{$\text{M}_2$} \\
       & {} & \textcolor{blue}{$X_{13}$} & \textcolor{orange}{${X_{14}}$} & \textcolor{mygreen}{$X_{15}$} & ${X_{16}}$ & ${X_{17}}$ & \textcolor{orange}{$X_{18}$} & ${X_{19}}$ & $X_{20}$ & \textcolor{orange}{${X_{21}}$} & \textcolor{mygreen}{$X_{22}$} & \textcolor{blue}{$X_{23}$} & \textcolor{orange}{$X_{24}$} \\
    \midrule
    $\beta$ & {} & 1 & 0 & 0 & 1 & 0 & 1 & 0 & 0 & 0 & 0 & 0 & 0 \\[0.05cm] 
    $\alpha$ & {} & 0.5 & 1 & 0.5 & 1 & 0 & 0 & 0.5 & 1 & 0 & 0 & 0 & 0 \\
    \bottomrule
     \multicolumn{14}{l}{\footnotesize \textcolor{blue}{E = Exponential}, \textcolor{mygreen}{B = Bernoulli}, N = independent normal,} \\
  \multicolumn{14}{l}{\footnotesize \textcolor{red}{$\text{M}_1$ = multivariate normal (first set of correlated covariates)},}\\
  \multicolumn{14}{l}{\footnotesize \textcolor{orange}{$\text{M}_2$ = multivariate normal (second set of correlated covariates)}.}
    \end{tabular}
    \end{center}
    \end{table}
\endgroup

\begin{table}[h!]

\begin{center}
\caption{\label{tab:var_selection_morecovariates_scen2}Simulation results: model selection metrics for the scenario with additional covariates}%
\begin{tabular}{@{}ccc@{~~}c@{~~}c@{~~}c@{}}
  \toprule
 {} & {} & \multicolumn{4}{c}{MPR-SIC}\\
 \cmidrule(l){3-6} 
 {}     & $n$   & C(12)  & IC(0) & PT & MSE \\ 
 \midrule
 $\beta$ & 500  & 11.73 & 0.00 & 0.77 & 0.01  \\
 {}      & 1000 & 11.85 & 0.00 & 0.86 & 0.00  \\
 {}      & 2000 & 11.91 & 0.00 & 0.92 & 0.00  \\[0.2cm]

 $\alpha$ & 500     & 11.78 & 0.00 & 0.81 & 0.08  \\ 
 {}       & 100     & 11.90 & 0.00 & 0.91 & 0.03  \\ 
 {}       & 2000    & 11.93 & 0.00 & 0.94 & 0.01  \\
 \bottomrule
   \multicolumn{6}{p{.35\textwidth}}{\footnotesize C, average correct zeros; IC, average incorrect zeros; PT, the probability of choosing the true model; MSE, the average mean squared error.}\\
\end{tabular}
\end{center}
\end{table}

\begin{table}[h!]

\begin{center}
\caption{\label{tab:parameter_inference_morecovariates_scen2}Simulation results: estimation and inference metrics for the scenario with additional covariates}%
\begin{tabular}{@{}c@{~~}c@{~~~} c@{~~}c@{~~}c@{~~}c@{~~} c@{~~~} c@{~~}c@{~~}c@{~~}c@{~~} c@{~~~} c@{~~}c@{~~}c@{~~}c@{}}
\toprule
\multicolumn{16}{l}{MPR-SIC}\\
{} & {} & \multicolumn{4}{c}{$n = 500$} && \multicolumn{4}{c}{$n = 1000$} && \multicolumn{4}{c}{$n = 2000$} \\
\cmidrule(r){3-6} \cmidrule(r){8-11} \cmidrule(){13-16}
{} & $\theta$ & $\hat{\theta}$ & SE & SEE & CP && $\hat{\theta}$ & SE & SEE & CP && $\hat{\theta}$ & SE & SEE & CP \\
\midrule
  $\beta_{0}$   & 0.0 & 0.00 & 0.06 & 0.05 & 0.92 && -0.00 & 0.04 & 0.04 & 0.92 && 0.00 & 0.03 & 0.02 & 0.94 \\ 
  $\beta_{1}$   & 1.0 & 1.00 & 0.04 & 0.03 & 0.93 && 1.00 & 0.02 & 0.02 & 0.93 && 1.00 & 0.01 & 0.01 & 0.93 \\ 
  $\beta_{2}$   & 0.5 & 0.50 & 0.04 & 0.04 & 0.92 && 0.50 & 0.03 & 0.02 & 0.93 && 0.50 & 0.02 & 0.02 & 0.94 \\ 
  $\beta_{3}$   & 0.5 & 0.50 & 0.03 & 0.02 & 0.90 && 0.50 & 0.02 & 0.01 & 0.94 && 0.50 & 0.01 & 0.01 & 0.93 \\ 
  $\beta_{4}$   & 1.0 & 1.00 & 0.03 & 0.02 & 0.92 && 1.00 & 0.02 & 0.02 & 0.94 && 1.00 & 0.01 & 0.01 & 0.94 \\ 
  $\beta_{5}$   & 0.5 & 0.50 & 0.03 & 0.02 & 0.94 && 0.50 & 0.02 & 0.01 & 0.93 && 0.50 & 0.01 & 0.01 & 0.95 \\ 
  $\beta_{6}$   & 1.0 & 1.00 & 0.05 & 0.04 & 0.91 && 1.00 & 0.03 & 0.02 & 0.92 && 1.00 & 0.02 & 0.02 & 0.94 \\
  $\beta_{13}$  & 1.0 & 1.00 & 0.04 & 0.03 & 0.92 && 1.00 & 0.02 & 0.02 & 0.92 && 1.00 & 0.01 & 0.01 & 0.95 \\ 
  $\beta_{14}$  & 0.5 & 0.50 & 0.05 & 0.04 & 0.92 && 0.50 & 0.03 & 0.03 & 0.93 && 0.50 & 0.02 & 0.02 & 0.93 \\ 
  $\beta_{15}$  & 0.5 & 0.50 & 0.03 & 0.02 & 0.90 && 0.50 & 0.02 & 0.01 & 0.94 && 0.50 & 0.01 & 0.01 & 0.95 \\ 
  $\beta_{16}$  & 1.0 & 1.00 & 0.03 & 0.02 & 0.91 && 1.00 & 0.02 & 0.02 & 0.93 && 1.00 & 0.01 & 0.01 & 0.95 \\ 
  $\beta_{17}$  & 0.5 & 0.50 & 0.03 & 0.02 & 0.92 && 0.50 & 0.02 & 0.01 & 0.93 && 0.50 & 0.01 & 0.01 & 0.94 \\ 
  $\beta_{18}$  & 1.0 & 1.00 & 0.05 & 0.04 & 0.92 && 1.00 & 0.03 & 0.02 & 0.92 && 1.00 & 0.02 & 0.02 & 0.94 \\[0.2cm]
  
  $\alpha_{0}$  & 0.0 & -0.08 & 0.14 & 0.13 & 0.89 && -0.04 & 0.09 & 0.09 & 0.91 && -0.02 & 0.07 & 0.06 & 0.91 \\
  $\alpha_{1}$  & 0.5 & 0.50 & 0.07 & 0.07 & 0.94 && 0.50 & 0.05 & 0.05 & 0.95 && 0.50 & 0.03 & 0.03 & 0.94 \\ 
  $\alpha_{2}$  & 1.0 & 1.02 & 0.08 & 0.07 & 0.92 && 1.01 & 0.05 & 0.05 & 0.94 && 1.01 & 0.03 & 0.03 & 0.94 \\ 
  $\alpha_{3}$  & 0.5 & 0.52 & 0.08 & 0.08 & 0.95 && 0.51 & 0.05 & 0.05 & 0.93 && 0.50 & 0.04 & 0.04 & 0.95 \\ 
  $\alpha_{4}$  & 1.0 & 1.02 & 0.07 & 0.07 & 0.92 && 1.01 & 0.05 & 0.05 & 0.94 && 1.00 & 0.03 & 0.03 & 0.94 \\ 
  $\alpha_{7}$  & 0.5 & 0.51 & 0.07 & 0.07 & 0.94 && 0.51 & 0.05 & 0.05 & 0.95 && 0.50 & 0.03 & 0.03 & 0.95 \\ 
  $\alpha_{8}$  & 1.0 & 1.03 & 0.07 & 0.07 & 0.92 && 1.01 & 0.05 & 0.05 & 0.95 && 1.01 & 0.03 & 0.03 & 0.95 \\
  $\alpha_{13}$ & 0.5 & 0.50 & 0.07 & 0.07 & 0.94 && 0.50 & 0.04 & 0.05 & 0.96 && 0.50 & 0.03 & 0.03 & 0.95 \\ 
  $\alpha_{14}$ & 1.0 & 1.02 & 0.08 & 0.07 & 0.90 && 1.01 & 0.05 & 0.05 & 0.94 && 1.01 & 0.03 & 0.03 & 0.94 \\ 
  $\alpha_{15}$ & 0.5 & 0.52 & 0.08 & 0.08 & 0.92 && 0.51 & 0.05 & 0.05 & 0.93 && 0.51 & 0.04 & 0.04 & 0.94 \\ 
  $\alpha_{16}$ & 1.0 & 1.03 & 0.07 & 0.07 & 0.92 && 1.01 & 0.05 & 0.05 & 0.93 && 1.00 & 0.03 & 0.03 & 0.95 \\ 
  $\alpha_{19}$ & 0.5 & 0.51 & 0.07 & 0.07 & 0.93 && 0.51 & 0.05 & 0.05 & 0.95 && 0.50 & 0.03 & 0.03 & 0.94 \\ 
  $\alpha_{20}$ & 1.0 & 1.03 & 0.07 & 0.07 & 0.92 && 1.01 & 0.05 & 0.05 & 0.93 && 1.01 & 0.03 & 0.03 & 0.94 \\
  \bottomrule
  \multicolumn{16}{p{.8\textwidth}}{\footnotesize SE, standard deviation of estimates over 1000 replications; SEE, average of estimated standard errors over 1000 replications; CP, the empirical coverage probability of a nominal 95\% confidence interval.}\\
\end{tabular}
\end{center}
\end{table}

\clearpage

\section{Real Data Analyses: Additional Results}\label{app:real_data_analyses}
This section contains additional results for the three real data analyses, where the standard errors and change in BIC are included for the BAMLSS, SPR-SIC and ALASSO-IC methods. (Note that BAMLSS does not produce standard errors).
\begin{itemize}
    \item Table~\ref{tab:dataset_estimates_app_pcancer} is analogous to Table 7 of the main paper for the prostate cancer data.
    \item Table~\ref{tab:dataset_estimates_app_sniffer} is analogous to Table 8 of the main paper for the sniffer data.
    \item Table~\ref{tab:dataset_estimates_app_hprice} is analogous to Table 9 of the main paper for the Boston house price data.
\end{itemize}

\begin{table}[h!]
\begin{center}
\caption{\label{tab:dataset_estimates_app_pcancer}Prostate Cancer Data: estimation metrics}%
\resizebox{\textwidth}{!}{
\begin{tabular}{@{}l@{~} >{}r@{~}>{}r@{~}>{}r@{~}>{}r@{}  c@{~}  r@{}c@{}r@{} r@{}c@{} c@{~} r@{}c@{}r@{} r@{}c@{}@{}} %
\toprule
{} & \multicolumn{4}{c}{{BAMLSS}} && \multicolumn{5}{c}{SPR-SIC} && \multicolumn{5}{c}{ALASSO-IC}\\
\cmidrule(){2-5} \cmidrule(){7-11} \cmidrule(){13-17}
{} & $\hat\beta_j$ & $\Delta\text{BIC}$ & $\hat\alpha_j$ & $\Delta\text{BIC}$ && \multicolumn{2}{c}{$\hat\beta_j$} & $\Delta\text{BIC}$ & \multicolumn{2}{c}{$\hat\alpha_j$} && \multicolumn{2}{c}{$\hat\beta_j$} & $\Delta\text{BIC}$ & \multicolumn{2}{c}{$\hat\alpha_j$} \\
\midrule
  \texttt{inter}   &-0.79         &       & 2.66           &       && -0.78          & (0.61) &        & -0.73 & (0.14)  && -0.27          & (0.63) &        & -0.68 & (0.15)  \\ 
  \texttt{lcavol}  &\textbf{0.52} & 35.56 & -0.19          &       && \textbf{0.53}  & (0.07) & 36.71  &       &         && \textbf{0.54}  & (0.08) & 36.71  &       &         \\ 
  \texttt{lweight} &\textbf{0.81} & 11.82 & \textbf{-0.93} & -1.23 && \textbf{0.66}  & (0.17) & 9.21   &       &         && \textbf{0.52}  & (0.18) & 9.21   &       &         \\ 
  \texttt{svi}     &\textbf{0.73} & 5.09  & 0.77           &       && \textbf{0.67}  & (0.20) & 5.64   &       &         && \textbf{0.53}  & (0.21) & 5.64   &       &         \\
  \texttt{age}     &-0.01         &       & 0.02           &       &&                &        &        &       &         &&                &        &        &       &         \\
  \texttt{lbph}    &0.06          &       & 0.05           &       &&                &        &        &       &         &&                &        &        &       &         \\
  \texttt{lcp}     &-0.16         &       & 0.45           &       &&                &        &        &       &         &&                &        &        &       &         \\
  \texttt{gleason} &0.02          &       & -0.12          &       &&                &        &        &       &         &&                &        &        &       &         \\
  \texttt{pgg45}   &0.01          &       & -0.01          &       &&                &        &        &       &         &&                &        &        &       &         \\
  \bottomrule
    \multicolumn{16}{p{.90\textwidth}}{\footnotesize Significant effects indicated in bold.}\\
\end{tabular}}
\end{center}
\end{table}

\begin{table}[h!]
    \begin{center}
    \caption{\label{tab:dataset_estimates_app_sniffer}Sniffer Data: estimation metrics}%
    \resizebox{\textwidth}{!}{
    \begin{tabular}{@{}l@{~} >{}r@{~}>{}r@{~}>{}r@{~}>{}r@{}  c@{~}  r@{}c@{}r@{} r@{}c@{} c@{~} r@{}c@{}r@{} r@{}c@{}@{}} %
    \toprule
    {} & \multicolumn{4}{c}{{BAMLSS}} && \multicolumn{5}{c}{SPR-SIC} && \multicolumn{5}{c}{ALASSO-IC}\\
    \cmidrule(){2-5} \cmidrule(){7-11} \cmidrule(){13-17}
    {} & $\hat\beta_j$ & $\Delta\text{BIC}$ & $\hat\alpha_j$ & $\Delta\text{BIC}$ && \multicolumn{2}{c}{$\hat\beta_j$} & $\Delta\text{BIC}$ & \multicolumn{2}{c}{$\hat\alpha_j$} && \multicolumn{2}{c}{$\hat\beta_j$} & $\Delta\text{BIC}$ & \multicolumn{2}{c}{$\hat\alpha_j$} \\
    \midrule
        \texttt{inter}      &   -1.20         &       & -0.96          &       && 0.45           & (1.01)  &       & 2.01 & (0.13) && 0.21             & (1.04) &        & 2.03  & (0.13)  \\ 
        \texttt{gaspres}    &   3.34          &       & \textbf{-3.46} & -1.51 && \textbf{10.84} & (1.51)  & 38.43 &      &        && \textbf{9.79}    & (1.63) & 28.45  &       &       \\ 
        \texttt{gastemp}    &   \textbf{0.26} & 70.75 & \textbf{0.09}  & -4.81 && \textbf{0.15}  & (0.04)  & 12.97 &      &        && \textbf{0.19}    & (0.04) & 15.37  &       &       \\ 
        \texttt{tanktemp}   &   \textbf{-0.15}& 8.45  & 0.01           &       &&                &         &       &      &        && -0.07            & (0.05) & -2.00  &       &       \\ 
        \texttt{tankpres}   &   2.69          &       & \textbf{2.72}  & -3.24 && \textbf{-5.73} & (1.23)  & 15.33 &      &        && \textbf{-4.08}   & (1.58) & 1.71   &       &       \\
        \bottomrule
        \multicolumn{16}{p{.90\textwidth}}{\footnotesize Significant effects indicated in bold.}\\
\end{tabular}}
\end{center}
\end{table}

\begin{table}[h!]
    \begin{center}
    \caption{\label{tab:dataset_estimates_app_hprice}Boston House Price Data: estimation metrics}%
    \resizebox{\textwidth}{!}{
    \begin{tabular}{@{}l@{~} >{}r@{~}>{}r@{~}>{}r@{~}>{}r@{}  c@{~}  r@{}c@{}r@{} r@{}c@{} c@{~} r@{}c@{}r@{} r@{}c@{}@{}} %
    \toprule
    {} & \multicolumn{4}{c}{{BAMLSS}} && \multicolumn{5}{c}{SPR-SIC} && \multicolumn{5}{c}{ALASSO-IC}\\
    \cmidrule(){2-5} \cmidrule(){7-11} \cmidrule(){13-17}
    {} & $\hat\beta_j$ & $\Delta\text{BIC}$ & $\hat\alpha_j$ & $\Delta\text{BIC}$ && \multicolumn{2}{c}{$\hat\beta_j$} & $\Delta\text{BIC}$ & \multicolumn{2}{c}{$\hat\alpha_j$} && \multicolumn{2}{c}{$\hat\beta_j$} & $\Delta\text{BIC}$ & \multicolumn{2}{c}{$\hat\alpha_j$} \\
    \midrule
        \texttt{inter}      & 10.51          &         & -2.39          &       && 13.26          & (0.36) &        & -3.30 & (0.06)  && 13.18          & (0.36) &        & -3.28 & (0.06)  \\     
        \texttt{rooms}      & \textbf{0.26}  & 160.18  & \textbf{-0.20} & -2.52 && \textbf{0.10}  & (0.02) & 29.60  &       &         && \textbf{0.10}  & (0.02) & 29.45  &       &       \\
        \texttt{lowstat}    & \textbf{-0.02} & 94.40   & \textbf{0.03}  & -1.26 && \textbf{-0.03} & (0.00) & 193.18 &       &         && \textbf{-0.03} & (0.00) & 193.02 &       &       \\
        \texttt{stratio}    & \textbf{-0.02} & 53.39   & -0.01          &       && \textbf{-0.04} & (0.00) & 55.96  &       &         && \textbf{-0.04} & (0.00) & 55.81  &       &       \\
        \texttt{lproptax}   & \textbf{-0.16} & 40.46   & 0.45           &       && \textbf{-0.26} & (0.05) & 24.78  &       &         && \textbf{-0.25} & (0.05) & 24.62  &       &       \\
        \texttt{ldist}      & \textbf{-0.11} & 35.59   & -1.21          & 25.06 && \textbf{-0.28} & (0.03) & 62.36  &       &         && \textbf{-0.27} & (0.03) & 62.20  &       &       \\
        \texttt{crime}      & \textbf{-0.01} & 25.81   & -0.01          &       && \textbf{-0.01} & (0.00) & 76.71  &       &         && \textbf{-0.01} & (0.00) & 76.55  &       &       \\
        \texttt{lnox}       & \textbf{-0.28} & 18.01   & -1.19          &       && \textbf{-0.62} & (0.09) & 34.87  &       &         && \textbf{-0.60} & (0.10) & 34.72  &       &       \\
        \texttt{radial}     & \textbf{0.00}  & 10.76   & \textbf{0.05}  & 21.61 && \textbf{0.01}  & (0.00) & 24.54  &       &         && \textbf{0.01}  & (0.00) & 24.38  &       &       \\
        \bottomrule
        \multicolumn{16}{p{.90\textwidth}}{\footnotesize Significant effects indicated in bold.}\\
    \end{tabular}}
    \end{center}
    \end{table}

\end{document}